\documentclass[aps,physrev,groupedaddress]{revtex4-2}
\usepackage{graphicx} 
\usepackage{caption}
\usepackage{subcaption}
\usepackage{amssymb, amsthm, amsmath}
\usepackage{braket}
\usepackage{multirow}
\usepackage{dcolumn}
\usepackage{bm}
\usepackage{bbm}
\usepackage{natbib}
\usepackage{url}
\usepackage{comment}
\usepackage{hyperref}
\begin{document}
\title{Exceptional point structure and chiral state transfer in a rate-operator-engineered $\mathcal{PT}$-symmetric Bose-Hubbard dimer}
\author{Jaakko Luomala}
\email{jaakko.s.luomala@utu.fi}
\affiliation{
Department of Physics and Astronomy,
University of Turku,
FI-20014 Turun yliopisto,
Finland
}

\author{Federico Settimo}
\affiliation{
Department of Physics and Astronomy,
University of Turku,
FI-20014 Turun yliopisto,
Finland
}

\author{Kimmo Luoma}
\affiliation{
Department of Physics and Astronomy,
University of Turku,
FI-20014 Turun yliopisto,
Finland
}

\author{Iiro Vilja}
\affiliation{
Department of Physics and Astronomy,
University of Turku,
FI-20014 Turun yliopisto,
Finland
}

\author{Jyrki Piilo}
\affiliation{
Department of Physics and Astronomy,
University of Turku,
FI-20014 Turun yliopisto,
Finland
}

\begin{abstract}
The rate operator quantum jumps (ROQJ) is a general method to unravel a master equation describing the dynamics of an open quantum system into stochastic pure state trajectories. By post-selecting on no jumps occurring one obtains a trajectory generated by a non-Hermitian Hamiltonian. The ROQJ allows for versatile modification of this non-Hermitian evolution, with each non-Hermitian Hamiltonian corresponding to a different continuous measurement scheme. We apply the ROQJ to modify the non-Hermitian dynamics of the open three-particle Bose-Hubbard dimer model under environmental non-local dephasing to construct a $\mathcal{PT}$-symmetric non-Hermitian Bose-Hubbard Hamiltonian with a general two parameter $\mathcal{PT}$-symmetric imaginary diagonal part, which generalizes the well-known gain-loss Bose-Hubbard dimer. We characterize the symmetry structure of the spectrum for different on-site interaction strengths. For vanishing on-site interaction strength we find the unbroken $\mathcal{PT}$-symmetry region enclosed by order two exceptional lines of different character and four order four exceptional points, two of which are not found in the usual gain-loss Bose-Hubbard dimer. Additionally, we find regions with degenerate imaginary parts of the eigenvalues, also not found in the gain-loss Bose-Hubbard dimer. In this parameter space, we use time-dependent rate operator transformations to investigate state transfer under slowly varying parameter loops. Loops encircling an order four exceptional point exhibit chiral state transfer. Because the loops cross exceptional lines, the final states are superpositions of eigenstates. Chiral state transfer is also found with non-zero homogeneous on-site interaction when the trajectory crosses over broken $\mathcal{PT}$-symmetry regions of different character. Using inhomogeneous on-site interactions, we find direction dependent state transfer between eigenstates without crossing or encircling exceptional points. 
\end{abstract}

\maketitle

\section{Introduction \label{sec-introduction}}

Non-Hermitian Hamiltonian quantum dynamics can be realized by postselecting dynamics of open quantum systems (OQS) under continuous measurement \cite{Ashida2020}. In the theory of OQS's the quantum system of interest $S$ defined in Hilbert space $\mathcal{H}_S$ interacts with environment $E$ on Hilbert space $\mathcal{H}_E$. Thus, the full quantum system is a composite one with product space $\mathcal{H}_S \otimes \mathcal{H}_E$. The goal in the theory of OQS's is to derive an equation of motion for the system $S$, usually in the form of a master equation (ME) for the reduced density matrix $\rho_S$ and study its physical properties. By tracing out the degrees-of-freedom of the environment, different master equations governing the dynamics of the density matrix $\rho_S$ have been derived under various assumptions \cite{Breuer2007,vacchini2024open}.

By performing a continuous measurement in a suitable way on the system or environment, one ends up with a stochastic pure state evolution for the system $S$ governed by some effective non-Hermitian Hamiltonian $K$ with stochastic quantum jumps interrupting the deterministic evolution \cite{dalibard_wave-function_1992, gardiner_wave-function_1992, molmer_monte_1993, Plenio1998, settimo_quantum_2026}. By post-selecting the measurement sequence with no quantum jumps, one gets a deterministic state evolution governed by the effective non-Hermitian Hamiltonian $K$. In this article, we assume that the system dynamics is governed by a quantum dynamical semigroup, such that the full unconditional dynamics of the system density matrix is described by a time-independent Markovian Gorini-Kossakowski-Sudarshan-Lindblad (GKSL) ME \cite{Gorini1976, Lindblad1976}, and we focus on deterministic pure state evolutions given by an unraveling of the GKSL equation called the rate operator quantum jumps (ROQJ) \cite{Smirne2020,Chruciski2022,Settimo2024,Luomala2026}. We use the ROQJ method because of its flexibility, allowing one to engineer vastly different jump trajectories and non-Hermitian dynamics by changing the continuous measurement scheme.

Time-dependent $\mathcal{PT}$-symmetric Hamiltonians with parameter loops encircling exceptional points, points in the Hamiltonian's parameter space in which two or more eigenstates coalesce, have been subject to much attention in recent years \cite{Uzdin2011,Khandelwal2024,Kumar2026,Svegborn2026,Wu2026}. It has been found that by varying parameters slowly around a loop, the final state of the evolution depends on the direction in which the loop was traversed. Other studies have found that the presence of exceptional points is not necessary for this chirality to emerge \cite{Hassan2017,Svegborn2026}. Typically, the evolution takes one eigenstate of the initial Hamiltonian to another eigenstate or back to the initial eigenstate. For research including quantum jumps in the evolution see e.g. \cite{Khandelwal2024,Svegborn2026}.

In \cite{Luomala2026} some of the current authors applied the ROQJ formalism to modify time-independent three-site and few-particle Bose-Hubbard Hamiltonians under local and non-local environmental dephasing and investigated the effects of rate operator transformations on static properties such as the steady state of non-Hermitian dynamics and the steady state phase structure. Here, we expand it to time-dependent rate operator transformations and focus on their dynamical consequences. We apply the ROQJ formalism to the Bose-Hubbard model \cite{Gersch1963,Jaksch2005} with two sites and three particles under non-local environmental dephasing. We use the rate operator transformations to build $\mathcal{PT}$-symmetric non-Hermitian Hamiltonians with non-Hermitian diagonal part of more general form than non-Hermitian Bose-Hubbard Hamiltonians with $\mathcal{PT}$-symmetric imaginary potentials \cite{Graefe2008a,Jin2013}. For vanishing on-site interaction, we study parameter loops that encircle a higher order exceptional point and loops that do not, while also crossing lower order exceptional points. We also investigate parameter loops in the case of non-zero homogenous on-site interaction, when higher order exceptional points no longer exist. Additionally, we construct non-Hermitian Hamiltonians in which the $\mathcal{PT}$-symmetry is broken by inhomogeneous on-site interaction and focus on one-dimensional parameter trajectories in this case. Traversal direction dependent state transfer is found in all cases.

In Section \ref{sec-background} we present the background to our research including non-Hermitian dynamics, the open Bose-Hubbard model, and the ROQJ formalism. Section \ref{sec-dephasing} is devoted to the results of the application of the rate operator transformation on the open Bose-Hubbard model with the dynamics separated in three cases, zero on-site interaction, non-zero homogeneous on-site interaction and non-zero inhomogeneous on-site interaction. Finally, Section \ref{CandD} is devoted to the discussion on the results.

\section{Background \label{sec-background}}

We start by presenting a short review of concepts related to non-Hermitian quantum mechanics relevant to this article. In Section \ref{sec-non-herm} we recall some key elements of non-Hermitian dynamics, then in Section \ref{sec-bose-hubbard} we introduce the Bose-Hubbard model under study, and in \ref{sec-RO} we recall the rate operator formalism \cite{Smirne2020,Chruciski2022,Settimo2024}.

\subsection{Non-Hermitian dynamics, $\mathcal{PT}$-symmetry, exceptional points and state control \label{sec-non-herm}}

We suppose that the system density matrix $\rho_S(t)$ is governed by the time independent GKSL equation
\begin{align}
    \mathcal{L}[\rho_S(t)]=-i[H,\rho_S(t)]+\sum_{i=1}^{d^2-1}\left[\gamma_iL_i\rho_S(t) L_i^\dagger-\frac{1}{2}\gamma_i\left(L_i^\dagger L_i\rho_S(t) + \rho_S(t) L_i^\dagger L_i\right)\right], \label{eq-Lindblad-ME}
\end{align}
where $d$ is the dimension of $\mathcal{H}_S$ and the rates $\gamma_i$ are non-negative constants, $\gamma_i\ge0$, and we have set $\hbar=1$. 

In the present work we focus on the interpretation of non-Hermitian dynamics as originating from post-selecting no-jump trajectories of an open quantum system under continuous measurement. The system evolves deterministically under Hamiltonian dynamics governed by a non-Hermitian Hamiltonian $K$, with the norm of the state vector determining the probability of observing such trajectory. Non-Hermitian dynamics have been shown to have useful properties for quantum technologies such as state engineering \cite{Li2023,Teixeira2023} and quantum sensing \cite{Wiersig2020}. Therefore new ways to create non-Hermitian dynamics are useful for advancing these technologies.

A specific subset of non-Hermitian Hamiltonians that have been studied in large extent in past decades are $\mathcal{PT}$-symmetric ones. These Hamiltonians present unconventional physical properties near symmetry breaking \cite{Bender1998,Ashida2020}. While other symmetries can also lead to similar properties, $\mathcal{PT}$-symmetric systems have been found generally to be simplest to realize experimentally. The $\mathcal{PT}$-operation is a combined operation of a parity operation $\mathcal{P}$ and time reversal operation $\mathcal{T}$ that acts as complex conjugation $i\mapsto -i$. The parity operator $\mathcal{P}$ is usually defined as the quantum mechanical parity operator that acts as $\mathcal{P}(a)=-a$ and $\mathcal{P}(a^\dagger)=-a^\dagger$ for annihilation operator $a$ and creation operator $a^\dagger$, or as a permutation matrix in Fock basis such that $\mathcal{P}(\ket{i_1,i_2,...,i_n})=\ket{i_n,i_{n-1},...,i_1}$, or as a combination of both. In this article we focus on a two-site lattice and define the parity operation as the permutation $\mathcal{P}(\ket{ij})=\ket{ji}$, such that for any local operator $O_1$ that acts on site 1 and any local operator $O_2$ that acts on site 2, the parity operation $P$ leads to mapping $O_1\mapsto O_2$ and $O_2\mapsto O_1$. 

$\mathcal{PT}$-symmetric Hamiltonians have an interesting feature that their eigenvalues are real when the $\mathcal{PT}$-symmetry is unbroken. From the dynamical aspect this means that the Hamiltonian has no unique steady state, since the imaginary parts of the eigenvalues of the Hamiltonian are all zero and thus equal. Moreover, the unbroken and broken symmetry regimes of the parameter space of the Hamiltonian may be separated by exceptional points (EP), in which two or more eigenvalues and their corresponding eigenstates coalesce. EPs have dramatic effects on many aspects of quantum dynamics and have further been found to have applications in quantum technologies \cite{Nasari2022,Yang2026}. Recently, a significant amount of studies have investigated chiral state transfer resulting from time-dependent trajectories in the parameter space of a non-Hermitian Hamiltonian with EPs \cite{Uzdin2011,Khandelwal2024,Kumar2026,Svegborn2026,Wu2026}, although the presence of EPs has been found to not be necessary for chiral state transfer \cite{Hassan2017,Svegborn2026}. In chiral state transfer the final state after performing a slow (in comparison to the timescales of the system dynamics) loop in parameter space depends on the direction of the traversal around the loop but not, usually, on the initial state. Contributions leading to state transfer in closed loops include adiabatic contribution, non-adiabatic contribution and perturbations and they are discussed e.g. in \cite{Svegborn2026,Kumar2026}. A common method to generate time-dependent non-Hermitian Hamiltonians is to use time-dependent rates $\gamma_i(t)$. Here we instead use time-dependent rate operator transformations while keeping the ME time independent.

\subsection{Open Bose-Hubbard model\label{sec-bose-hubbard}}

For simplicity, we focus on the 2-site Bose-Hubbard model with particle number $N=3$ and assume there is no potential difference between the two sites, so that the potential term $\omega(n_1+n_2)$, $\omega\in \mathbb{R}$, is omitted. The Hamiltonian of the system can then be written as \cite{Gersch1963}
\begin{align}
    H=-J(a_1^\dagger a_2 + a_2^\dagger a_1) + \frac{U}{2}(a_1^\dagger a_1^\dagger a_1 a_1 + a_2^\dagger a_2^\dagger a_2 a_2),
\end{align}
where the $a_i$ and $a_i^\dagger$ are the bosonic annihilation and creation operators, respectively. Hereafter we fix $J=1$ and all other parameters and time are given in units of $J$. The open Bose-Hubbard model can be described using the GKSL equation with suitable jump operators $L_i$ and rates $\gamma_i$. The Bose-Hubbard model has a simple structure, while still presenting complex dynamics originating from the non-linear interaction. It also describes many experimentally realizable systems with considerable control over the model's parameters, such as cold atoms and transmon arrays. For these reasons the open Bose-Hubbard model has been investigated extensively \cite{Ruostekoski1998,Witthaut2008,Witthaut2009,Kordas2012,Kordas2013,Kordas2015,Busel2023} (for a review on the open Bose-Hubbard model see eg. \cite{Kordas2015}). Related non-Hermitian extensions of the Bose-Hubbard model have also been the object of many studies \cite{Graefe2008a,Graefe2008b,Graefe2010,Jin2013,Znojil2019}. $\mathcal{PT}$-symmetric non-Hermitian Bose-Hubbard Hamiltonians have been studied in \cite{Graefe2008a,Jin2013,Znojil2019}.

\subsection{Rate operator quantum jumps \label{sec-RO}}

The rate operator quantum jumps unravels the GKSL ME (\ref{eq-Lindblad-ME}) into stochastic pure state evolutions consisting of deterministic evolution governed by a non-Hermitian Hamiltonian interrupted by sudden discontinuous jumps \cite{Smirne2020,Chruciski2022,Settimo2024}. The power of the formalism relies on the vast amount of different unravelings it contains. To describe these unravelings we first introduce the un-transformed scheme in which the stochastic jumps are performed to the eigenstates $\ket{\phi_i}$ of the rate operator
\begin{align}
    \mathcal J[\psi]=\sum_{i=1}^{d^2-1} \gamma_i L_i\ket{\psi}\bra{\psi}L_i^\dagger
\end{align}
\noindent with probabilities
\begin{align}\label{eq:p_i_RO}
    p_i = \lambda_idt
\end{align}
during time interval $dt$, where $\lambda_i$ is the eigenvalue corresponding to the eigenstate $\ket{\phi_i}$. In the case of negative eigenvalues, one can still unravel the ME using the reverse jump method \cite{Piilo2008, Piilo2009, Settimo-SSE-RO}. The deterministic evolution is governed by the effective Hamiltonian
\begin{align}
    K = H - \frac{i}{2}\sum_{i=1}^{d^2-1}\gamma_iL_i^\dagger L_i.
\end{align}

It has been proven \cite{Settimo2024}, that one still gets a valid unraveling of the ME, when rate operator $R$ and the effective Hamiltonian $K$ are transformed as
\begin{align}
    R_{t,\psi} & = \mathcal J[\psi] + \frac{1}{2} \left(C_{t,\psi}\ket{\psi}\bra{\psi}+\ket{\psi}\bra{\psi}C_{t,\psi}^\dagger\right), \label{eq-Rate-op-transformation} \\
    K_{t,\psi} & = K - \frac{i}{2}C_{t,\psi}, \label{eq-effective-H-transformation}
\end{align}
where the possibly time and state dependent operator $C_{t,\psi}$ can take any form. For an interpretation of the trajectories as originating from a continuous measurement scheme, it is required that the transformation $C_{t,\psi}$ keeps the rate operator $R_{t,\psi}$ positive semi-definite, so that the eigenvalues define proper probabilities. In this article we condition the evolution on no jumps happening. The state is evolved using the non-unitary evolution operator $\exp(-idtK_{t,\psi})$; for small $dt$ it can be expanded as $\ket{\psi(t+dt)}=(I-idtK_{t,\psi})\ket{\psi(t)}$ and then normalized. The value of $dt$ used in this work is $dt=10^{-4}$.

In the next section, we study the rate operator transformations on the open Bose-Hubbard model, in particular environmental dephasing, and describe the structure of the eigenspectrum of the non-Hermitian Hamiltonian in the parameter space of $\mathcal{PT}$-symmetric rate operator transformations for different interaction profiles.

\section{Results \label{sec-dephasing}}

We focus in this section on a non-local dephasing mechanism given by jump operators
\begin{align}
    L_{ij,kl}=\ket{ij}\bra{ij}-\ket{kl}\bra{kl} \label{dephasing-ops}
\end{align}
written in the number basis. In general the set of jump operators includes all pairs $(ij,\,kl)$ from the set of unordered pairs of number states relevant to the Hilbert space in question. In this article, the relevant number states are the three particle states $\{\ket{30},\,\ket{21},\,\ket{12},\,\ket{03}\}$ and the jump operators include the six unordered pairs of these states. We also set the rates to be equal, $\gamma_{ij,kl}\equiv\gamma=0.1$. We use rate operator transformations of the form
\begin{align}
    C=l\,L_{30,03} + k\,L_{21,12} + hI, \label{eq-dephasing-transformation}
\end{align}
where the parameters $k$, $l$ and $h$ are real, so that the non-Hermitian Hamiltonian after the transformation is $\mathcal{PT}$-symmetric up to irrelevant terms proportional to the identity. The term $hI$ is only included to ensure non-negativity of the rate operator and thus allow a continuous measurement interpretation for the rate operator formalism. We note that for each pair $(l,\,k)$ such value of $h$ can be found and in the above transformation the minimal such value is $h=(|l|^2+|k|^2)/(8\gamma)$. The non-Hermitian Hamiltonian can then be written as
\begin{align}
    K=-J(a_1^\dagger a_2 + a_2^\dagger a_1) + \frac{U}{2}(a_1^\dagger a_1^\dagger a_1 a_1 + a_2^\dagger a_2^\dagger a_2 a_2) - \frac{i}{2}l(\ket{30}\bra{30}-\ket{03}\bra{03}) - \frac{i}{2}k(\ket{21}\bra{21}-\ket{12}\bra{12}) - \frac{i}{2} pI, \label{eq-dephasing-hamiltonian}
\end{align}
\noindent where $p\in \mathbb{R}$ includes contributions from the term $\gamma\sum_{ij,kl} L_{ij,kl}^\dagger L_{ij,kl}\propto I$ (for equal rates $\gamma_{ij,kl}$) and the term $hI$ in the transformation $C$. In the following the last term is ignored as it does not affect the dynamics or conclusions from the plotted spectrum of the $\mathcal{PT}$-symmetric part of the effective Hamiltonian $K$. The effective Hamiltonian in Eq. (\ref{eq-dephasing-hamiltonian}) can be seen as a generalization, for $N=3$, of the much studied $\mathcal{PT}$-symmetric gain-loss Bose-Hubbard dimer Hamiltonian \cite{Graefe2008a,Jin2013,Znojil2019}
\begin{align}
    K'=-J(a_1^\dagger a_2 + a_2^\dagger a_1) + \frac{U}{2}(a_1^\dagger a_1^\dagger a_1 a_1 + a_2^\dagger a_2^\dagger a_2 a_2) - \frac{i}{2}\gamma(n_1-n_2), \label{eq-reference-dissipative-hamiltonian}
\end{align}
where the last term can be seen as originating from a GKSL equation with, for example, one jump operator $L=a_1$ with rate $2\gamma$. Then the effective Hamiltonian (\ref{eq-reference-dissipative-hamiltonian}) is constructed by scaling the Hamiltonian by an appropriate imaginary scalar, assuming that the state is in a fixed total particle number sector. Similar $\mathcal{PT}$-symmetric non-Hermitian Hamiltonians to \eqref{eq-dephasing-hamiltonian} can be constructed for all particle numbers $N$, but this results to a large number of parameters to control so for simplicity we focus only on the case $N=3$.

\subsection{Symmetry structure \label{sec-dephasing-symmetry}}

\begin{figure}[t]
    \includegraphics[width=0.3\linewidth]{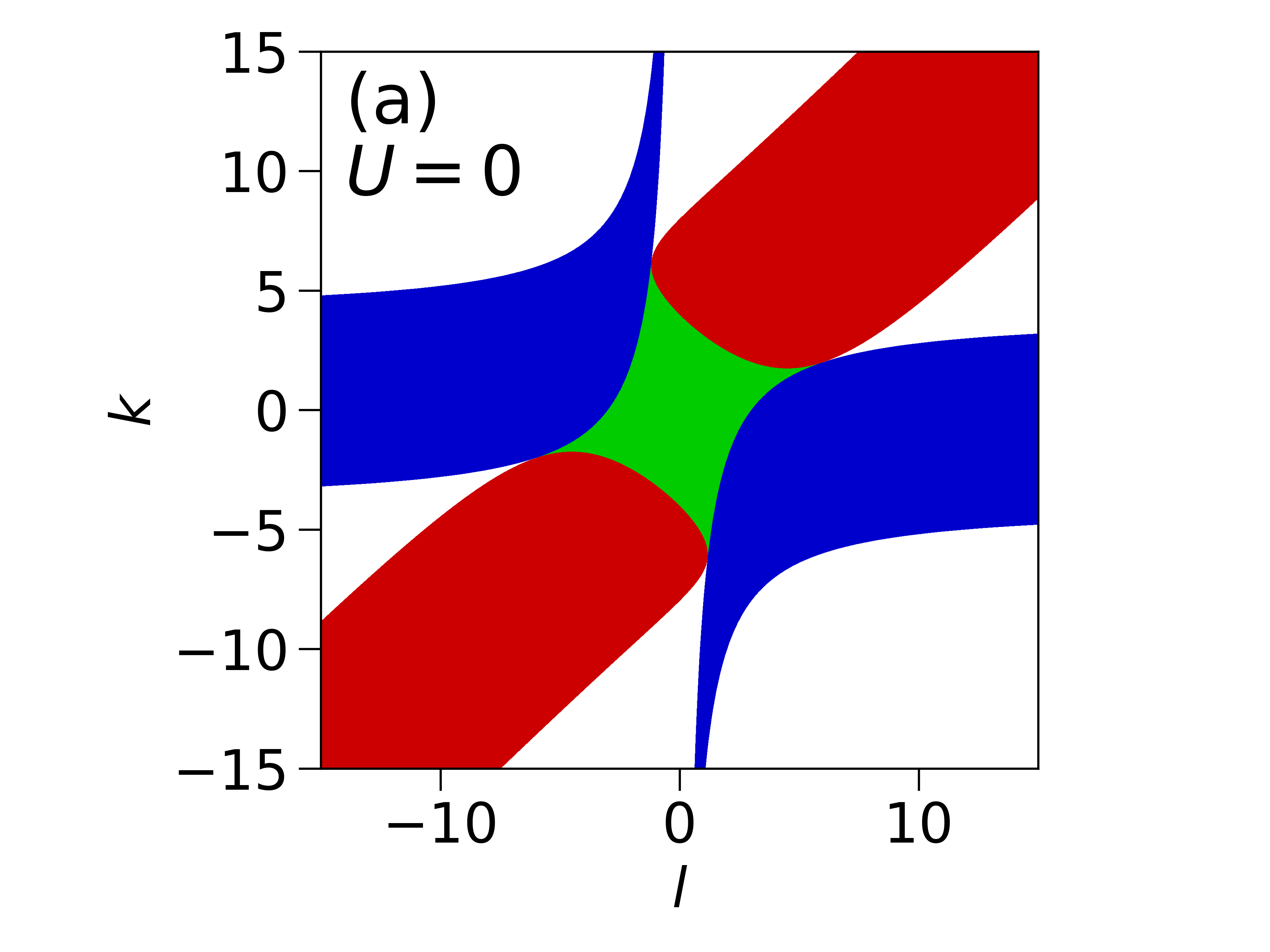}
    \includegraphics[width=0.3\linewidth]{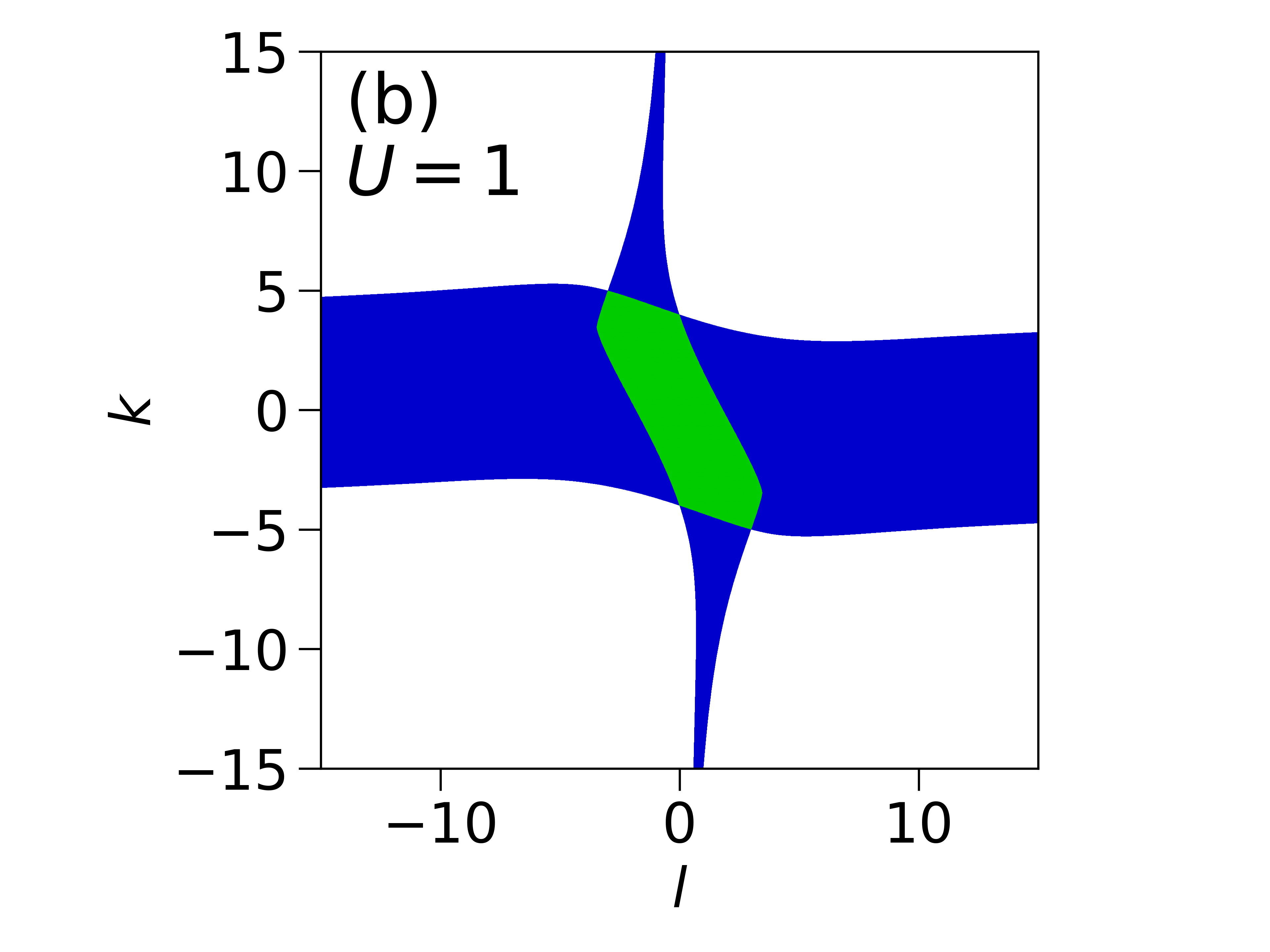}
    \includegraphics[width=0.3\linewidth]{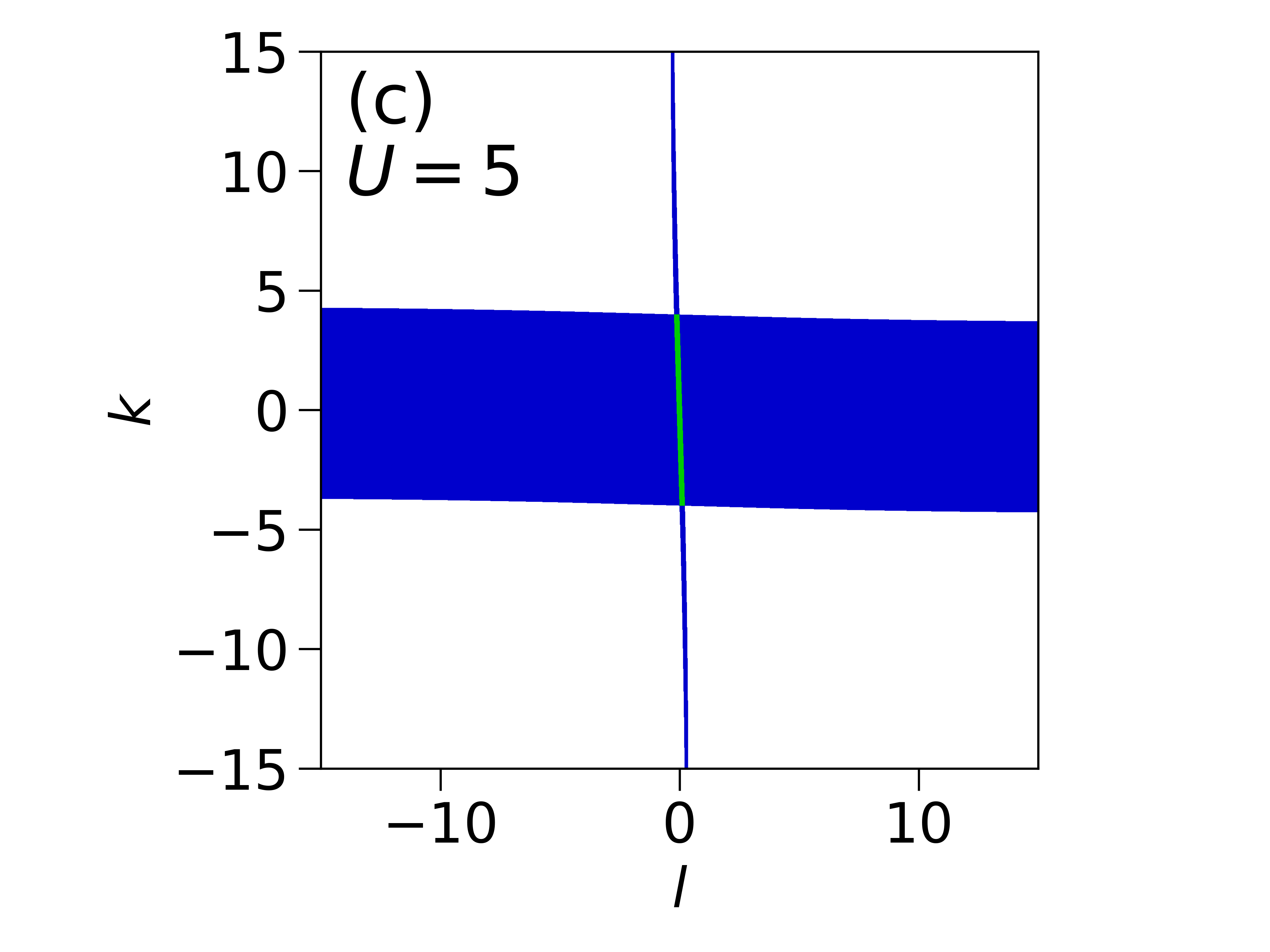}
    \caption{Regions with four (green), two (blue) and no $\mathcal{PT}$-symmetric eigenstates (white) of the non-Hermitian Hamiltonian \eqref{eq-dephasing-hamiltonian}. Red regions show where the effective Hamiltonian has two eigenvalue pairs with equal imaginary parts. On-site interaction strengths are in Figs. (a-c) $U=0$, $U=1$ and $U=5$, respectively.}
    \label{fig-dephasing-phasemaps-bin}
\end{figure}
\begin{figure}[t]
    \includegraphics[width=0.3\linewidth]{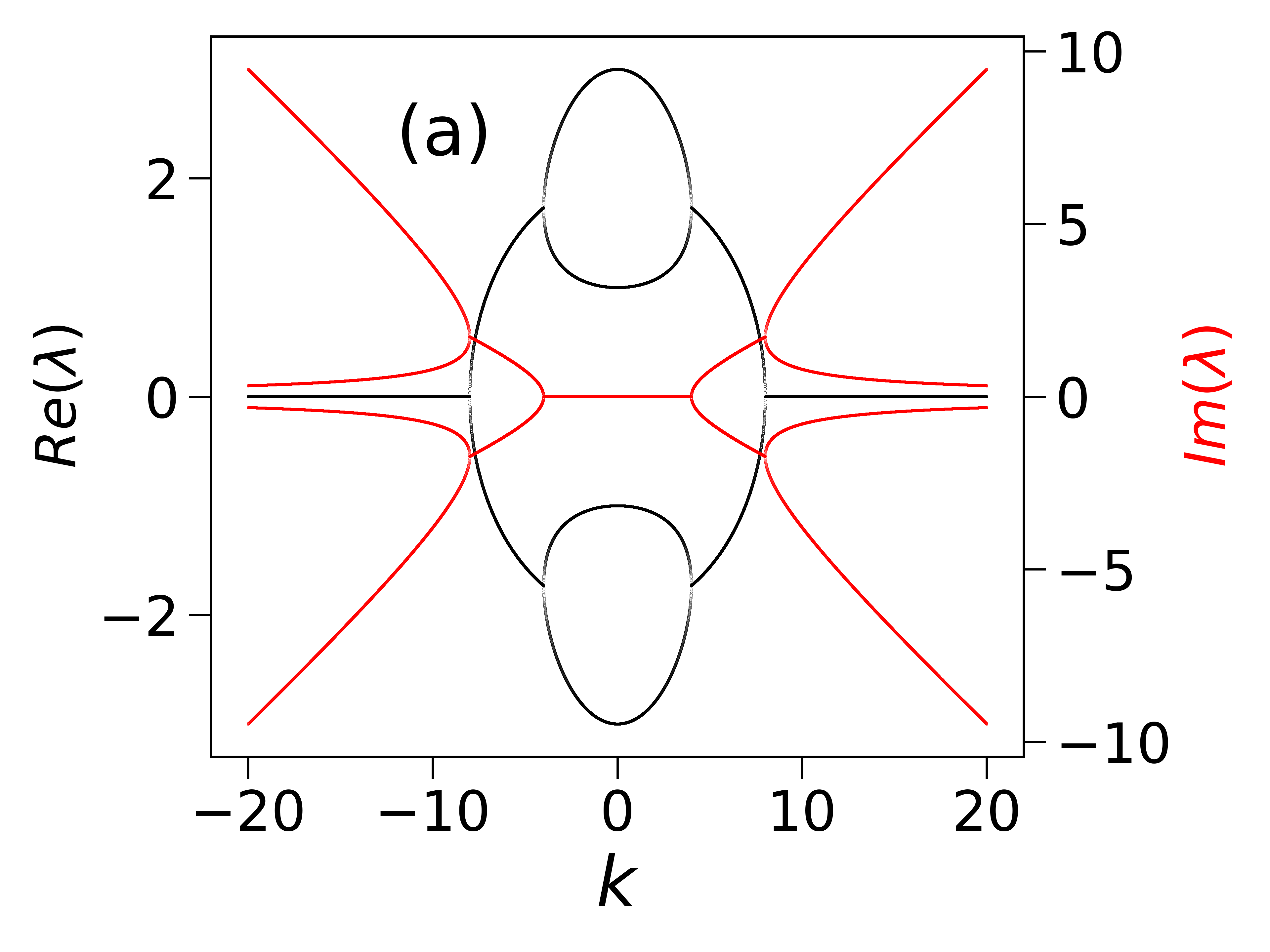}
    \includegraphics[width=0.3\linewidth]{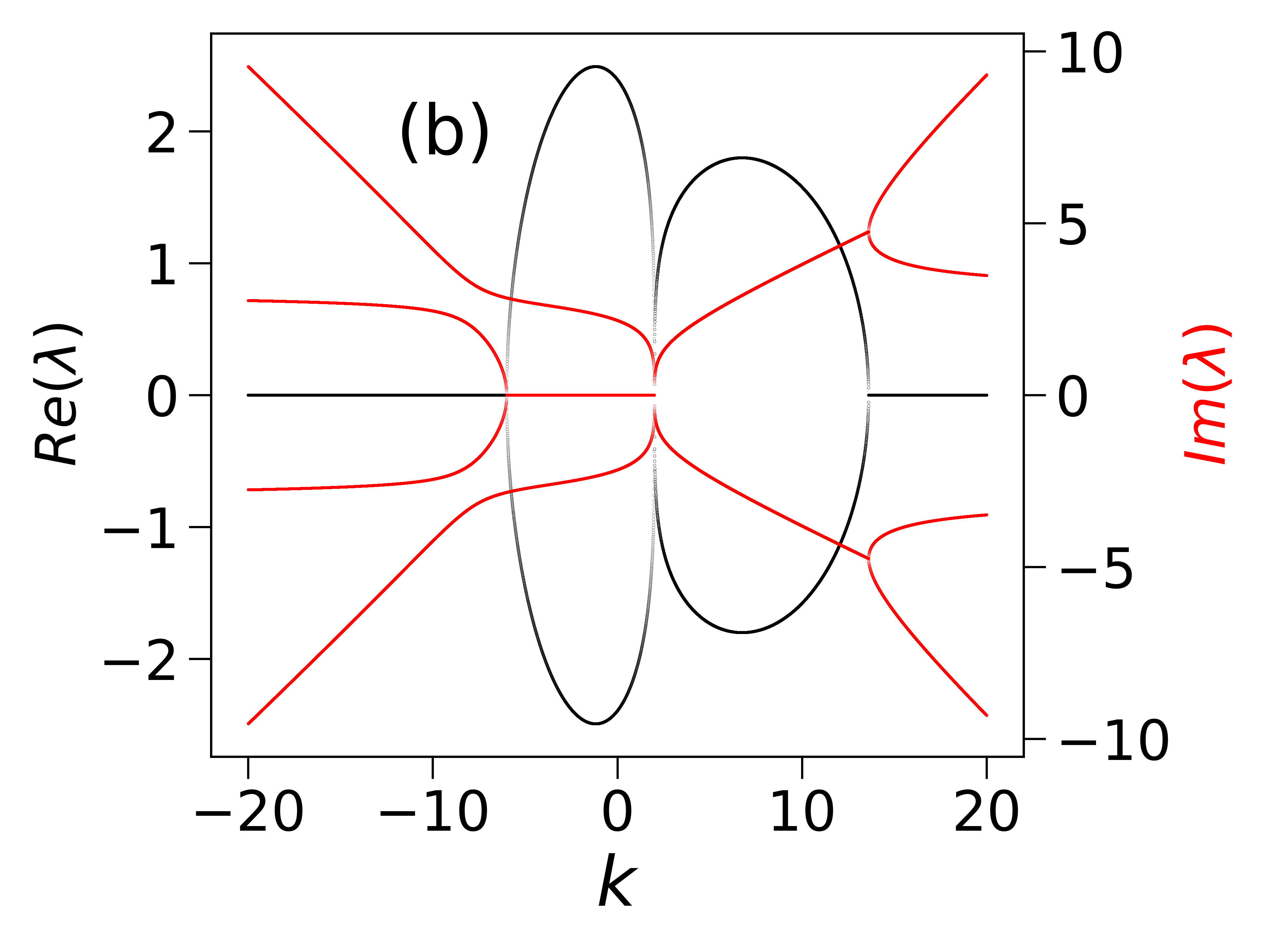}
    \\
    \includegraphics[width=0.3\linewidth]{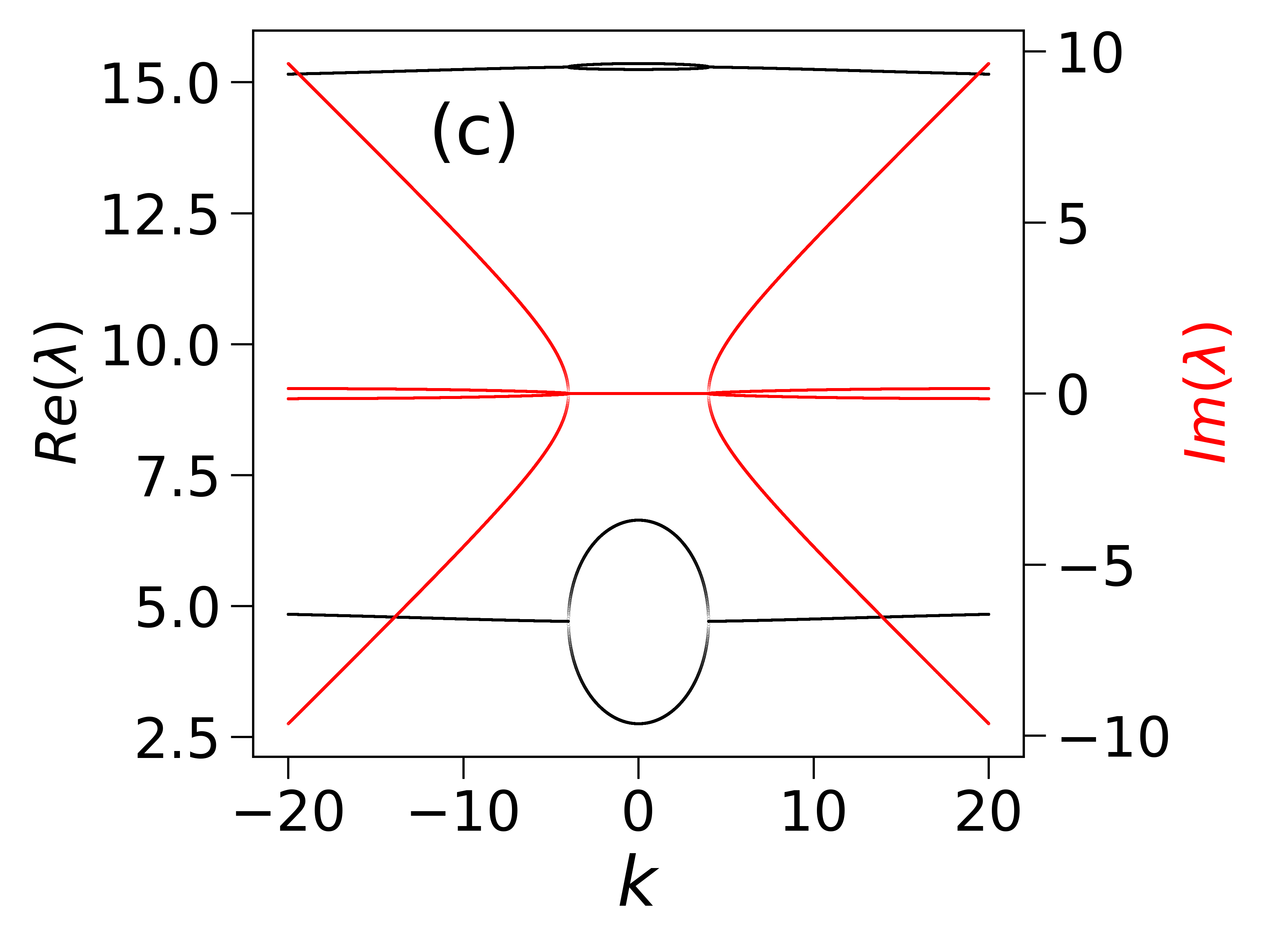}
    \includegraphics[width=0.3\linewidth]{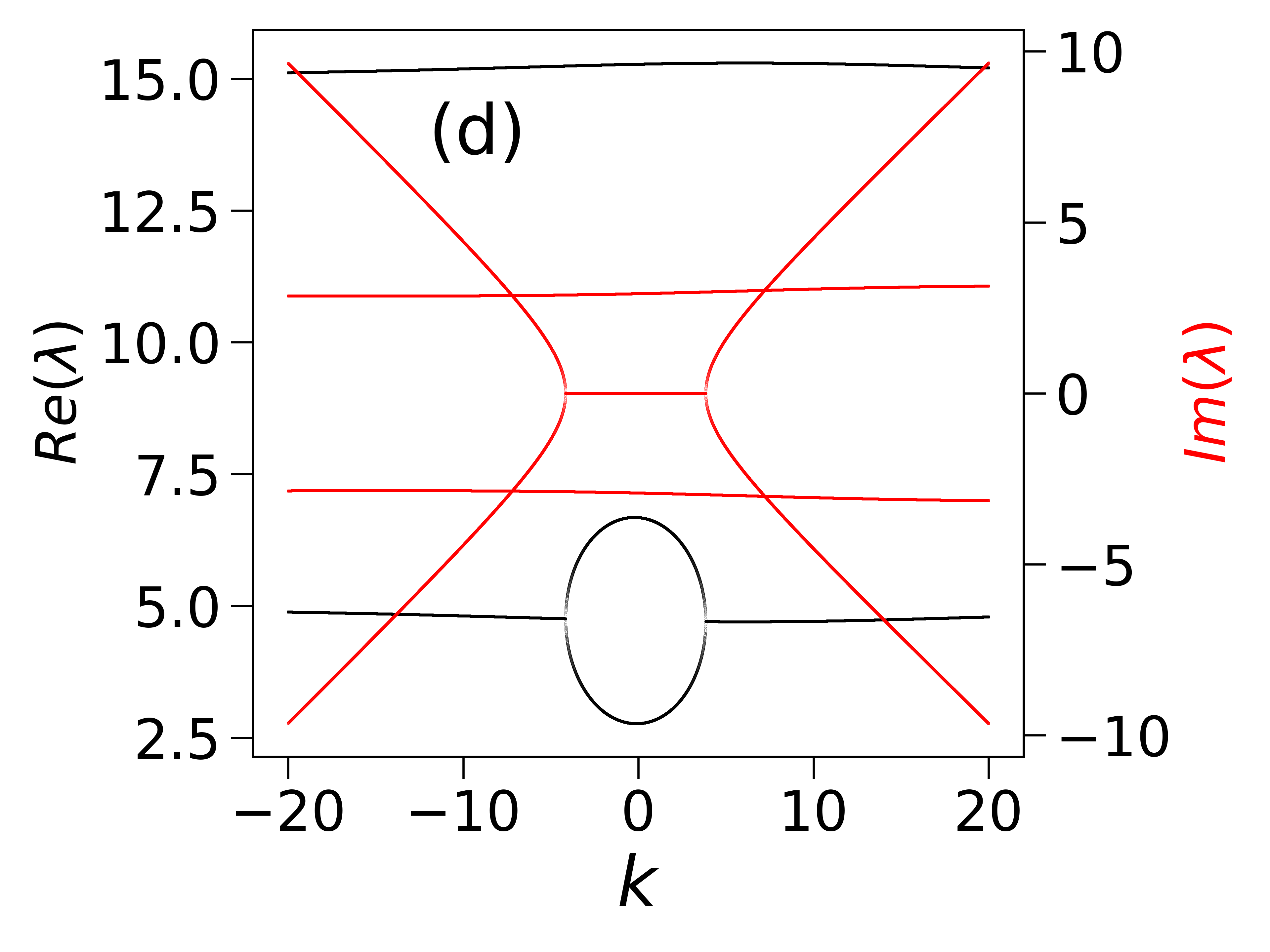}
    \caption{Imaginary and real parts of the eigenvalues of the effective Hamiltonian \eqref{eq-dephasing-hamiltonian} as a function of parameter $k$. (a): $U=0$ and $l=0$. (b): $U=0$ and $l=6$. (c): $U=5$ and $l=0$. (d): $U=5$ and $l=6$.}
    \label{fig-U-k-table}
\end{figure}

The effect of the interaction term $U$ on the spectrum and $\mathcal{PT}$-symmetry breaking of the effective Hamiltonian (\ref{eq-dephasing-hamiltonian}) in parameter space $(l,k)$, can be described by presenting the phase diagrams of broken/unbroken $\mathcal{PT}$-symmetry for different interaction strengths $U$. In Fig.~\ref{fig-dephasing-phasemaps-bin} we plot the regions of unbroken $\mathcal{PT}$-symmetry (green), partially unbroken $\mathcal{PT}$-symmetry (blue) and fully broken $\mathcal{PT}$-symmetry (white and red) with respect to the transformation parameters $k$ and $l$. Additionally, for the non-interacting case, we have also colored regions where the spectrum has degenerate imaginary parts as red. For non-zero interaction this degeneracy is lifted. In the rest of the paper the regions will be named based on the number of $\mathcal{PT}$-symmetric eigenvectors ($n$PTS) as green -- 4PTS, blue -- 2PTS, white -- 0PTS and red -- DIP (degenerate imaginary parts).

The phase diagram for $U=0$ in Fig.~\ref{fig-dephasing-phasemaps-bin}(a) was plotted by counting the number of pairs of imaginary parts of the eigenvalues that are within some threshold of each others. For two eigenvalues $\lambda_i$ and $\lambda_j$, if the condition $|\text{Im}\lambda_i-\text{Im}\lambda_j|<0.01$ was satisfied, the pair was counted as equal imaginary parts. In the case of the $\mathcal{PT}$-symmetric Hamiltonian studied here this information is enough to determine the number of $\mathcal{PT}$-symmetric eigenstates. For $U\neq0$ the $\mathcal{PT}$-symmetry of the eigenstates was tested explicitly. If $|\braket{\phi|\mathcal{PT}|\phi}|<1-10^{-4}$ for an eigenvector $\ket{\phi}$, the eigenvector was counted as not $\mathcal{PT}$-symmetric.

Figure~\ref{fig-dephasing-phasemaps-bin} reveals that when the interaction strength $U$ grows the area with fully unbroken $\mathcal{PT}$-symmetry narrows to a vertical line at $l=0$ and between $k\approx-4$ and $k\approx4$ (green line surrounded by blue). Also, for nonzero $U$ the region of completely unbroken $\mathcal{PT}$-symmetry is not in general contained inside of the region of that with vanishing interaction strength $U=0$. For all $U$, large $|k|$ and large $|l|$ have similar symmetry regions. For large $|k|$ there is a narrow partially unbroken $\mathcal{PT}$-symmetry at $l=0$. For large $|l|$, on the other hand, the region of partially unbroken $\mathcal{PT}$-symmetry extends from $k\approx-4$ to $k\approx4$.

In Fig.~\ref{fig-U-k-table} we plot the spectrum of the effective Hamiltonian $K$ as a function of parameter $k$ with fixed values of parameters $l$ and $U$. The values of $l$ used are $l=0$ and $l=6$, where the latter was chosen so that the line $(k,6)$ passes over an order 4 EP known in literature for the effective Hamiltonian (\ref{eq-reference-dissipative-hamiltonian}) with vanishing interaction $U=0$ and $\gamma=2$ \cite{Graefe2008a}. When $U=0$, we can see how the EP unravels in a different way on either side of $k=2$. For larger values of $k$ the $\mathcal{PT}$-symmetry is fully broken, but the imaginary parts of the spectrum stay pairwise degenerate. For smaller values of $k$ on the other hand, there is still partially unbroken $\mathcal{PT}$-symmetry. 

\begin{figure}[t]
    \includegraphics[width=0.3\linewidth]{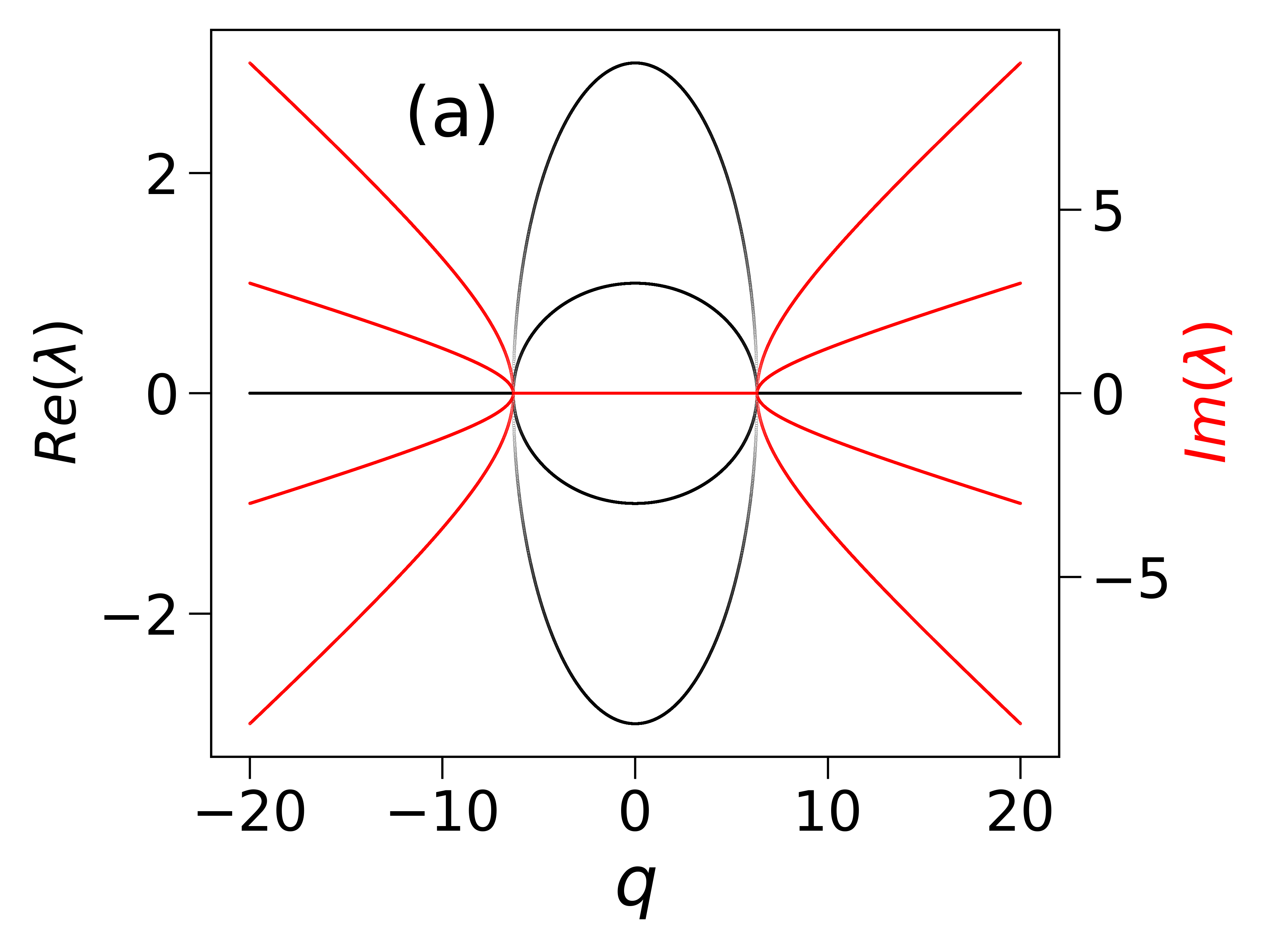}
    \includegraphics[width=0.3\linewidth]{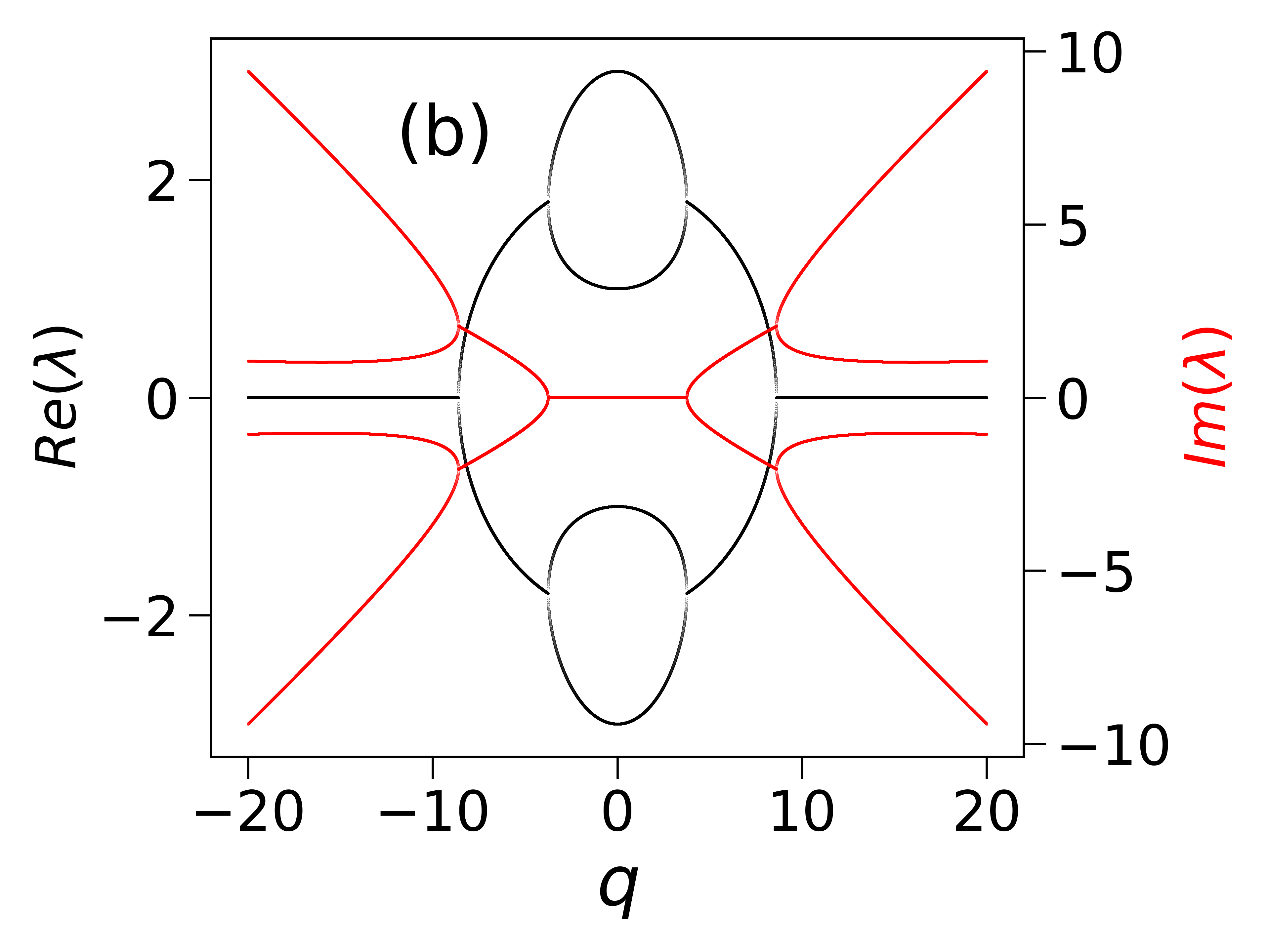}
    \includegraphics[width=0.3\linewidth]{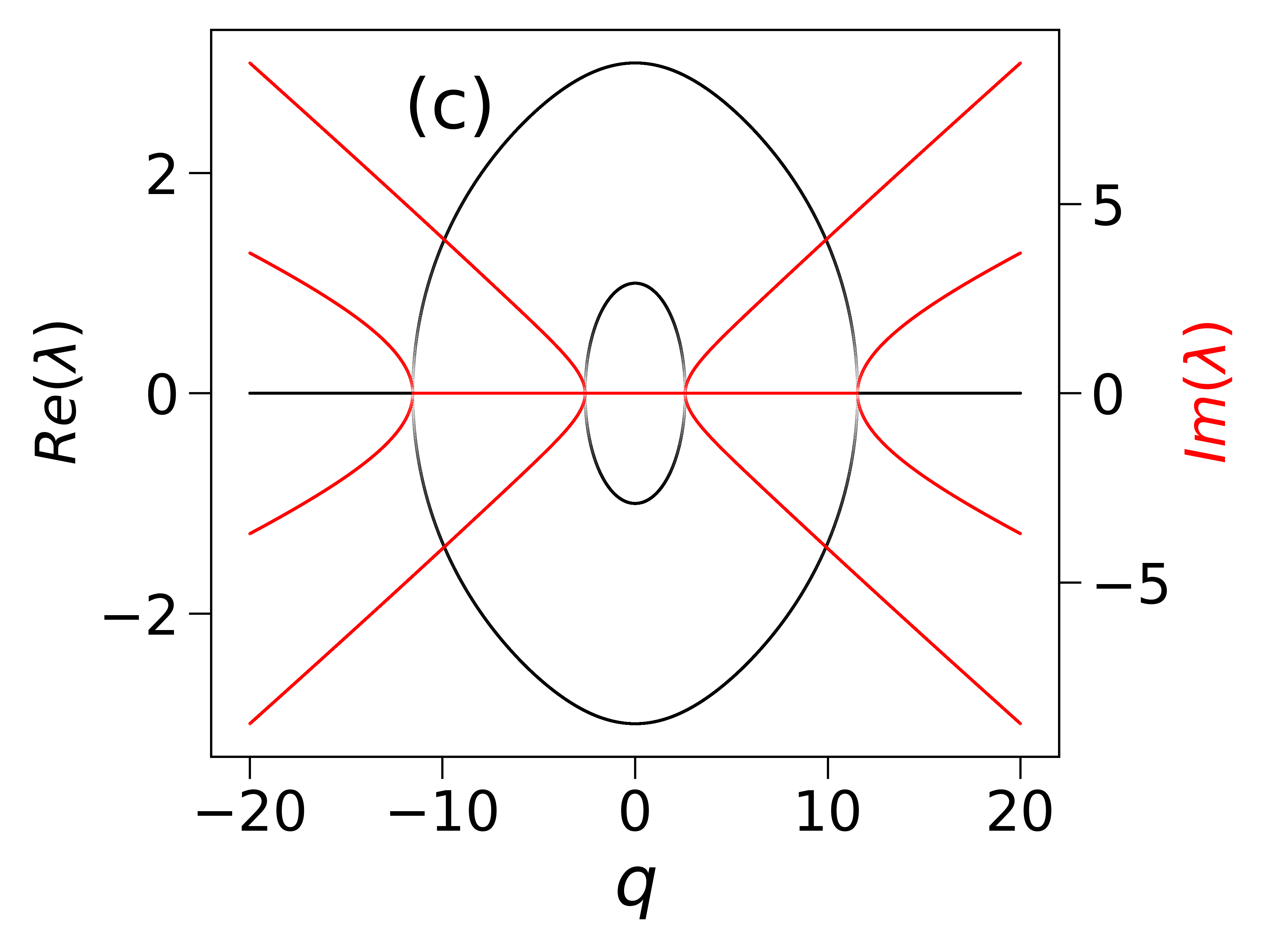}
    \caption{Real and imaginary parts of the eigenvalues of the effective Hamiltonian of Eq.~\eqref{eq-dephasing-hamiltonian} (omitting the term proportional to the identity) with $U=0$ parametrized by $q$ in transformation Eq.~\eqref{eq-lines-transform} with $\varphi\in\{0,~3\pi/8,~6\pi/8\}$ in figures (a), (b) and (c), respectively.}
    \label{fig-lines-through-origin}
\end{figure}

In Fig.~\ref{fig-lines-through-origin} we plot the spectrum of the effective Hamiltonian \eqref{eq-dephasing-hamiltonian} using transformations
\begin{align}
    C_{q,\varphi}=q\left[\cos(\varphi)\left(\frac{3}{\sqrt{10}}L_{30,03}+\frac{1}{\sqrt{10}}L_{21,12}\right)+\sin(\varphi)\left(-\frac{1}{\sqrt{10}}L_{30,03}+\frac{3}{\sqrt{10}}L_{21,12}\right)\right] + h_{q,\varphi},\qquad q\in\mathbb{R},\quad \varphi\in[0,\pi), \label{eq-lines-transform}
\end{align}
and omitting the term proportional to the identity in the effective Hamiltonian. The above transformations correspond to lines crossing the origin of the parameter space $(l,k)$. The transformation is chosen such that the angle $\varphi$ refers to the angle in parameter space $(l,k)$ with respect to the line corresponding to transformation $C_q=q/\sqrt{10}(n_1-n_2)$. Figure \ref{fig-lines-through-origin} illustrates more accurately the structure of the spectrum that is depicted in Fig.~\ref{fig-dephasing-phasemaps-bin} only in terms of symmetries of the eigenstates, or equalities between the imaginary parts of the eigenvalues. All of the colored regions in Fig.~\ref{fig-dephasing-phasemaps-bin} are separated by exceptional lines. Figures \ref{fig-dephasing-phasemaps-bin} and \ref{fig-lines-through-origin} show that the parameter space includes different kinds of branching points of the spectrum. For vanishing interaction $U=0$, branching points related to $\mathcal{PT}$-symmetry breaking can be found to form two types of exceptional lines. The first type is lines consisting of two order two exceptional points (4PTS to DIP in Fig.~\ref{fig-dephasing-phasemaps-bin}(a) and first branching point in Fig.~\ref{fig-lines-through-origin}(b)). The second type is lines consisting of order two exceptional points (4PTS to 2PTS and 2PTS to 0PTS in Fig.~\ref{fig-dephasing-phasemaps-bin}(a) and both branching points in Fig.~\ref{fig-lines-through-origin}(c)). Additionally, there are four order four exceptional points (sharp corners where all colors meet in Fig.~\ref{fig-dephasing-phasemaps-bin}(a) and branching point in Fig.~\ref{fig-lines-through-origin}(a)). Two of these (left and right corners) are known from the Hamiltonian \eqref{eq-reference-dissipative-hamiltonian} while the two others have not been presented in literature before, as far as the authors are aware. Whether or not these two new EPs or more are found in models analogous to Eq. \eqref{eq-dephasing-hamiltonian} (with $U=0$) with higher particle numbers is left as an open question. 

To show that the boundaries discussed above are genuine exceptional points, we write the characteristic equation of $K_\mathcal{PT}=K+(i/2)pI$, i.e. the $\mathcal{PT}$-symmetric part of the effective Hamiltonian \eqref{eq-dephasing-hamiltonian}, with $U=0$:
\begin{align}
    \det (\lambda I-K_\mathcal{PT})=\lambda^4 + \frac{k^2 + l^2 - 40}{4}\lambda^2 + \frac{k^2l^2 + 24kl - 16l^2 + 144}{16}.
\end{align}
A direct calculation shows that there is only one eigenvector corresponding to the eigenvalue $\lambda=0$, proving that points in the parameter space $(l,\,k)$ where $\lambda=0$ has an algebraic multiplicity larger than one correspond to genuine exceptional points. Assuming that $k^2l^2 + 24kl - 16l^2 + 144=0$ but $k^2 + l^2 - 40 \neq 0$ the eigenvalue $\lambda=0$ has an algebraic multiplicity of 2 but geometric multiplicity of 1. The curves defined by these equations correspond to the lines consisting of exceptional points of order two at the borders of the 2PTS regions. Imposing also that $k^2 + l^2 - 40 = 0$ we have a single eigenvalue $\lambda=0$ with algebraic multiplicity of 4, and we get an equation for the parameters $l$ and $k$ with four real solutions, which give the locations of order 4 exceptional points in the $(l,\,k)$-space. These are $\{\pm(6,\,2),\,\pm(2^{5/3}-2,\,4-12/(2^{5/3}-2))\}$. The $\mathcal{PT}$-symmetry breaking at the border between 4PTS and DIP regions happens at points where in the spectrum all eigenvalues simultaneously gain a non-zero imaginary part. This corresponds to parameter values for which the discriminant of the quadratic equation 
\begin{align}
    \mu^2 + \frac{k^2 + l^2 - 40}{4}\mu + \frac{k^2l^2 + 24kl - 16l^2 + 144}{16} &= 0,
\end{align}
where we have defined $\lambda^2=\mu$, goes to zero and $k^2 + l^2 - 40<0$. We then have eigenvalues $\pm\sqrt{(40 - k^2 - l^2 )/8}$ at the EPs, both of which have algebraic multiplicity of 2. For $k^2 + l^2 - 40>0$ we get the exceptional lines between the DIP and 0PTS regions with eigenvalues $\pm i\sqrt{(k^2 + l^2 - 40 )/8}$ again with algebraic multiplicities of 2. Direct calculation shows that there is only one eigenvector corresponding to each eigenvalue.

\begin{figure}[t]
    \includegraphics[width=0.27\linewidth]{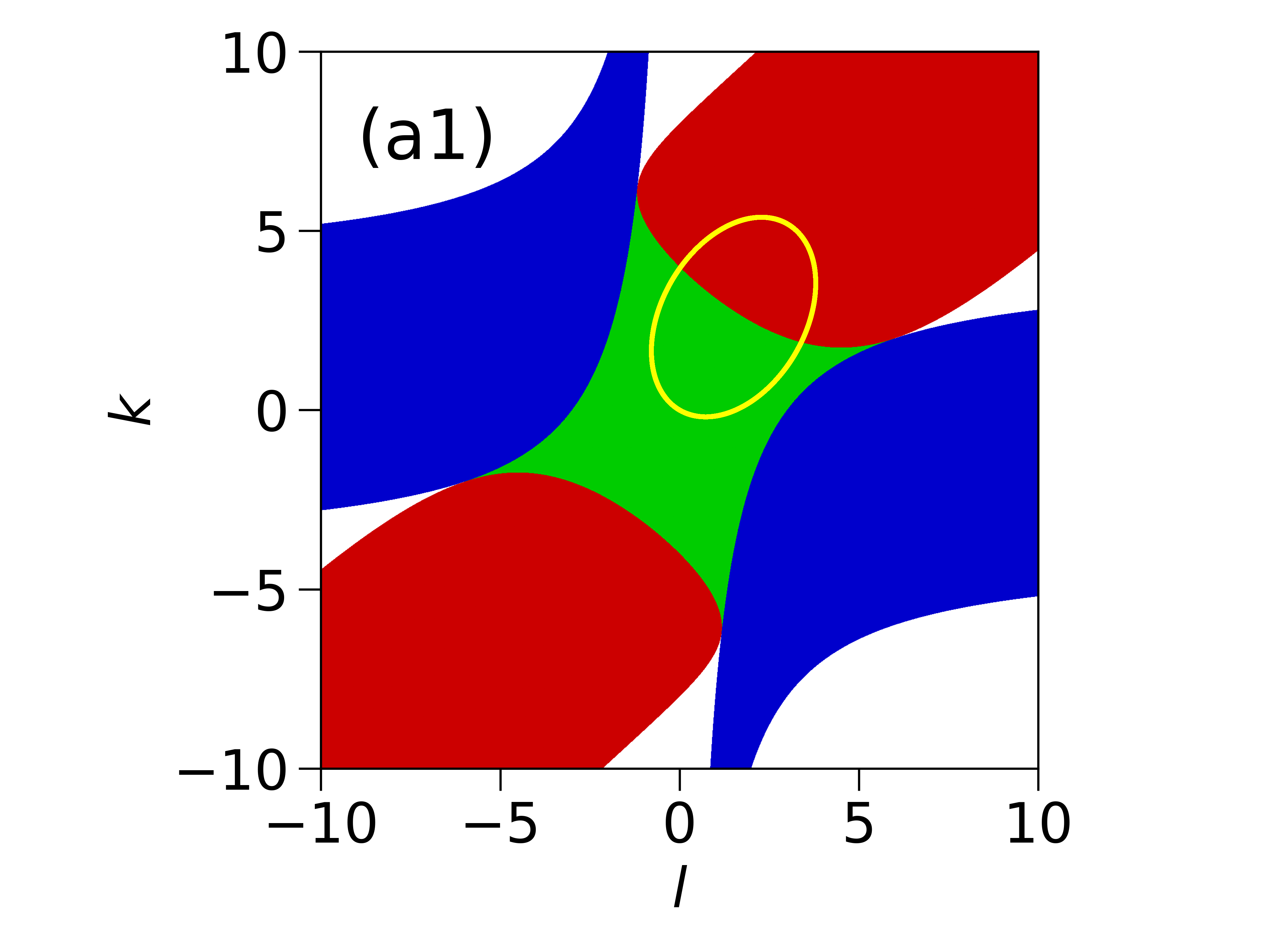}
    \includegraphics[width=0.27\linewidth]{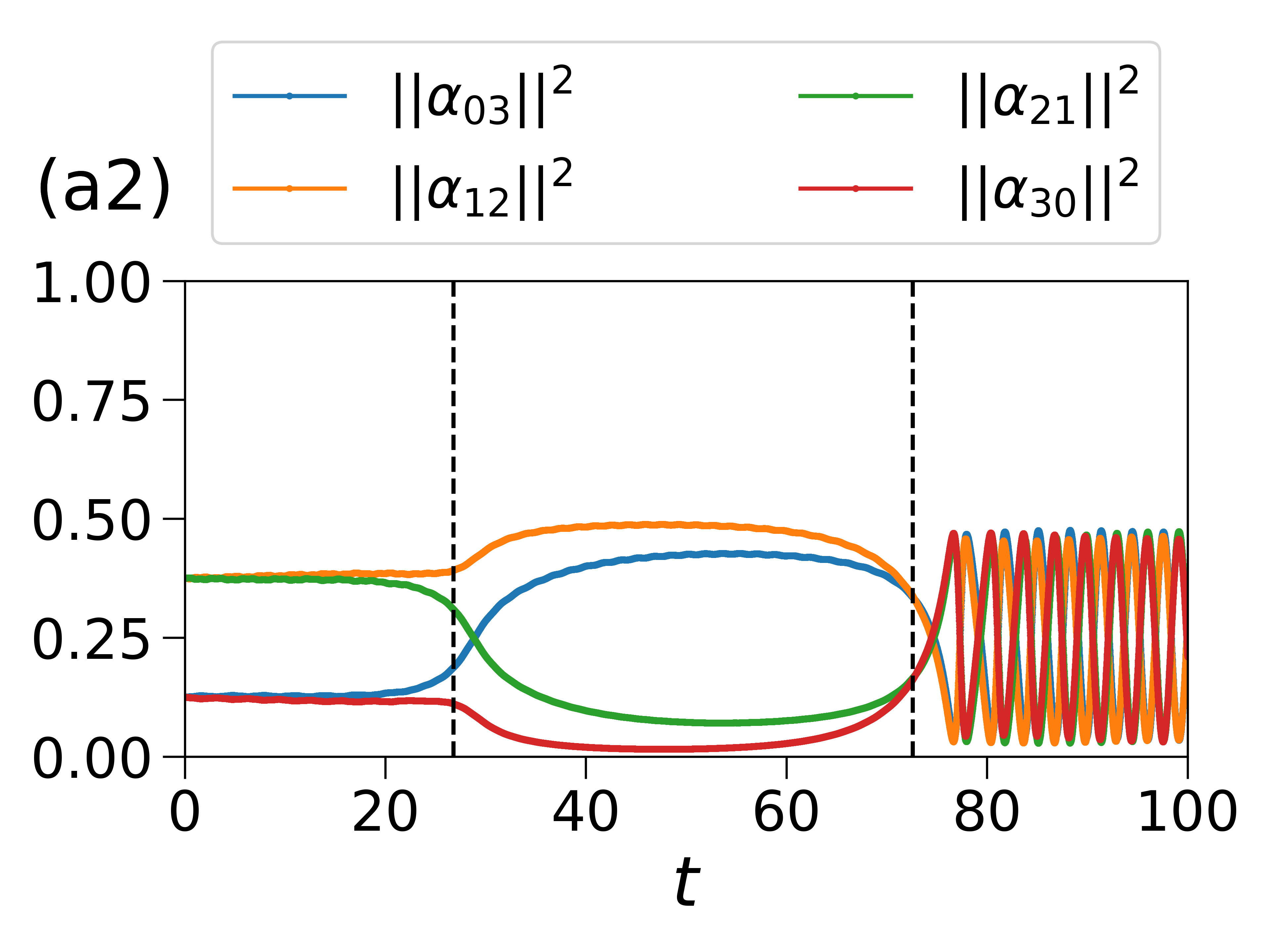}
    \includegraphics[width=0.27\linewidth]{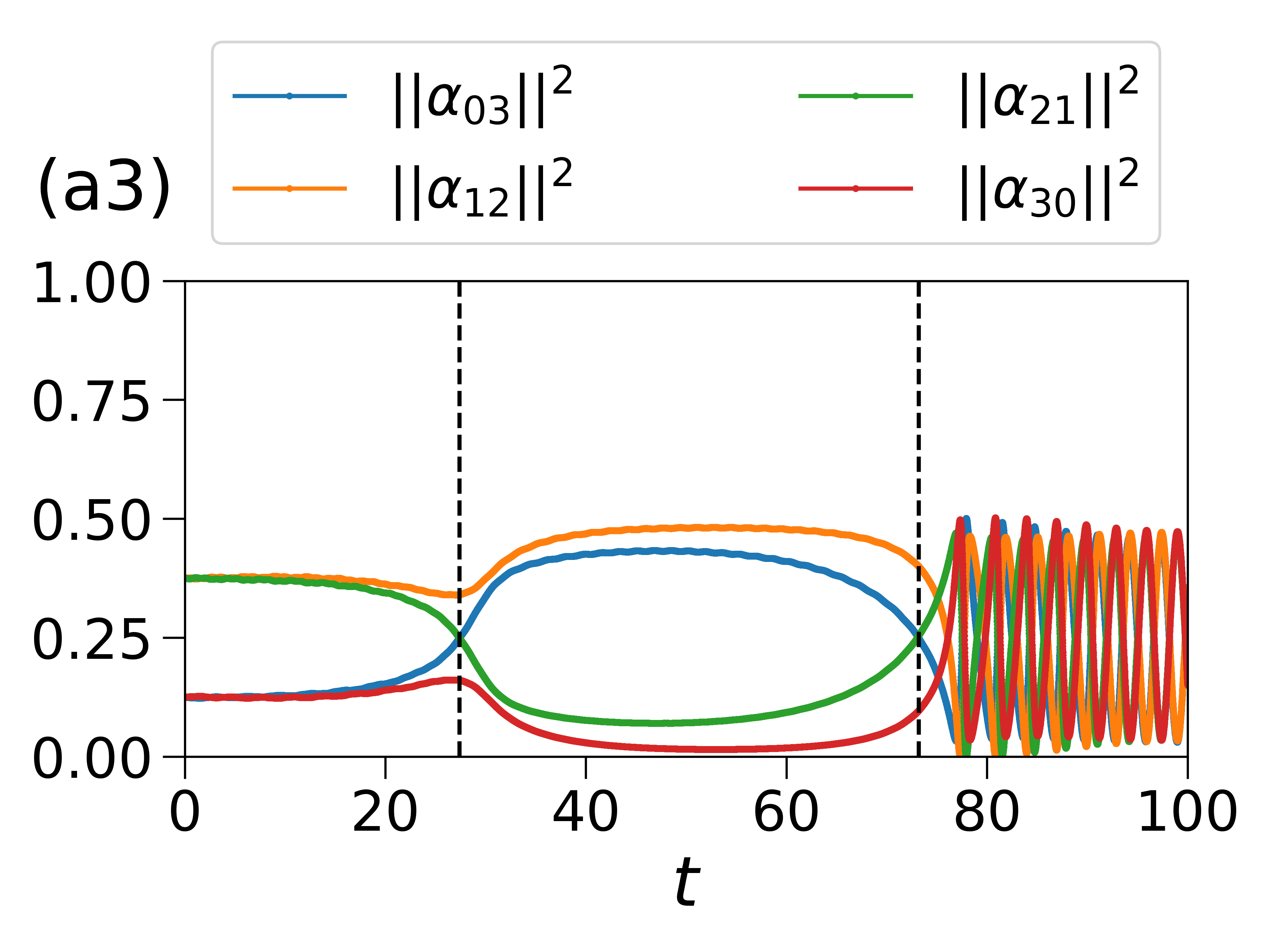}
    \\
    \includegraphics[width=0.27\linewidth]{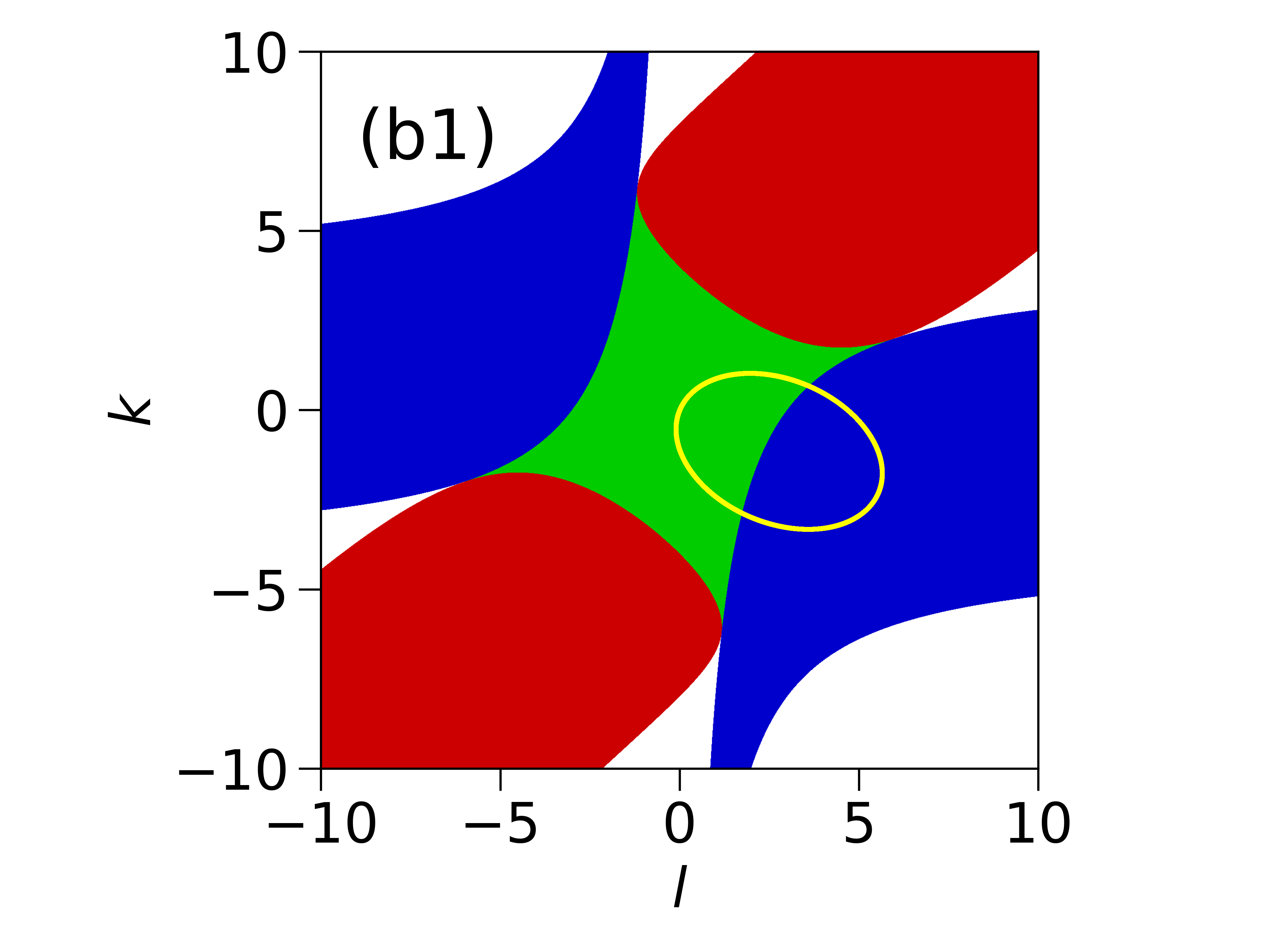}
    \includegraphics[width=0.27\linewidth]{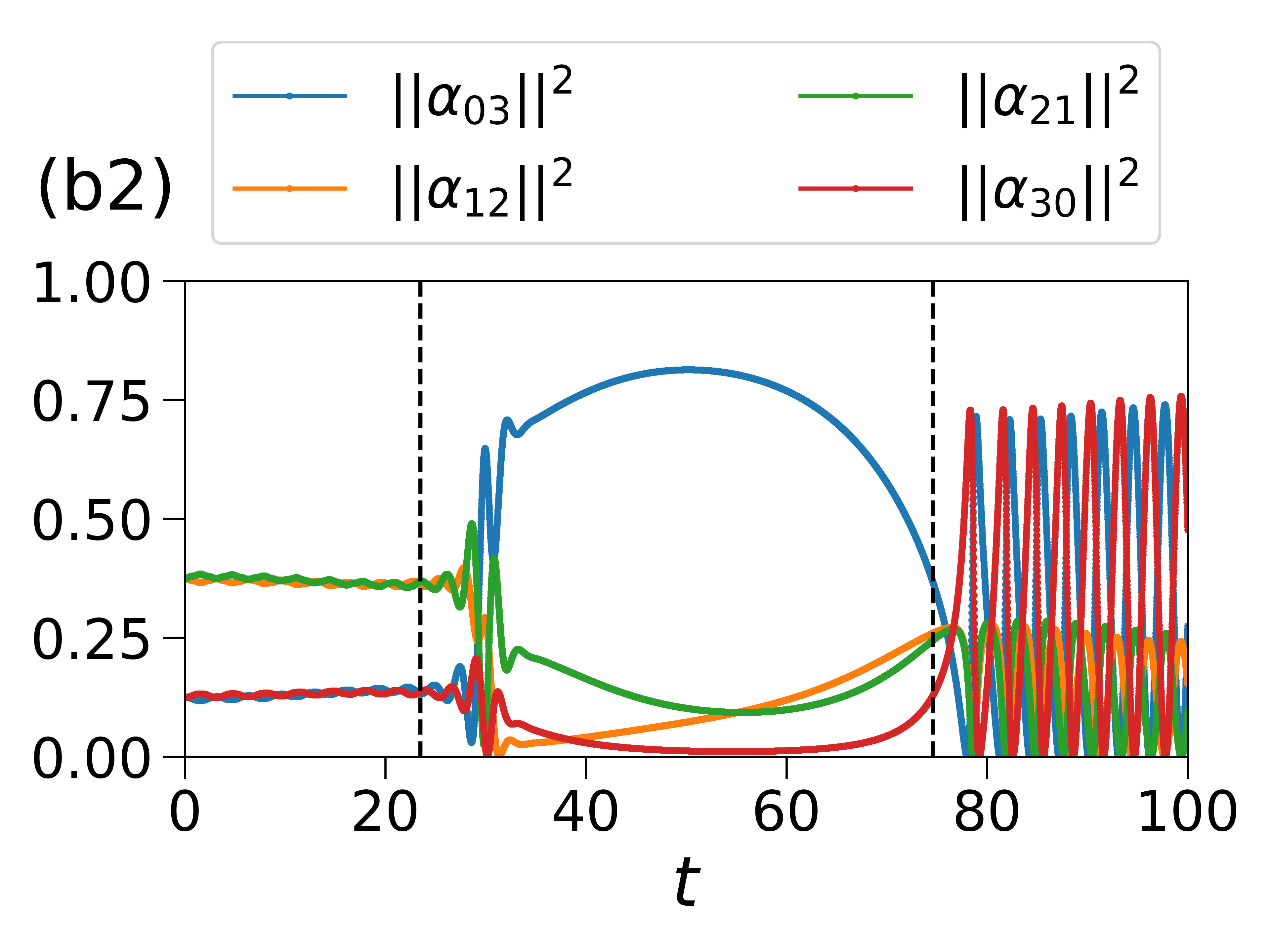}
    \includegraphics[width=0.27\linewidth]{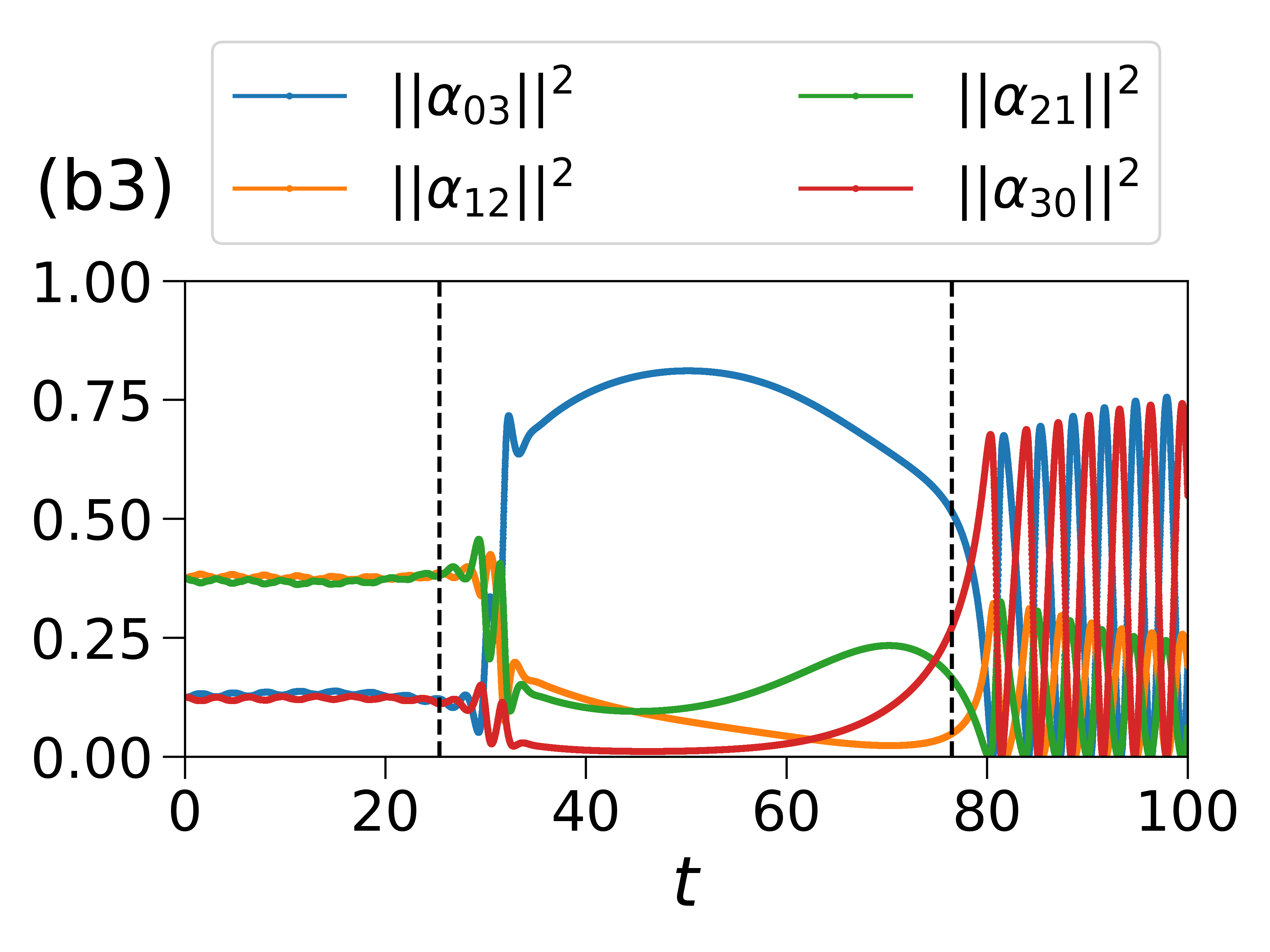}
    \\
    \includegraphics[width=0.27\linewidth]{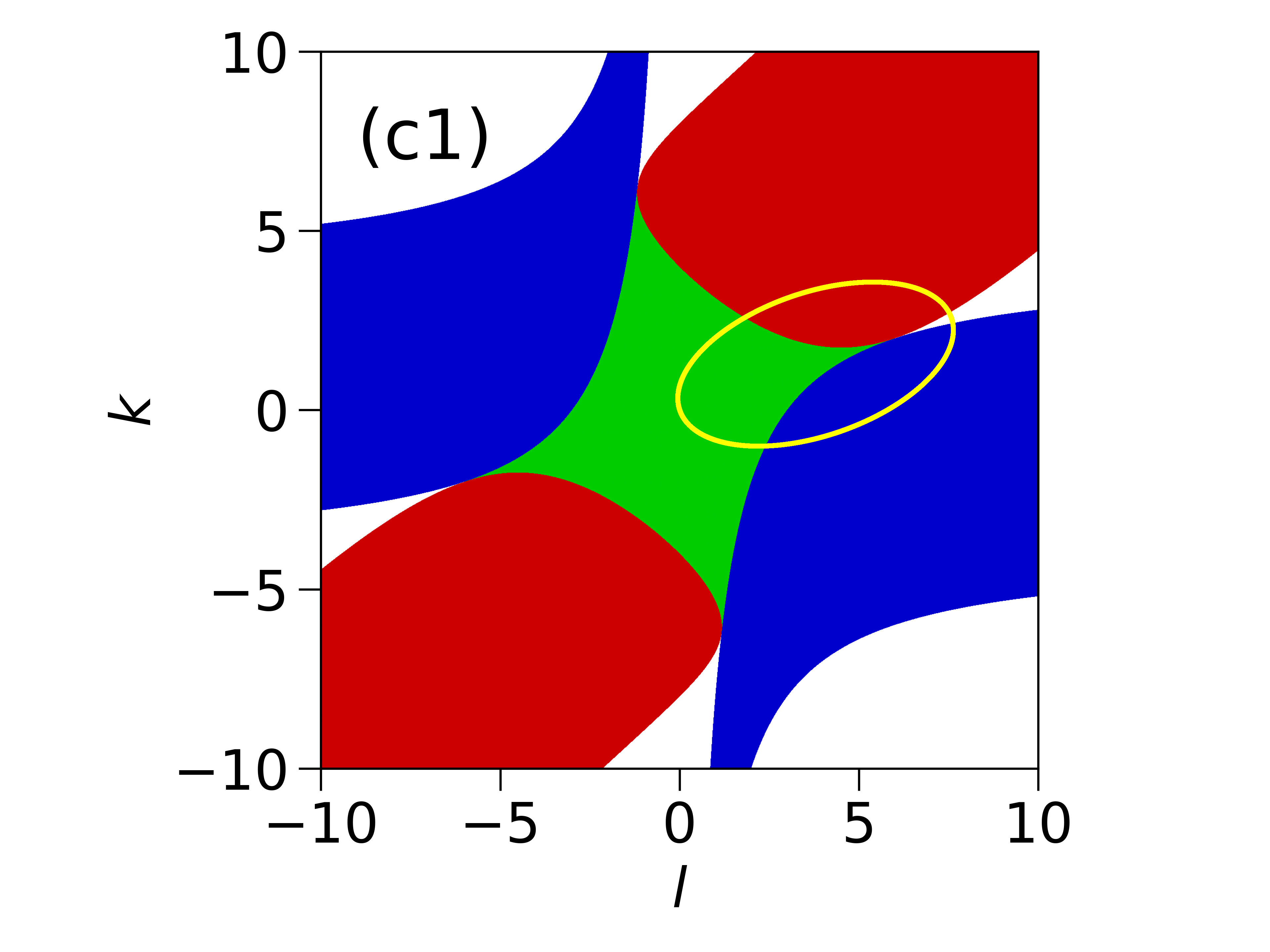}
    \includegraphics[width=0.27\linewidth]{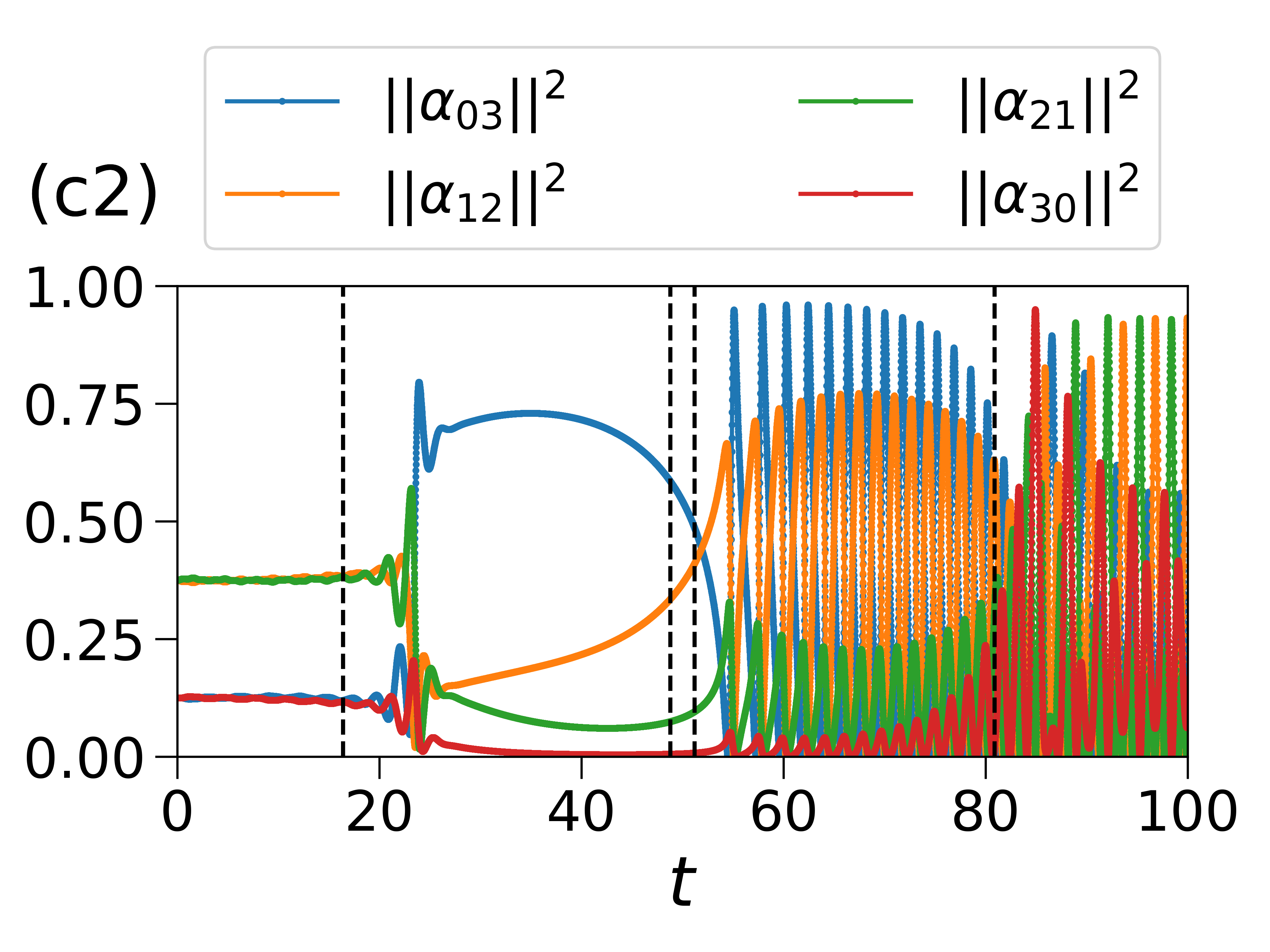}
    \includegraphics[width=0.27\linewidth]{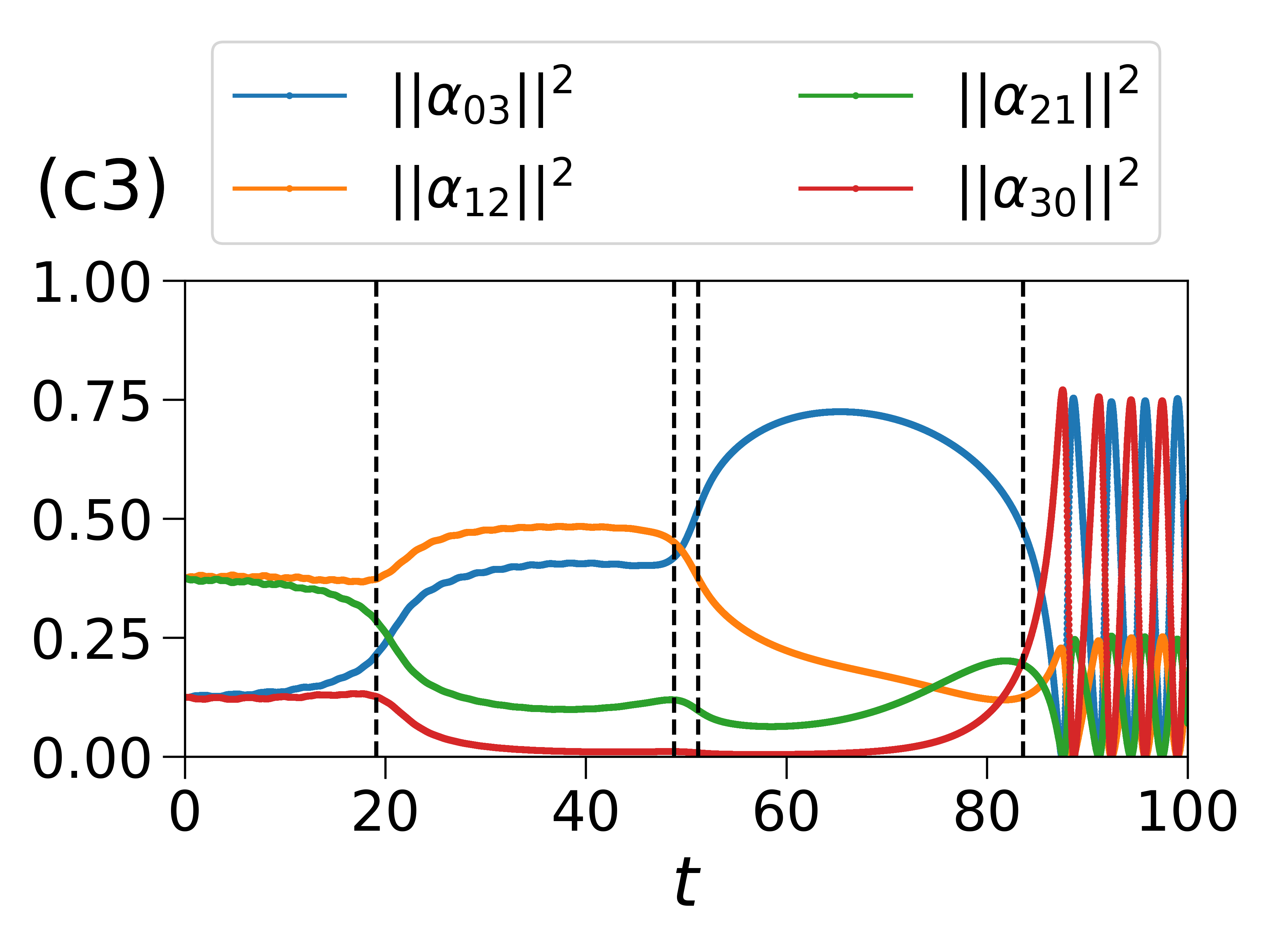}
    \caption{Deterministic evolution of populations $|\alpha_{ij}(t)|^2$ generated by the non-Hermitian Hamiltonian of Eq.~\eqref{eq-dephasing-hamiltonian-traj-U=0} with time-dependent transformations (\ref{ellipse-transformation}). The total time of the evolution is $T=100$. Figures (a1, b1, c1) show the parameter trajectory in yellow. The starting point is at $(l,\,k)=(0,\,0)$. Figures (a2, b2, c2) show the evolution for counterclockwise trajectories and figs. (a3, b3, c3) for clockwise trajectories. All plots are with $U=0$ and the initial state is the ground state of the Hermitian Hamiltonian. Vertical dashed lines correspond to the instants of times when an exceptional point is crossed.}
    \label{dephasing-detev-loops-T=all-U=0-ground}
\end{figure}

There are also exceptional points not related to $\mathcal{PT}$-symmetry breaking seen in Fig.~\ref{fig-lines-through-origin}(b) where imaginary parts of eigenvalues branch off at values of the imaginary part that are not zero (DIP to 0PTS). The degeneracy in imaginary parts of the eigenvalues in DIP region is due to an additional symmetry of the effective Hamiltonian $K$ with vanishing on-site interaction $U=0$. This symmetry can be written as 
\begin{align}
    \mathcal{PD}K_\mathcal{PT}(\mathcal{PD})^{-1}=-K_\mathcal{PT}, \label{eq-chiral-symmetry}
\end{align}
where $\mathcal{P}$ is the parity operator and $\mathcal{D}$ is $\mathcal{D}=\text{diag}(1,-1,1,-1)$ in the ordered basis $\{\ket{30},\,\ket{21},\,\ket{12},\,\ket{03}\}$. This symmetry leads to the characteristic equation $\det(\lambda I-K_\mathcal{PT})$ comprising only of terms of order 4, 2 and 0 in $\lambda$. The eigenvalues can then be given as $\{\pm\mu_1,\pm\mu_2\}$, where $\mu_1$ and $\mu_2$ are the roots of the characteristic equation after substitution $\lambda^2\to\mu$. The DIP region corresponds to $\mu_1=\mu_2^*$ and the set of eigenvalues is then $\{\lambda,-\lambda,\lambda^*,-\lambda^*\}$. For the general case with an arbitrary particle number $N$ we can write $\mathcal{D}=e^{i\pi n_2}$, which acts as 1 on basis states with an even number of particles on lattice site 2 and as $-1$ on basis states with an odd number of particles on lattice site 2. We then have relations
\begin{align}
    \mathcal{D} a_1 \mathcal{D}^{-1}&=a_1, \quad \mathcal{D} a_2 \mathcal{D}^{-1}=-a_2, \quad \mathcal{D} (\ket{ij}\bra{ij}-\ket{ji}\bra{ji}) \mathcal{D}^{-1}=(\ket{ij}\bra{ij}-\ket{ji}\bra{ji})~~\forall i,j, \\
    \mathcal{P} a_1 \mathcal{P}^{-1}&=a_2, \quad \mathcal{P} a_2 \mathcal{P}^{-1}=a_1, \quad \mathcal{P} (\ket{ij}\bra{ij}-\ket{ji}\bra{ji}) \mathcal{P}^{-1}=-(\ket{ij}\bra{ij}-\ket{ji}\bra{ji})~~\forall i,j.
\end{align}
These lead to the symmetry \eqref{eq-chiral-symmetry} to apply in the two site Bose-Hubbard model with any number of particles, zero on-site interaction and non-local dephasing operators $\ket{ij}\bra{ij}-\ket{ji}\bra{ji}$. It is then plausible that similar spectral properties as observed in this simple example can also be found for larger particle numbers $N$. This symmetry is broken by non-zero on-site interaction $U$.

From the point of view of the dynamics, the relevant eigenstate is the least decaying eigenstate, to which almost all initial conditions tend to asymptotically. We note that for vanishing interaction $U=0$ in the 2PTS region it is unique, as evident from Fig.~\ref{fig-lines-through-origin}(c), while in the DIP region there are two least decaying eigenstates, since the imaginary parts of the eigenvalues are degenerate. This has non-trivial consequences to the dynamics of the state when using time-dependent parameters in the transformation $C$, as we will see next.

\begin{figure}[t]
    \includegraphics[width=0.27\linewidth]{EqualIm-C=l_30-03_+k_21-12_-U=0-min_ax=2-maj_ax=3-theta=4pid12.png}
    \includegraphics[width=0.27\linewidth]{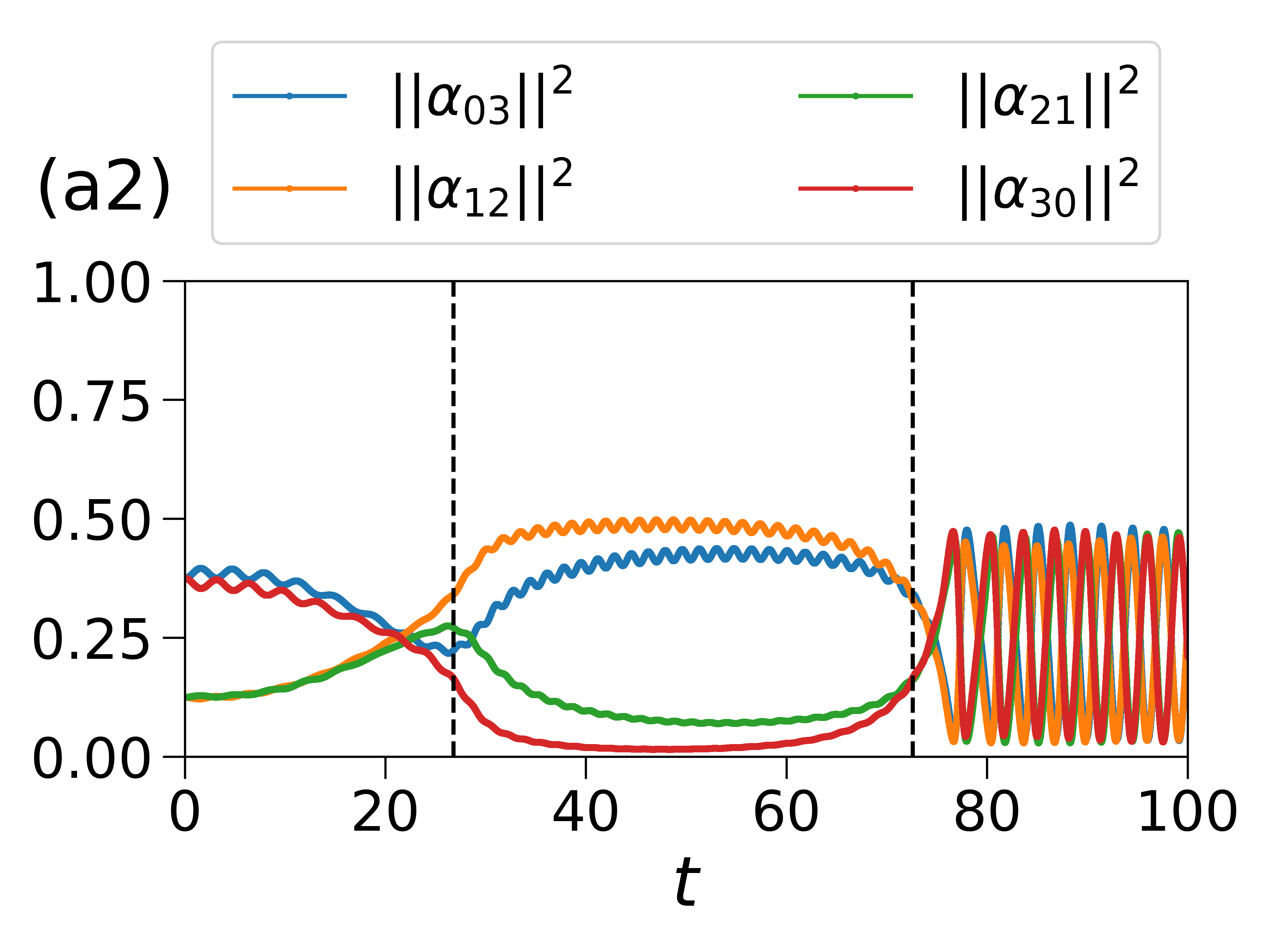}
    \includegraphics[width=0.27\linewidth]{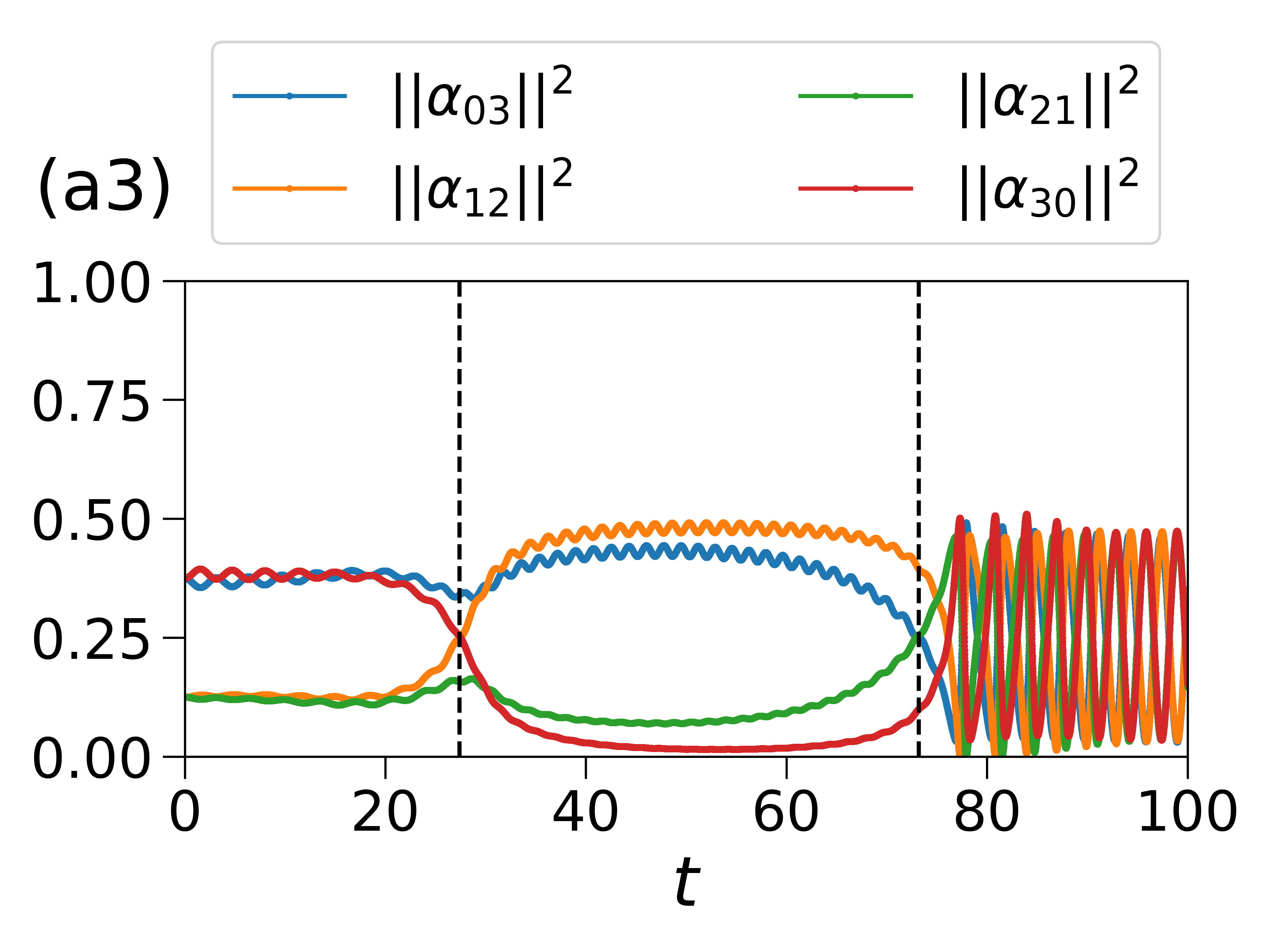}
    \\
    \includegraphics[width=0.27\linewidth]{EqualIm-C=l_30-03_+k_21-12_-U=0-min_ax=2-maj_ax=3-theta=-1p5pid12.png}
    \includegraphics[width=0.27\linewidth]{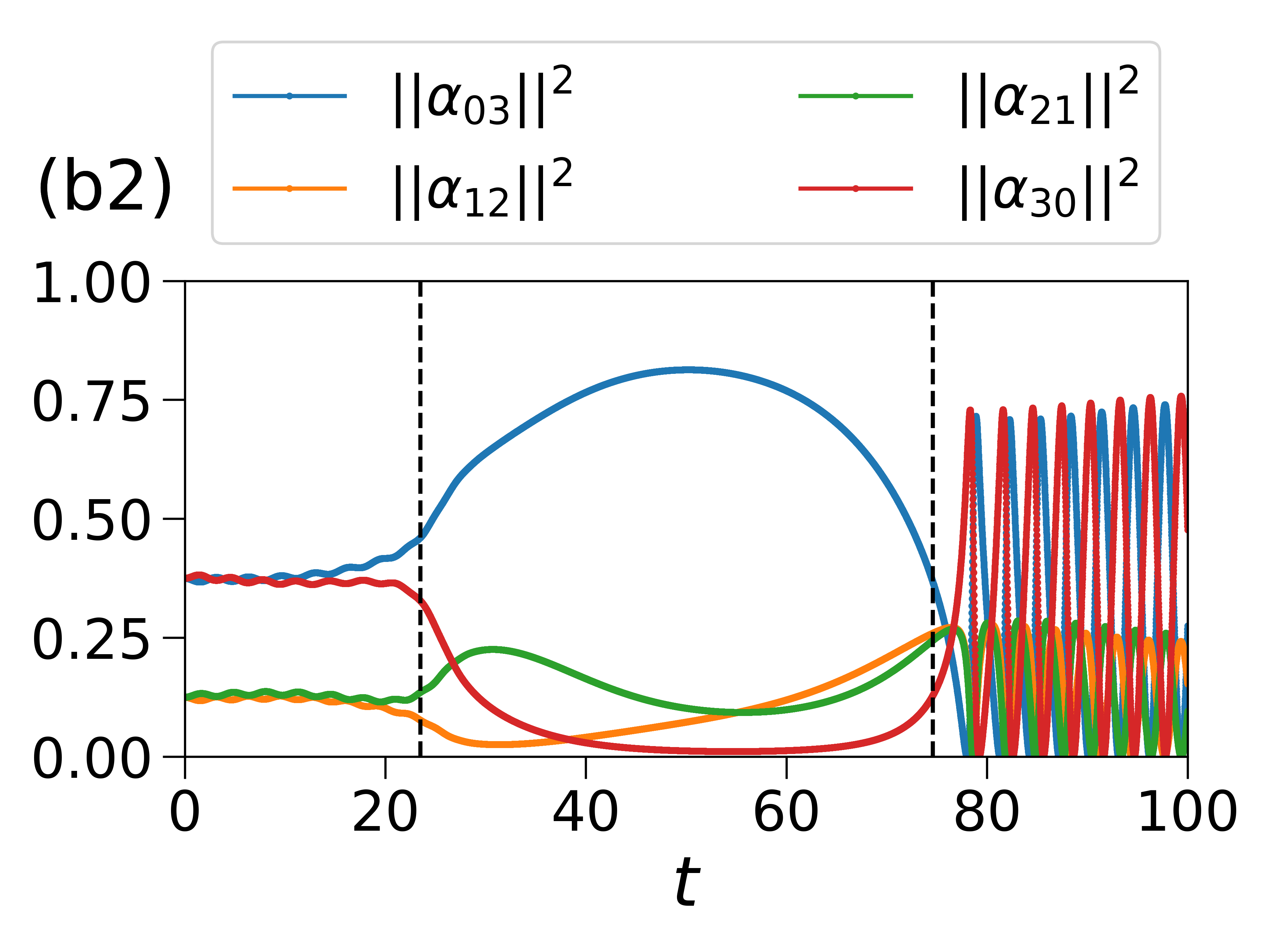}
    \includegraphics[width=0.27\linewidth]{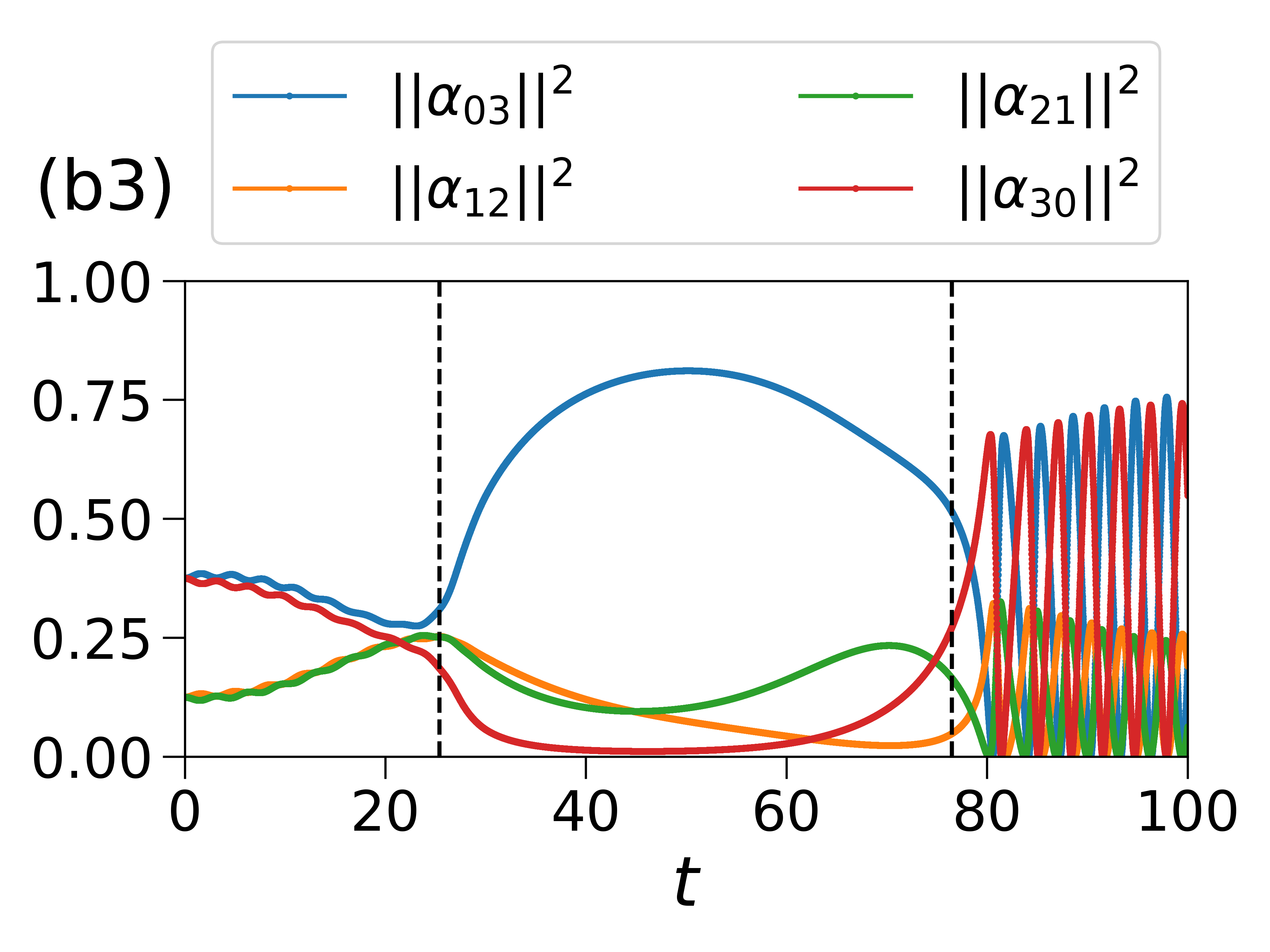}
    \\
    \includegraphics[width=0.27\linewidth]{EqualIm-C=l_30-03_+k_21-12_-U=0-min_ax=2-maj_ax=4-theta=1p25pid12.png}
    \includegraphics[width=0.27\linewidth]{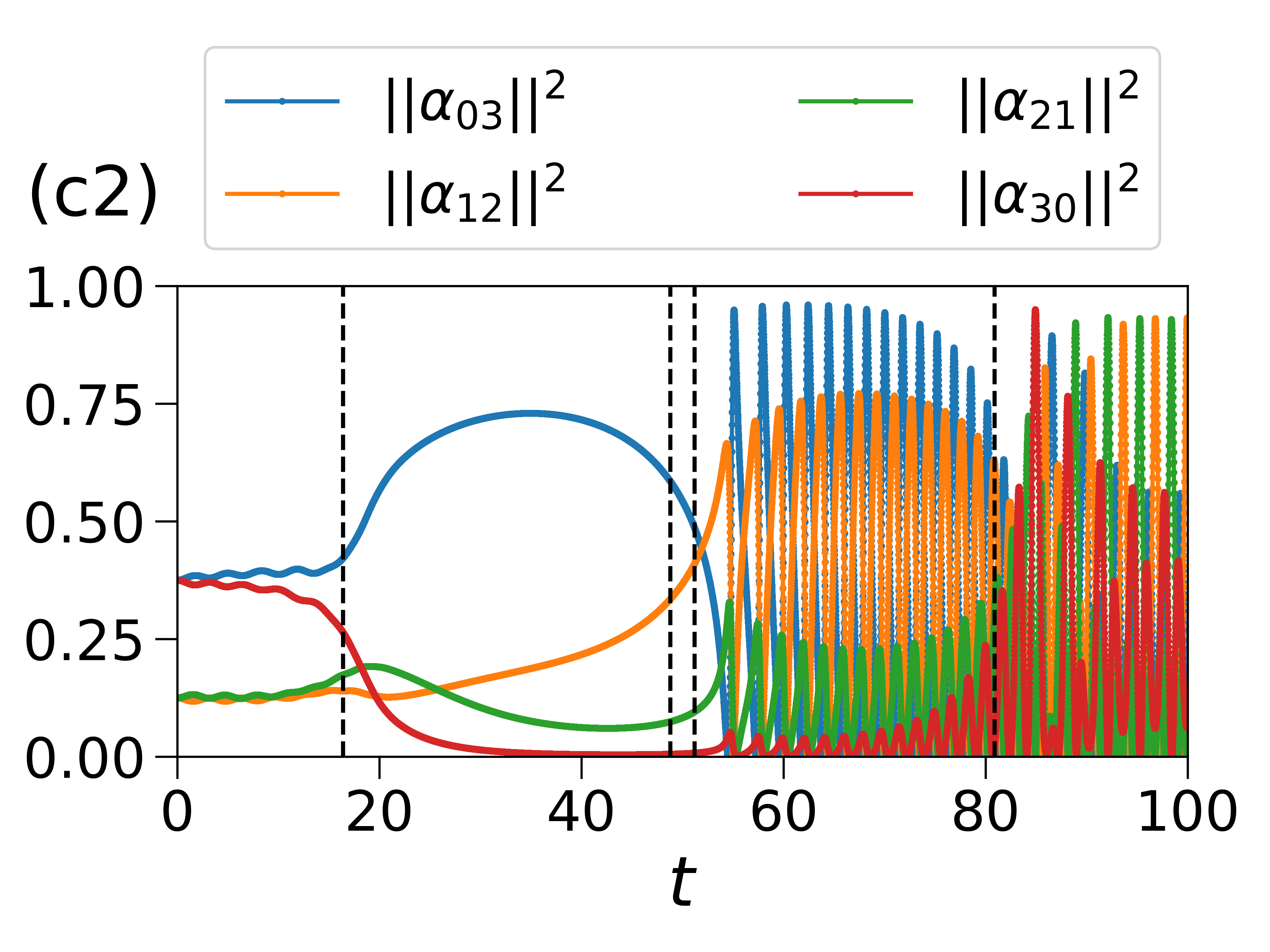}
    \includegraphics[width=0.27\linewidth]{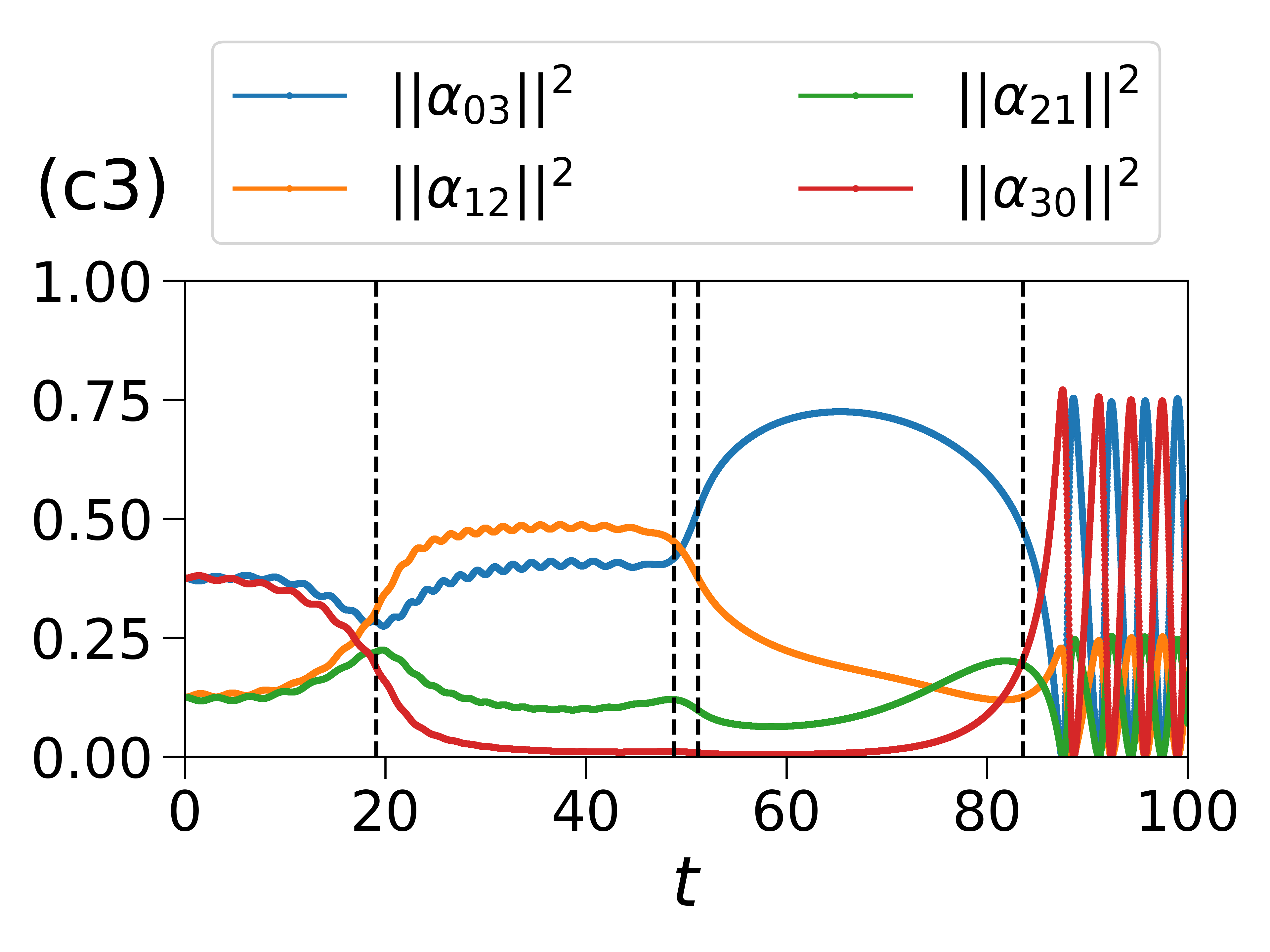}
    \caption{Deterministic evolution of populations $|\alpha_{ij}(t)|^2$ generated by the non-Hermitian Hamiltonian of Eq.~\eqref{eq-dephasing-hamiltonian-traj-U=0} with time-dependent transformations (\ref{ellipse-transformation}). The total time of the evolution is $T=100$. Figures (a1, b1, c1) show the parameter trajectory in yellow. The starting point is at $(l,\,k)=(0,\,0)$. Figures (a2, b2, c2) show the evolution for counterclockwise trajectories and figs. (a3, b3, c3) for clockwise trajectories. All plots are with $U=0$ and the initial state is the second excited state of the Hermitian Hamiltonian. Vertical dashed lines correspond to the instants of times when an exceptional point is crossed.}
    \label{dephasing-detev-loops-T=all-U=0-e2}
\end{figure}

\subsection{Dynamics of the population}

\begin{figure}[t]
    \includegraphics[width=0.27\linewidth]{EqualIm-C=l_30-03_+k_21-12_-U=0-min_ax=2-maj_ax=3-theta=4pid12.png}
    \includegraphics[width=0.27\linewidth]{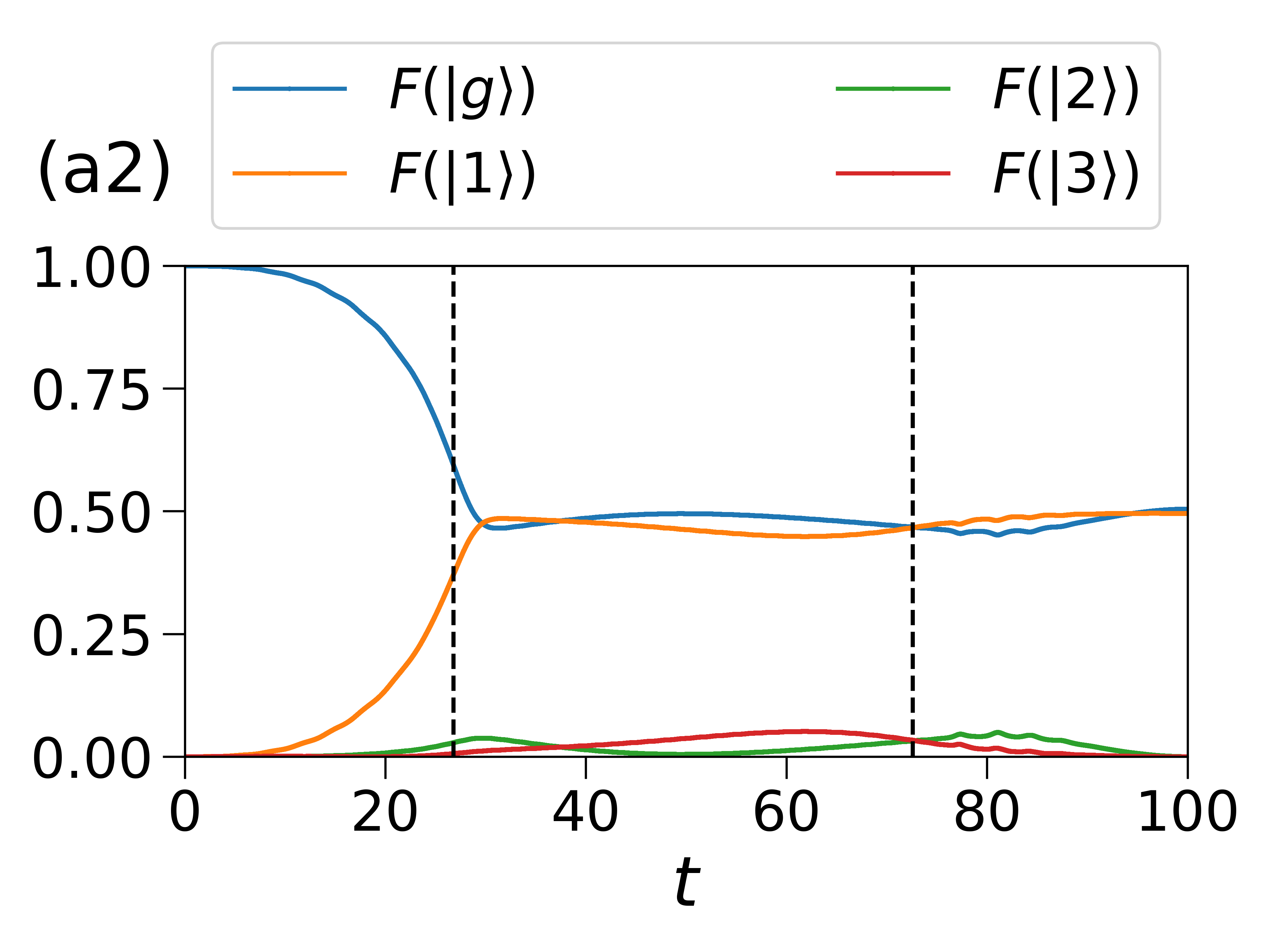}
    \includegraphics[width=0.27\linewidth]{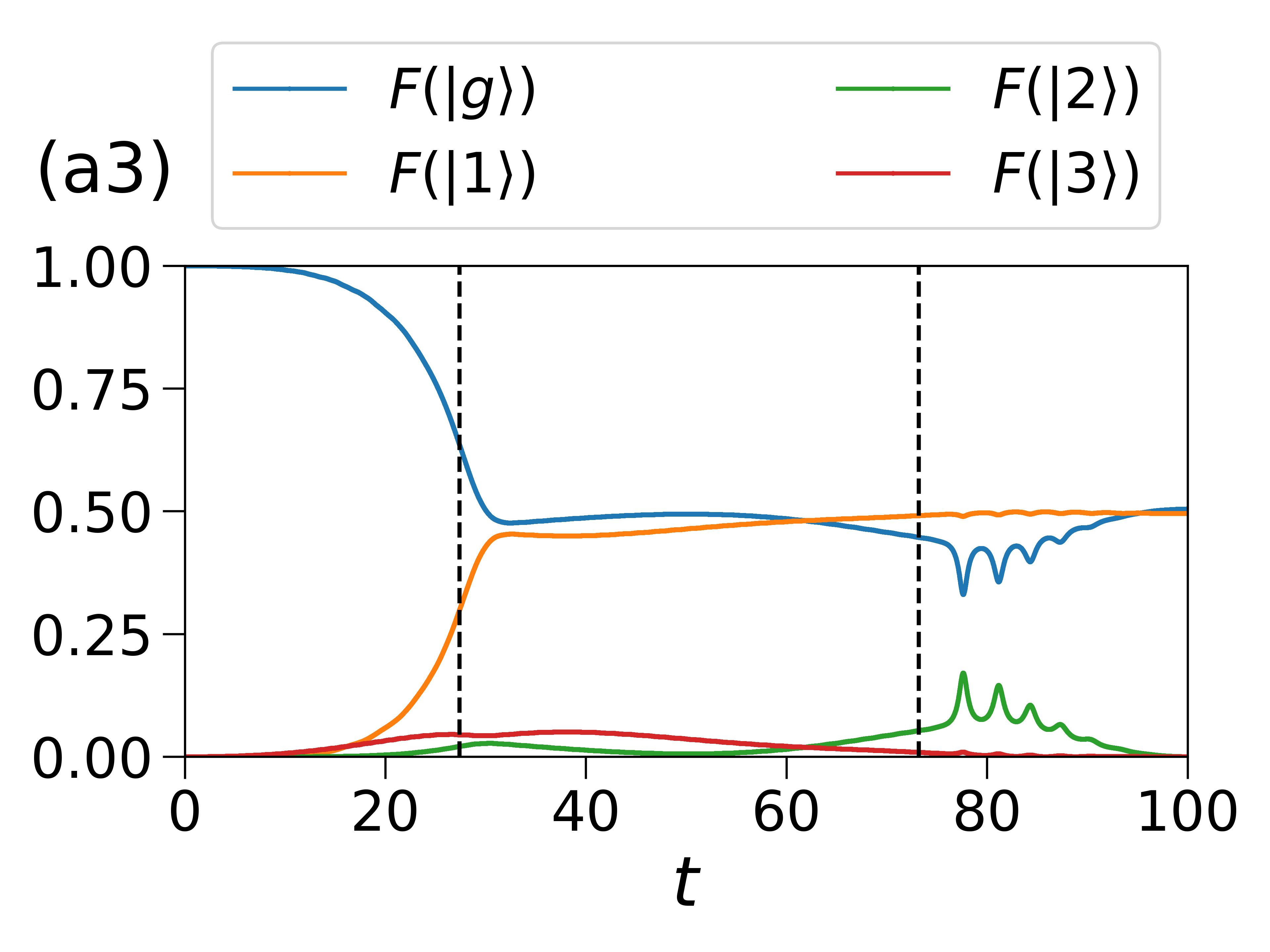}
    \\
    \includegraphics[width=0.27\linewidth]{EqualIm-C=l_30-03_+k_21-12_-U=0-min_ax=2-maj_ax=3-theta=-1p5pid12.png}
    \includegraphics[width=0.27\linewidth]{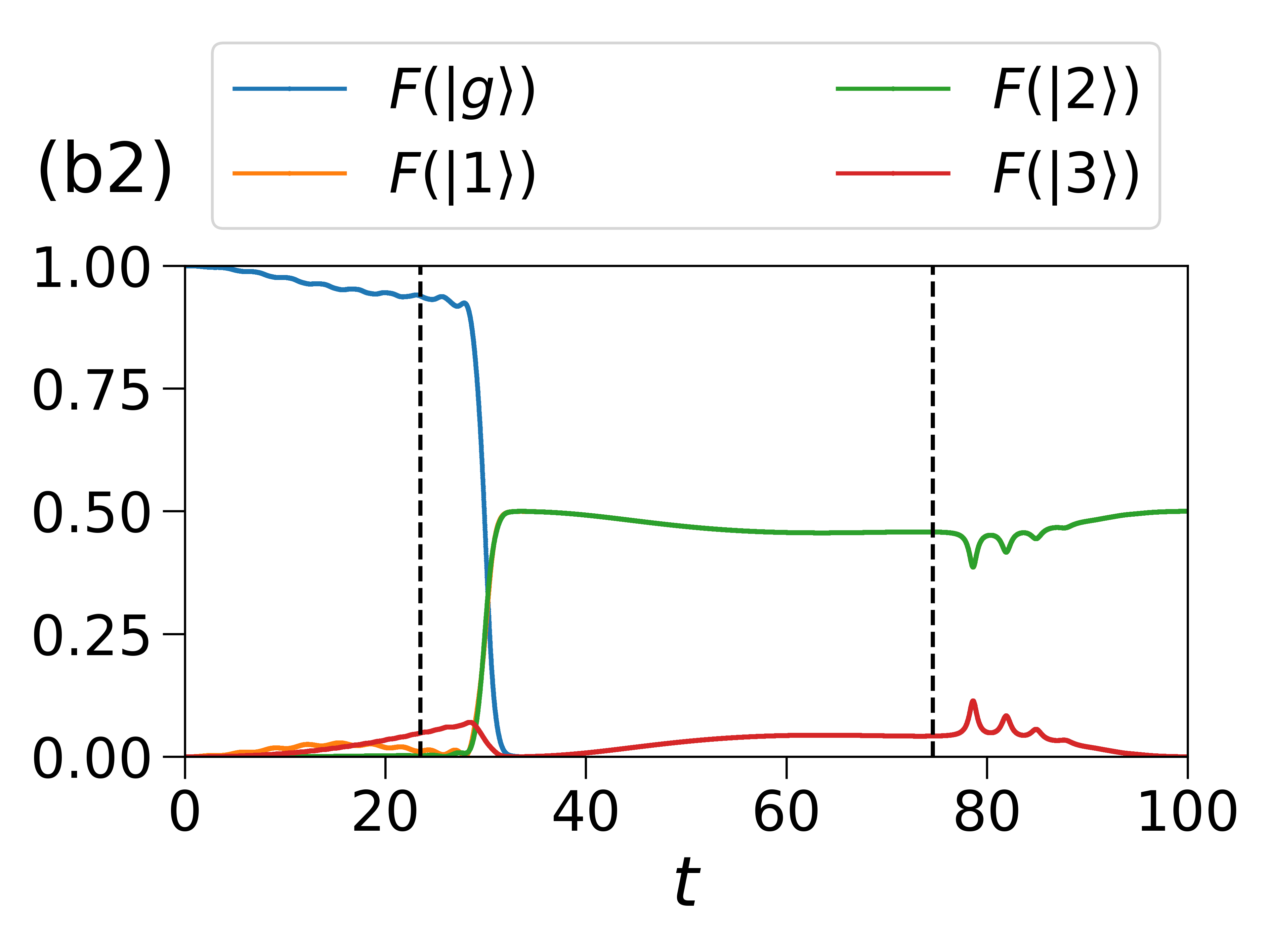}
    \includegraphics[width=0.27\linewidth]{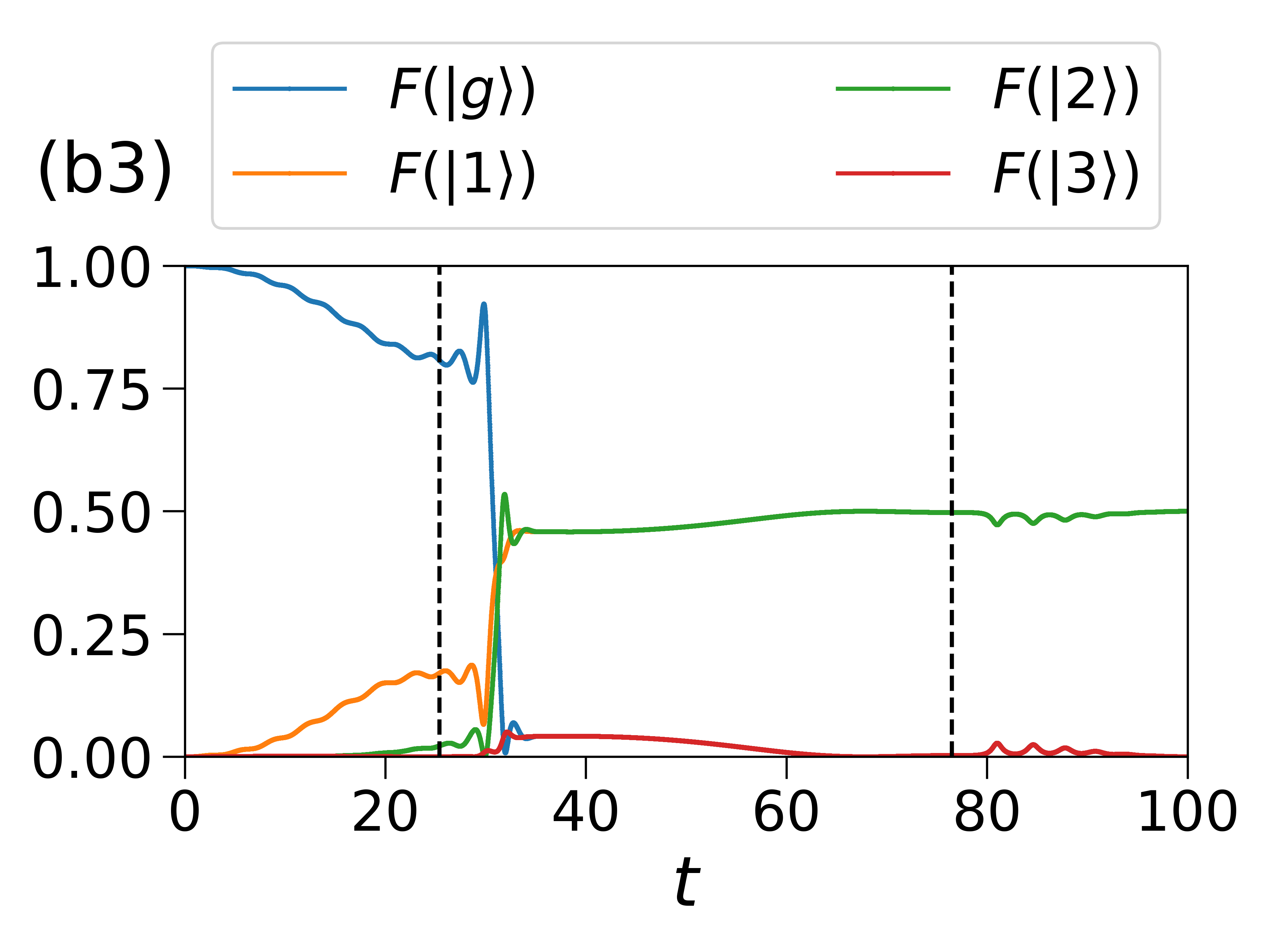}
    \\
    \includegraphics[width=0.27\linewidth]{EqualIm-C=l_30-03_+k_21-12_-U=0-min_ax=2-maj_ax=4-theta=1p25pid12.png}
    \includegraphics[width=0.27\linewidth]{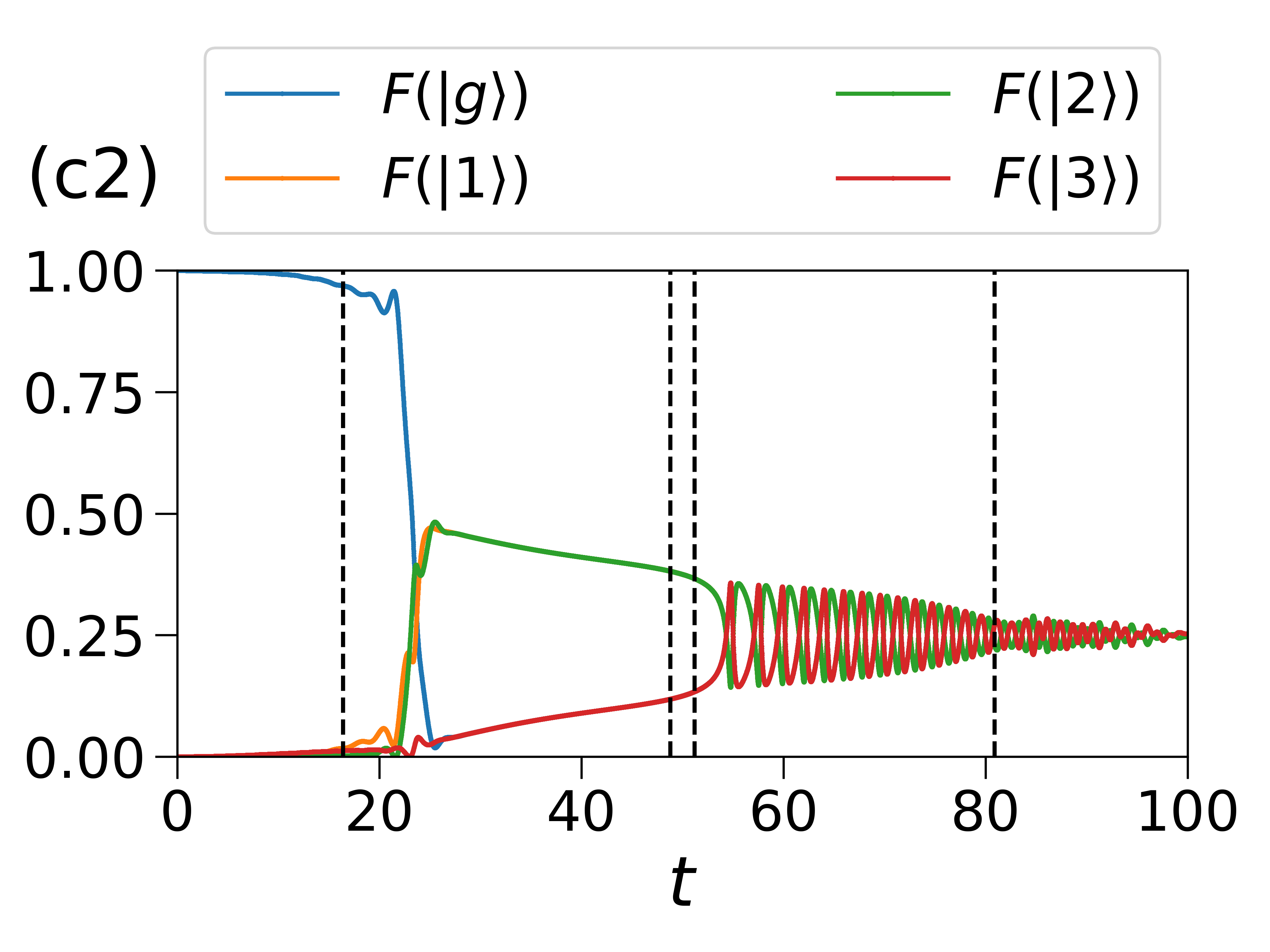}
    \includegraphics[width=0.27\linewidth]{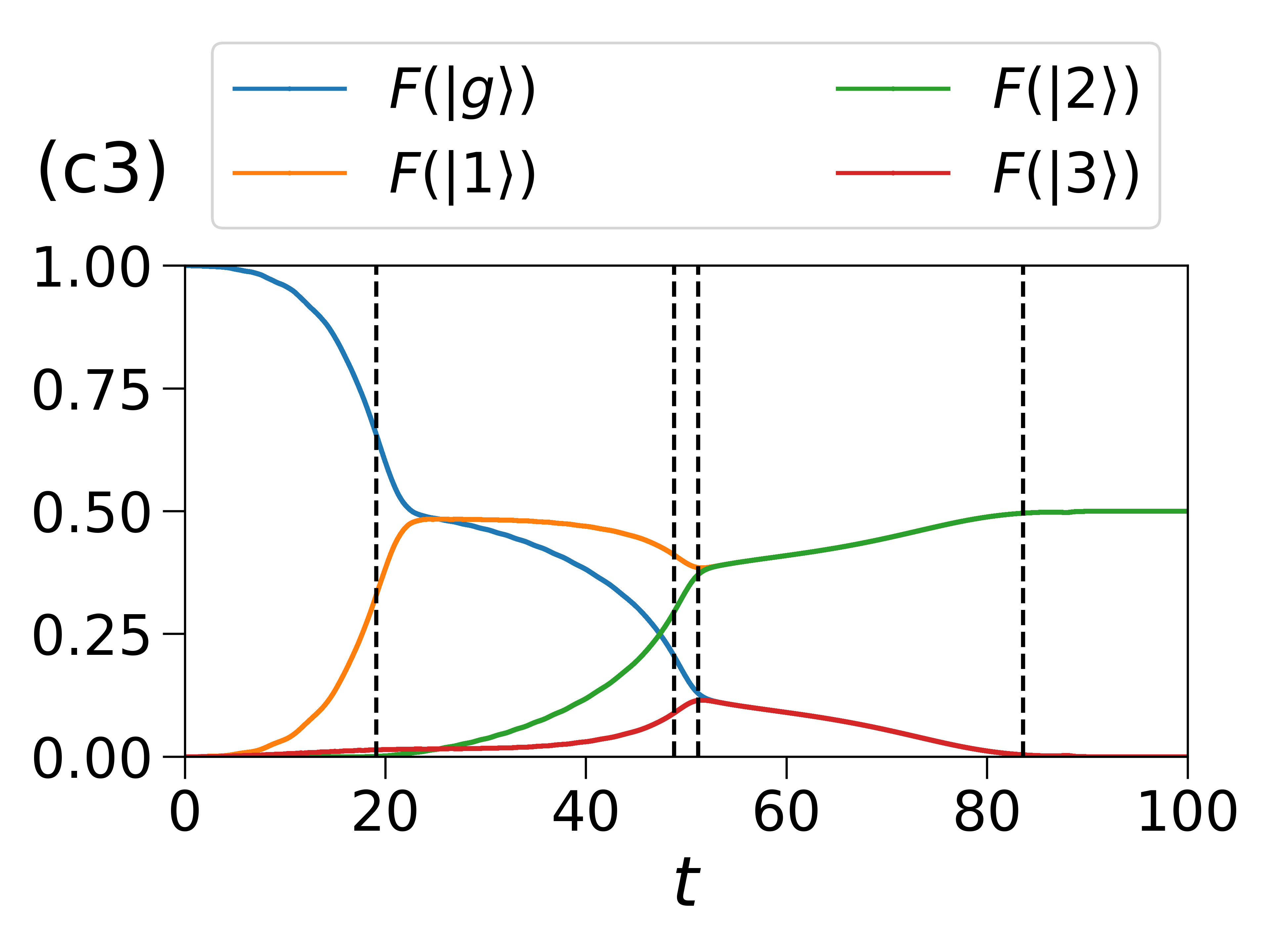}
    \caption{Deterministic evolution of the fidelities of the eigenstates of the Hermitian Hamiltonian generated by the non-Hermitian Hamiltonian of Eq.~\eqref{eq-dephasing-hamiltonian-traj-U=0} with time-dependent transformations (\ref{ellipse-transformation}). The total time of evolution is $T=100$. Figures (a1, b1, c1) show the parameter trajectory in yellow. The starting point is at $(l,\,k)=(0,\,0)$. Figures (a2, b2, c2) show the fidelities for counterclockwise trajectories and figs. (a3, b3, c3) for clockwise trajectories. All plots are with $U=0$ and the initial state is the ground state of the Hermitian Hamiltonian. Vertical dashed lines correspond to the instants of times when an exceptional point is crossed.}
    \label{dephasing-detev-loops-U=0-fids-in=ground}
\end{figure}

Next we investigate the effects of the rate operator transformation (\ref{eq-dephasing-transformation}) on the dynamics of the non-Hermitian Bose-Hubbard dimer. We study time-dependent parameter trajectories $(l(t),\,k(t))$ and find chiral state transfer for certain parameter trajectories of slowly varying transformation. We take elliptic $(l,\,k)$ space loop paths of the form
\begin{equation}
    \begin{split}
        l(t) &= -A\cos(2\pi t/T)\cos(\theta) \pm 2\sin(2\pi t/T)\sin(\theta) + A\cos(\theta), \\
        k(t) &= -A\cos(2\pi t/T)\sin(\theta) \mp 2\sin(2\pi t/T)\cos(\theta) + A\sin(\theta), \label{ellipse-transformation}
    \end{split}
\end{equation}
where $A$ and $\theta$ are constants used to control the shape of the loop. The upper signs correspond to a counterclockwise (CCW) trajectories and the lower signs correspond to a clockwise (CW) trajectories. The parameter $T$ is the time to complete the loop starting and ending at $(l,k)=(0,0)$. Its value is chosen to be large enough for the dynamics to be adiabatic in the $\mathcal{PT}$-symmetric phase and to reveal chiral behavior. The effective Hamiltonian $K(t)$ reads now
\begin{align}
    K(t)=-(a_1^\dagger a_2 + a_2^\dagger a_1) - \frac{i}{2}l(t)(\ket{30}\bra{30}-\ket{03}\bra{03}) - \frac{i}{2}k(t)(\ket{21}\bra{21}-\ket{12}\bra{12}) - \frac{i}{2} p(t)I, \label{eq-dephasing-hamiltonian-traj-U=0}
\end{align}
where we have $J=1$ and the time dependence in $p(t)=3\gamma+1/(8\gamma)\left(|k(t)|^2+|l(t)|^2\right)$ comes from maintaining the rate operator positive semi-definite while keeping the imaginary part of the term proportional to the identity as small as possible. Note, that the eigenstates of the effective Hamiltonian and its Hermitian part $H$ coincide at $t=0$ and $t=T$ as the homothety $ - \frac{i}{2} p(t)I$ does not affect the eigenstates.

In the following we show results on parameter trajectories that extend outside of the $\mathcal{PT}$-symmetric phase surrounded by EPs. Sufficiently slow trajectories fully contained inside the $\mathcal{PT}$-symmetric phase were found to lead to adiabatic evolution of the system state with no state transfer. In our investigation of state transfer the types of boundary crossing between different symmetry regions during the evolution and the least decaying eigenstate in each region are more relevant than the exact value of $A$ (see \eqref{ellipse-transformation}), so we use only one value of $A$ for each orientation of the loop. In section \ref{sec-U=0} we use values $A=4$ for the loop encircling an order 4 EP and $A=3$ for others. In section \ref{sec-U=5} we use $\theta=\pi/2$ and $A=1.5$. The point at which an exceptional line is crossed was found to have no influence on the final state. Trajectories shown in the following are different from the usual case in literature where the trajectories do not cross the EPs. This difference leads to final states that are superpositions of different eigenstates of the effective Hamiltonian instead of a specific eigenstate, and for some parameter trajectories they depend on the initial state. 

\subsubsection{No on-site interaction \label{sec-U=0}}

Figures \ref{dephasing-detev-loops-T=all-U=0-ground} and \ref{dephasing-detev-loops-T=all-U=0-e2} show the evolution of the populations of the number basis with time-dependent parameter trajectories for interaction strength $U=0$. In Fig.~\ref{dephasing-detev-loops-T=all-U=0-ground} the initial state is the ground state of the Hamiltonian $H$ and in Fig.~\ref{dephasing-detev-loops-T=all-U=0-e2} it is the second excited state of the Hamiltonian $H$. The behavior of the state towards the end of the evolution is in most cases independent of the initial state. This is due to the fact that in slow trajectories the state has sufficient time to reach the least decaying state. The final state is then decided in most cases only by the least decaying state(s) at late times before re-entering the phase of unbroken $\mathcal{PT}$-symmetry. The exception is trajectories passing the DIP region, where there are two least decaying states, which allows for some initial information to be conserved for some parameter trajectories.

The exceptions caused by the DIP region can be explained for the EP encircling trajectories as follows. When crossing from DIP to 0PTS to 2PTS or 2PTS to 0PTS to DIP, there is no difference in the final state, because the least decaying state is unique in the 2PTS region, and this state determines the evolution after these crossings in either direction. The only case where the initial state matters is in trajectories that cross from 4PTS to DIP and back to 4PTS. When crossing from 4PTS to DIP region, all instantaneous eigenstates coalesce with one other instantaneous eigenstate at the EP and after the crossing one of these will be least decaying and the state will quickly reach that eigenstate. When crossing back to the 4PTS region these eigenstates coalesce and separate again and the evolution continues in a superposition of two eigenstates that depends on the initial state.

The sustained quick oscillations for long times in Figs. \ref{dephasing-detev-loops-T=all-U=0-ground} and \ref{dephasing-detev-loops-T=all-U=0-e2} are due to two reasons. Firstly, the crossing of an exceptional point, in which some eigenstates coalesce and then separate again leads to many eigenstates being populated after the bifurcation. Secondly, the coalescing of imaginary parts of the eigenvalues at the exceptional point and them remaining equal in the 4PTS region leads to a subspace of dimension larger than one where oscillations persist due to interference between the multiple eigenstates populated during the crossing of the exceptional point. We can see chiral behavior in the trajectories that encircle the order 4 EP. This is due to this trajectory being the only one studied here that returns to the symmetry unbroken phase from different phases depending on the direction of traversal of the trajectory. For the other trajectories both traversal directions are qualitatively similar (4PTS to DIP/2PTS to 4PTS). Therefore the states after boundary crossings are related to the same eigenstate(s) independent of the traversal direction. For slow trajectories the evolution is approximately adiabatic in the unbroken $\mathcal{PT}$-symmetry phase such that both of these states reach the same final state.

\begin{figure}[t]
    \includegraphics[width=0.27\linewidth]{EqualIm-C=l_30-03_+k_21-12_-U=0-min_ax=2-maj_ax=3-theta=4pid12.png}
    \includegraphics[width=0.27\linewidth]{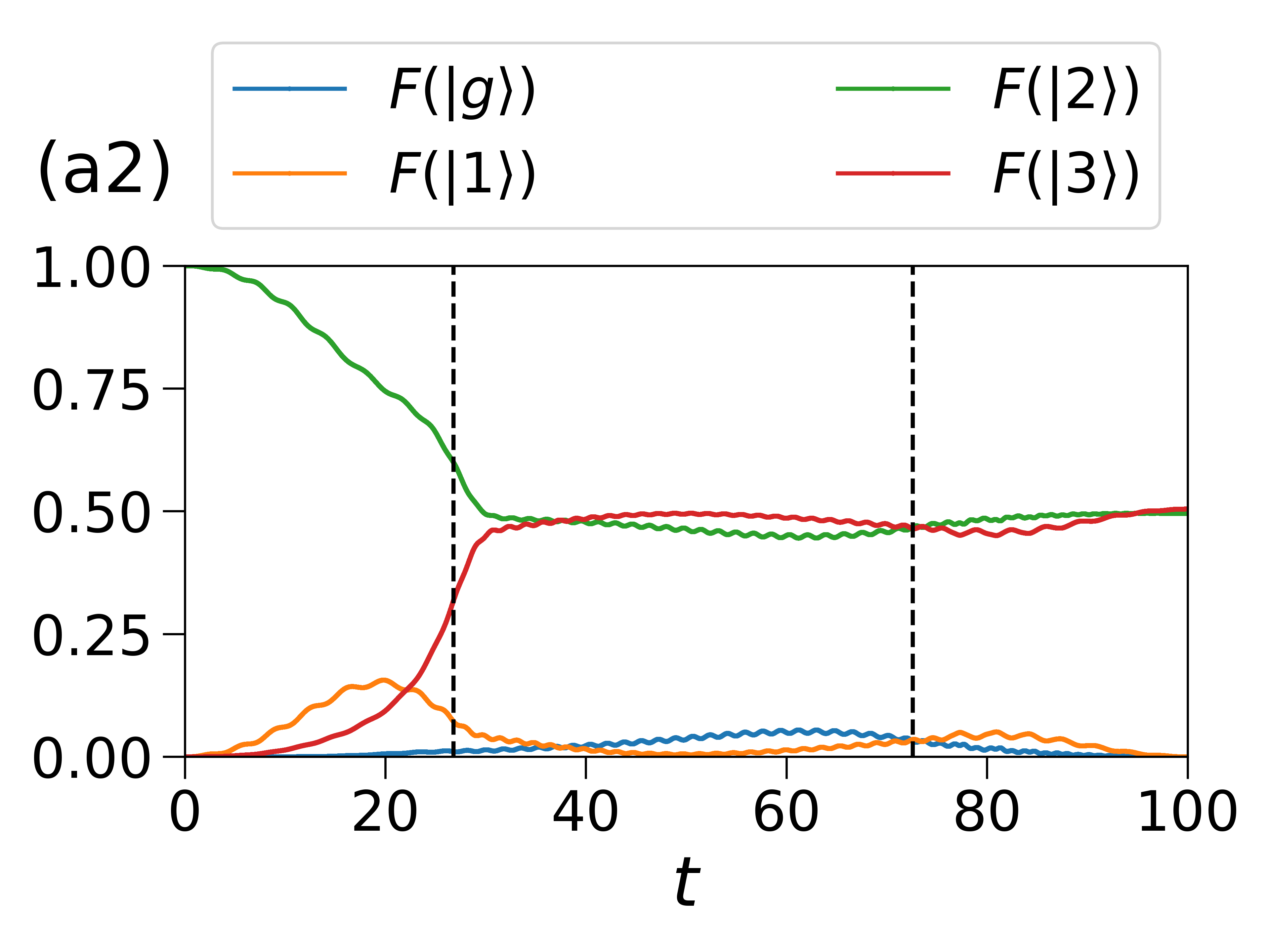}
    \includegraphics[width=0.27\linewidth]{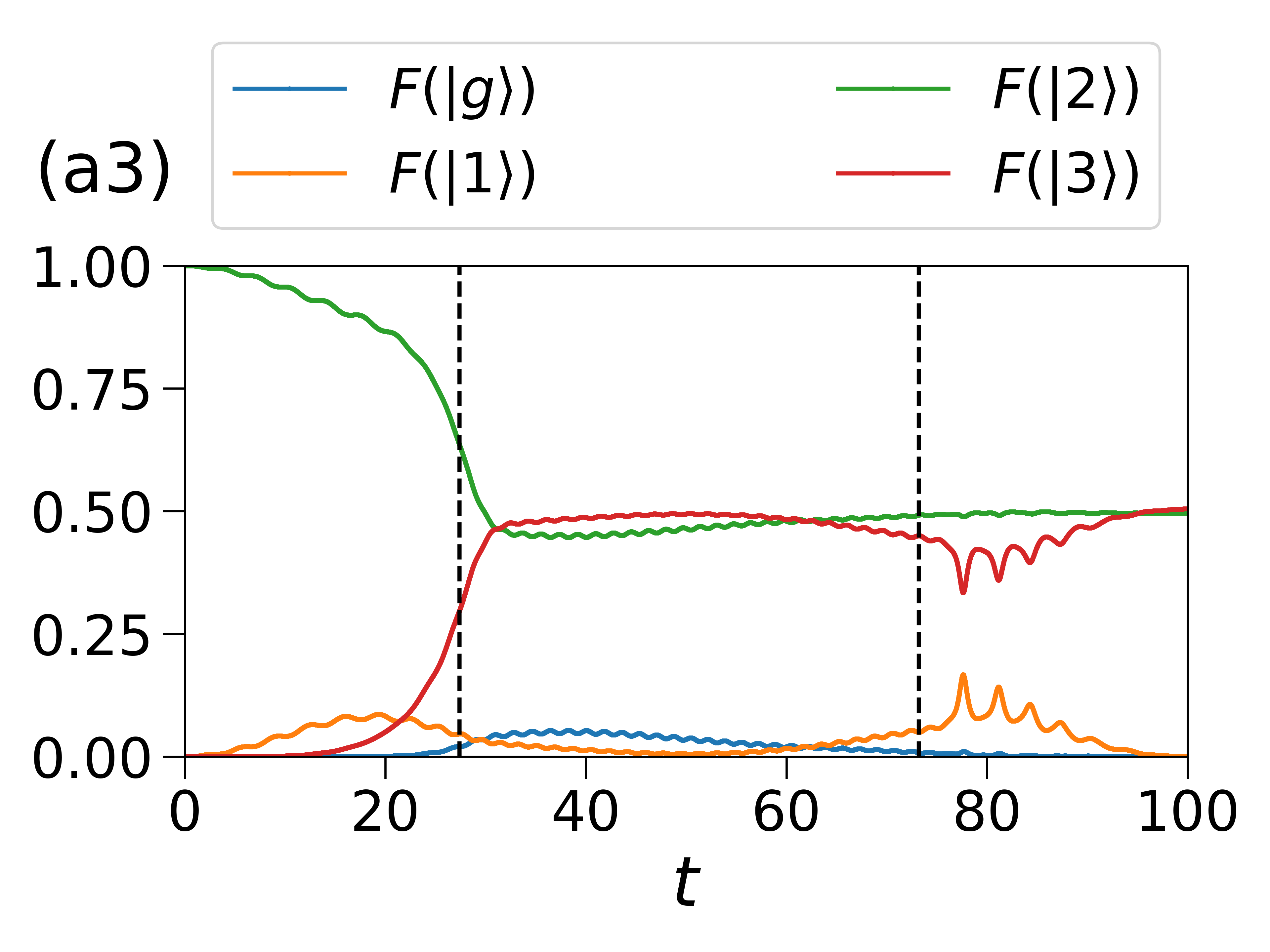}
    \\
    \includegraphics[width=0.27\linewidth]{EqualIm-C=l_30-03_+k_21-12_-U=0-min_ax=2-maj_ax=3-theta=-1p5pid12.png}
    \includegraphics[width=0.27\linewidth]{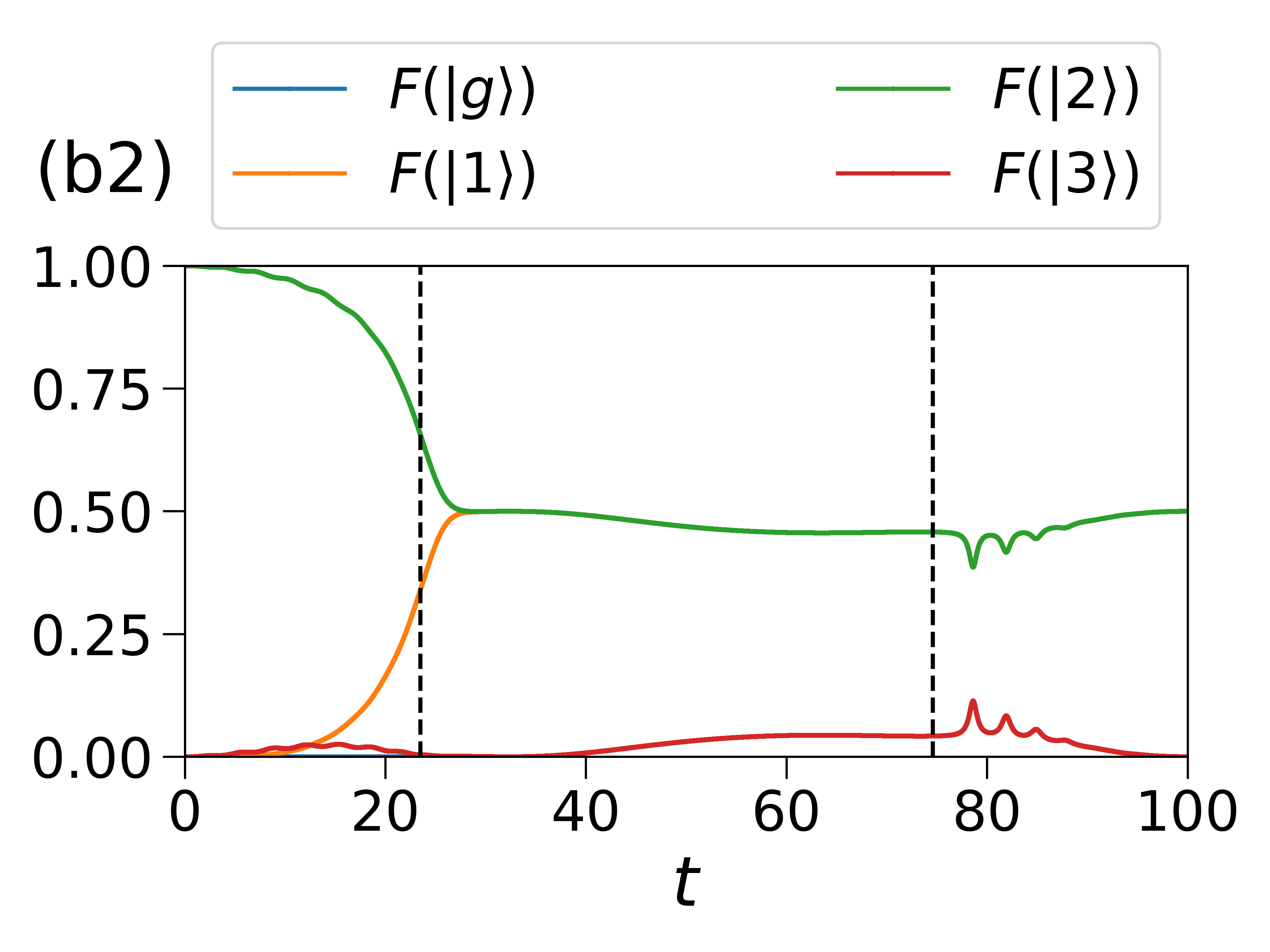}
    \includegraphics[width=0.27\linewidth]{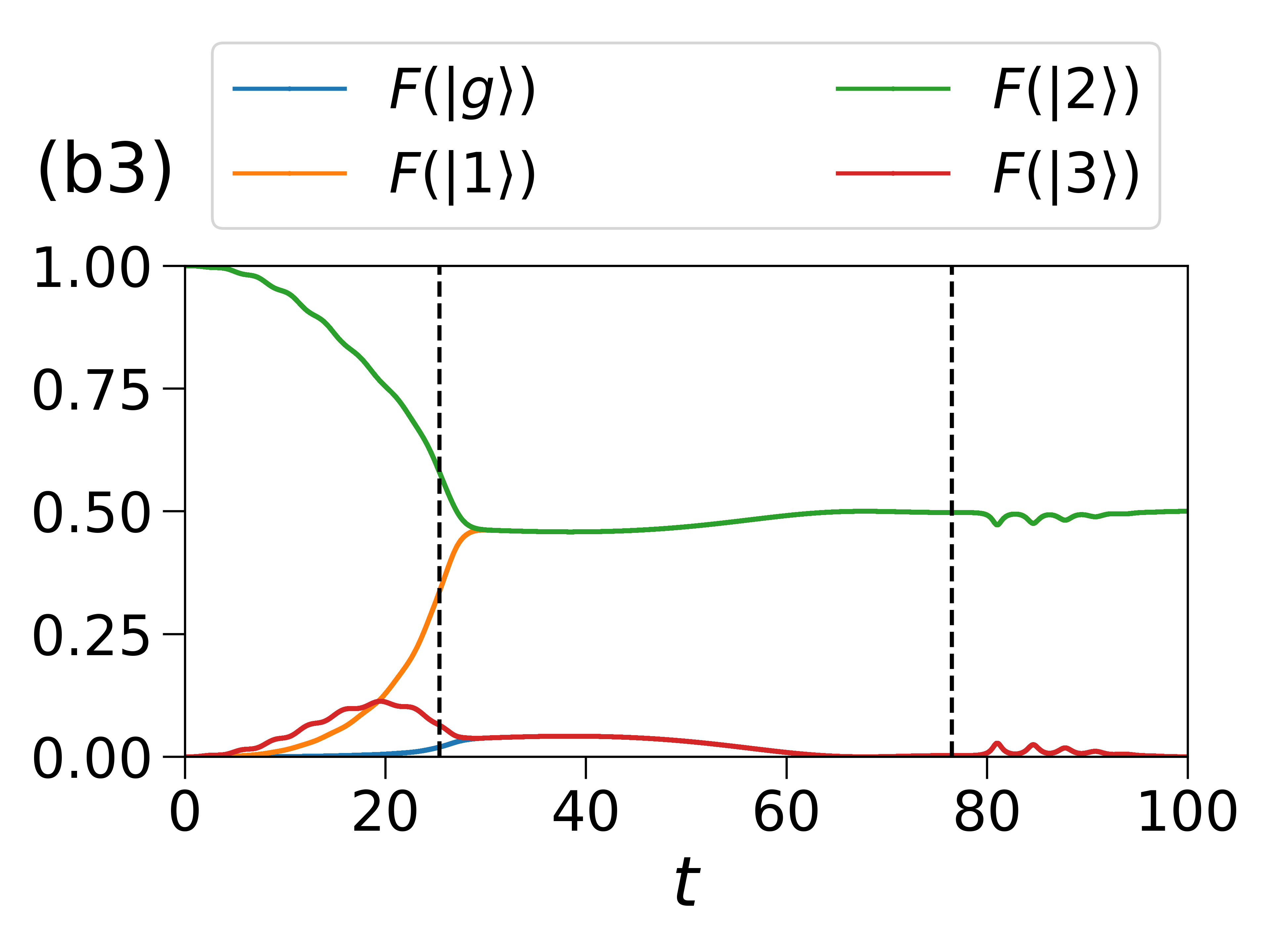}
    \\
    \includegraphics[width=0.27\linewidth]{EqualIm-C=l_30-03_+k_21-12_-U=0-min_ax=2-maj_ax=4-theta=1p25pid12.png}
    \includegraphics[width=0.27\linewidth]{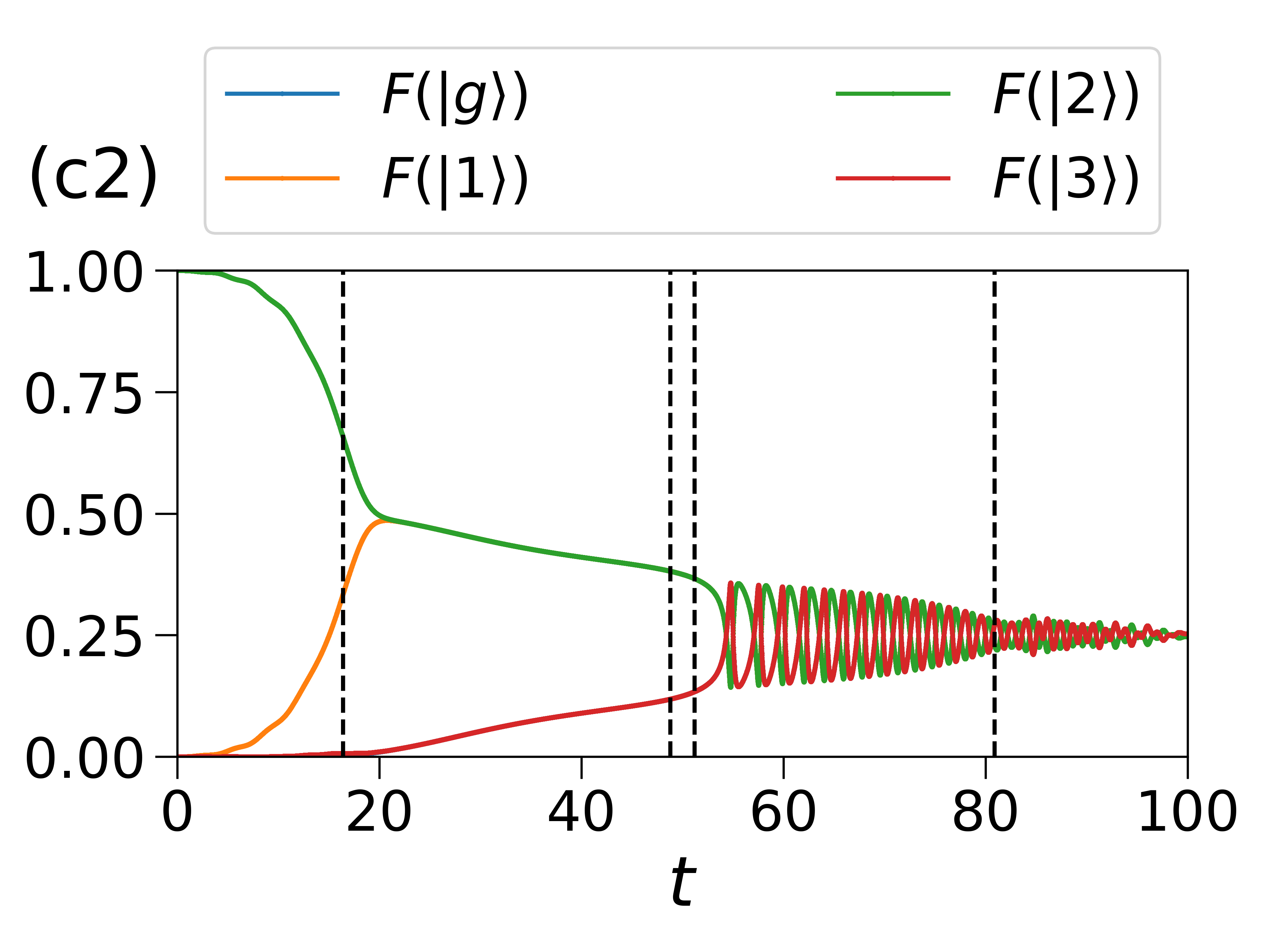}
    \includegraphics[width=0.27\linewidth]{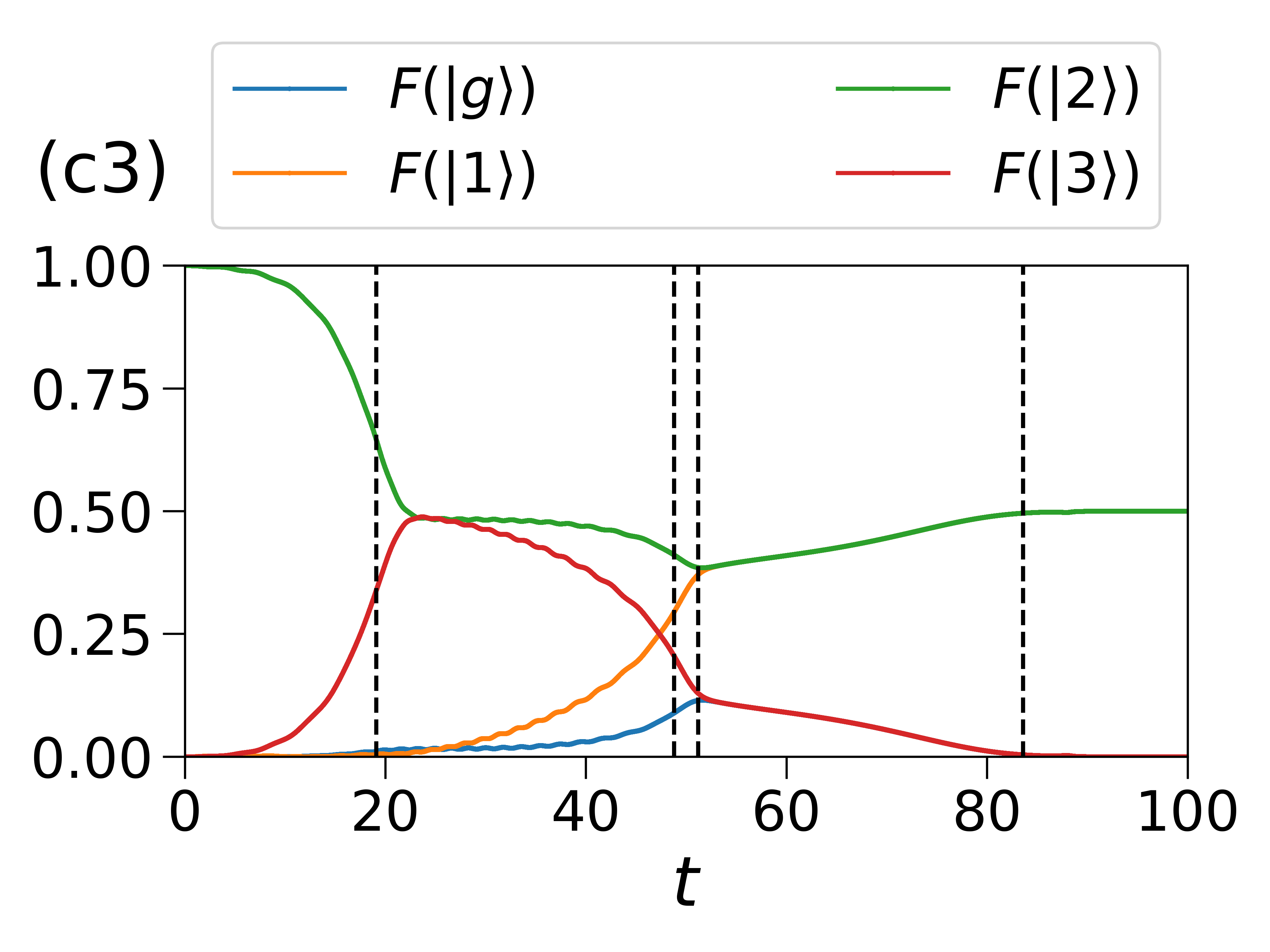}
    \caption{Deterministic evolution of the fidelities of the eigenstates of the Hermitian Hamiltonian generated by the non-Hermitian Hamiltonian of Eq.~\eqref{eq-dephasing-hamiltonian-traj-U=0} with time-dependent transformations (\ref{ellipse-transformation}). The total time of evolution is $T=100$. Figures (a1, b1, c1) show the parameter trajectory in yellow. The starting point is at $(l,\,k)=(0,\,0)$. Figures (a2, b2, c2) show the fidelities for counterclockwise trajectories and figs. (a3, b3, c3) for clockwise trajectories. All plots are with $U=0$ and the initial state is the second excited state of the Hermitian Hamiltonian. Vertical dashed lines correspond to the instants of times when an exceptional point is crossed.}
    \label{dephasing-detev-loops-U=0-fids-in=e2}
\end{figure}

The population of various eigenstates at the crossings of exceptional points is seen in a clearer way in Figs. \ref{dephasing-detev-loops-U=0-fids-in=ground} and \ref{dephasing-detev-loops-U=0-fids-in=e2}, where we plot the fidelities $F(\cdot,t)=|\braket{\cdot\,|\,\psi(t)}|^2$ of the eigenstates of the initial Hamiltonian $K(0)$, which has the same eigenstates as $H$, as a function of time. We see that the state ends up in a balanced superposition of eigenstates of $H$. When the parameter trajectory crosses from the blue region to the green region towards the end of the evolution, the final state is a superposition of two eigenstates. This happens because in the blue region the state is in the unique least decaying instantaneous eigenstate of the effective Hamiltonian, and at the crossing of the exceptional point, it coalesces with one other instantaneous eigenstate and after they separate again, the state evolves as a balanced superposition of them. During the rest of the evolution both of these components evolve adiabatically to the eigenstates of the Hamiltonian $K(T)=K(0)$. When encircling the order 4 EP in CCW direction, the least decaying eigenstate in the blue and white region coalesces with one other eigenstate at an EP when crossing to the red region. After the EP the state is in a superposition of two instantaneous eigenstates that are both least decaying. Each of these coalesce with one of the remaining eigenstates at an EP between the red and green regions and after the EP all four instantaneous eigenstates are populated. These again evolve adiabatically to the eigenstates of the Hamiltonian $K(0)$ and the final state is a balanced superposition of all of the eigenstates.

\begin{figure}[t]
    \includegraphics[width=0.27\linewidth]{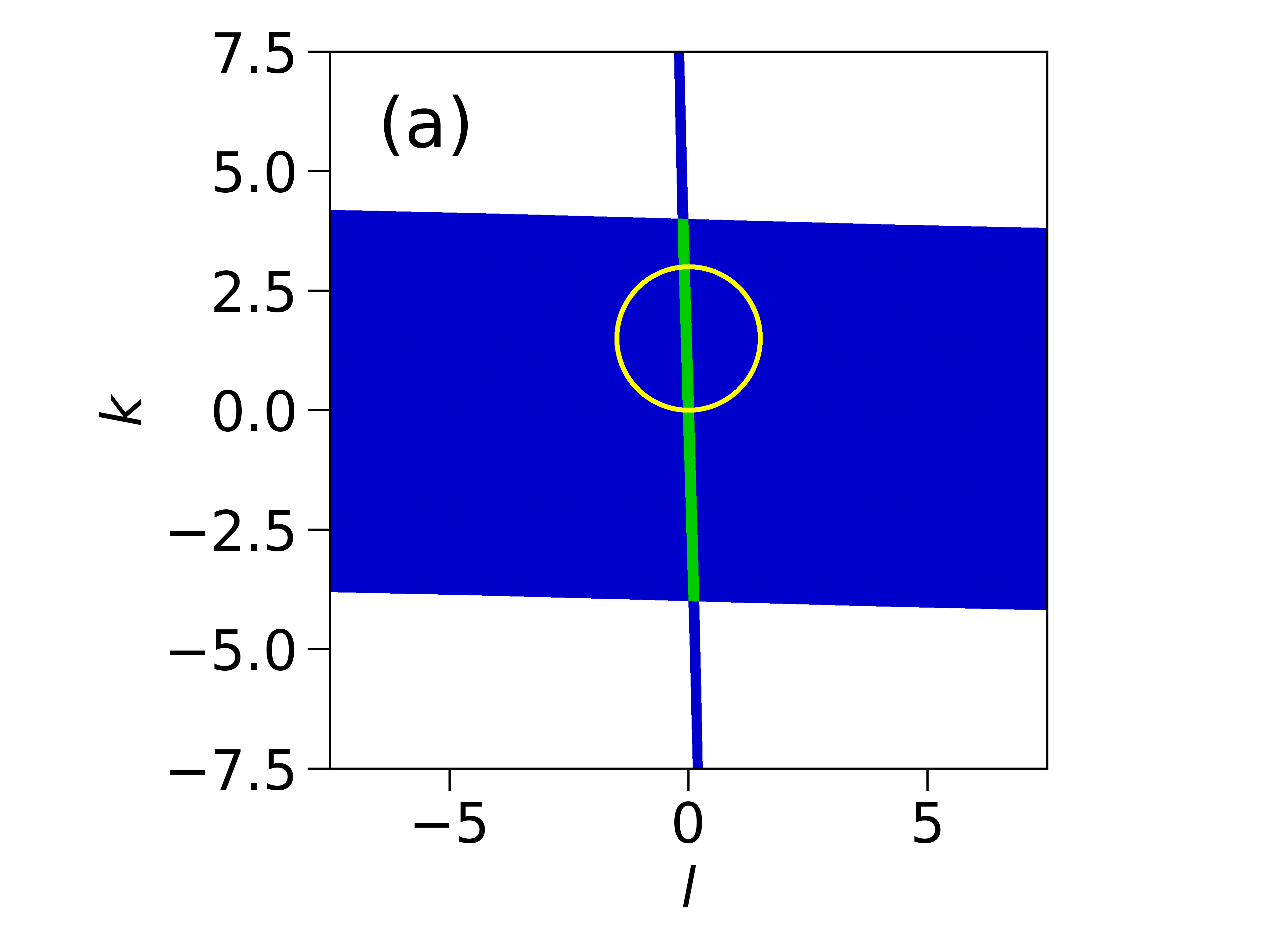}
    \includegraphics[width=0.27\linewidth]{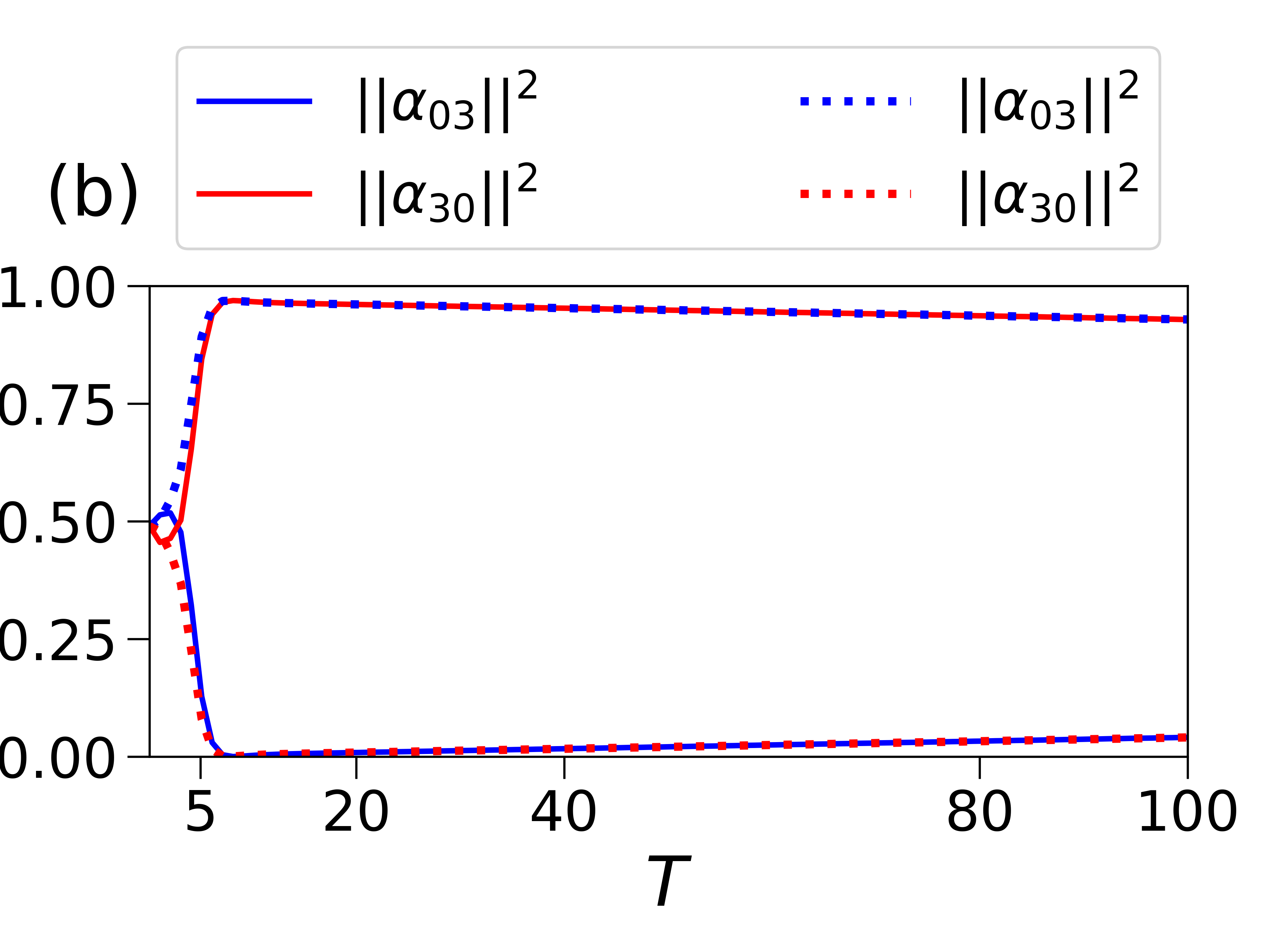}
    \\
    \includegraphics[width=0.27\linewidth]{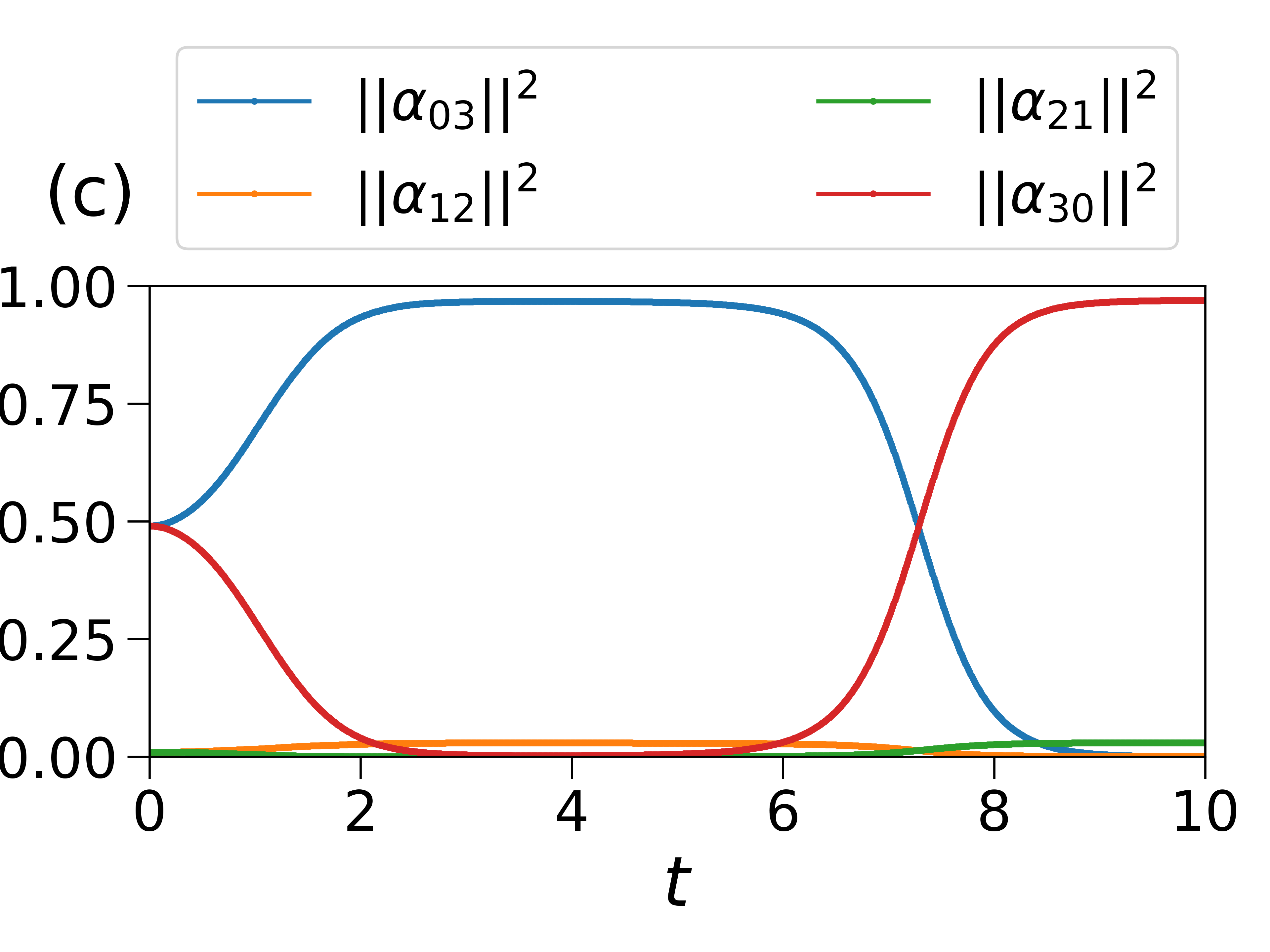}
    \includegraphics[width=0.27\linewidth]{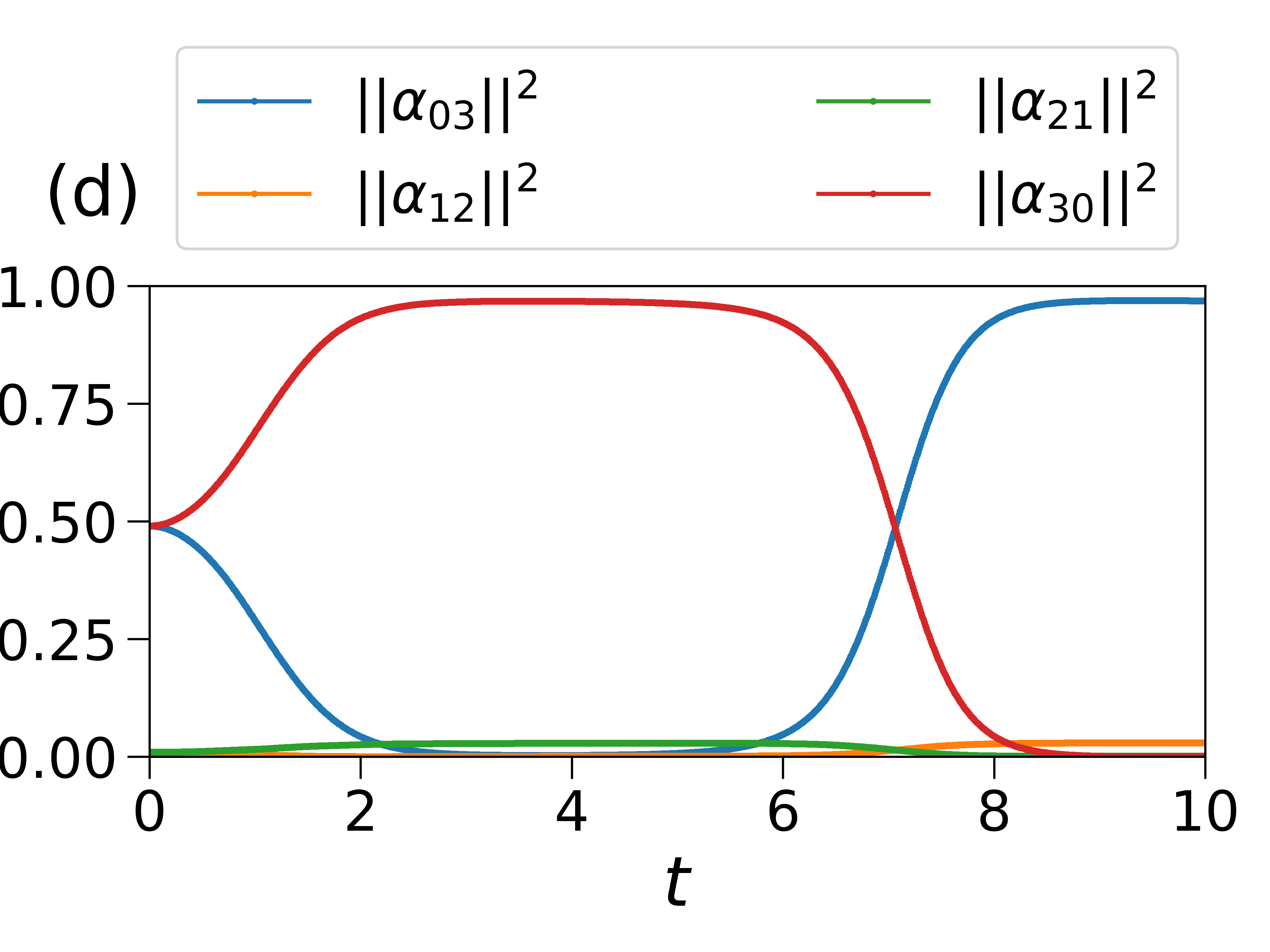}
    \caption{Deterministic evolution of populations $|\alpha_{ij}(t)|^2$ generated by the Hamiltonian of Eq.~\eqref{eq-dephasing-hamiltonian} with time-dependent transformations (\ref{ellipse-transformation-U=5}) and on-site interaction $U=5$. The initial state is the second excited state of the Hermitian Hamiltonian. Figure (a) shows the parameter trajectory in yellow. The starting point is at $(l,\,k)=(0,\,0)$. Figure (b) shows the final populations $|\alpha_{30}(T)|^2$ and $|\alpha_{03}(T)|^2$ at time $t=T$ for different values of $T$, starting at $T=0.1$, for CCW (solid lines) and CW (dashed lines) trajectories. Figure (c) and (d) show the evolution of the populations for CCW and CW trajectories, respectively. The total time of the evolution is $T=10$. Note the small difference in Figs. (c) and (d) in the times of the populations $|\alpha_{30}(T)|^2$ and $|\alpha_{03}(T)|^2$ switching due to the exceptional lines crossed not being exactly vertical.}
    \label{dephasing-detev-loops-T=all-U=5}
\end{figure}

\subsubsection{Non-zero, homogeneous on-site interaction \label{sec-U=5}}

We now proceed to study the non-Hermitian evolutions generated using time-dependent parameter trajectories with non-zero interaction $U=5$. Figures \ref{dephasing-detev-loops-T=all-U=5}(c) and \ref{dephasing-detev-loops-T=all-U=5}(d) show the evolution of the populations with a time-dependent circular parameter trajectory in CCW and CW directions, respectively, with initial values $(l(0),\,k(0))=(0,\,0)$ and
\begin{equation}
    \begin{split}
        l(t) &= \pm 1.5\sin(2\pi t/T), \\
        k(t) &= -1.5\cos(2\pi t/T) + 1.5. \label{ellipse-transformation-U=5}
    \end{split}
\end{equation}
The initial state is the second excited state of the initial Hamiltonian $K(0)$. The state evolution was found to have only a weak dependence on the shape of the trajectory, so we present here only one example. Here the oscillations relating to the crossing of exceptional points are not present, since the time scale of the oscillation increases along with $U$ and therefore the narrow region of unbroken $\mathcal{PT}$-symmetry is passed relatively quickly. The large timescale of the oscillations can be deduced from Fig. \ref{dephasing-detev-loops-T=all-U=5}(b), which shows the final population $|\alpha_{30}(T)|^2$ at time $t=T$ for different values of $T$. The local maximum is reached for $T\approx 7$ after which the final population lowers slowly as $T$ increases.

The trajectories were found to exhibit chiral behavior in the final state, as CCW and CW trajectories have different least decaying state in the long time limit. The least decaying states in the blue regions are states with high population of the state $\ket{30}$ for $l<0$ and $\ket{03}$ for $l>0$. We see that for sufficiently fast trajectories we can get high fidelity of these states even when we close the loop at $(l(T),k(T))=(0,0)$, though the states are not stable as they are not eigenstates of the final Hamiltonian.

\subsubsection{Non-zero, inhomogeneous on-site interaction}

\begin{figure}[t]
    \includegraphics[width=0.23\linewidth]{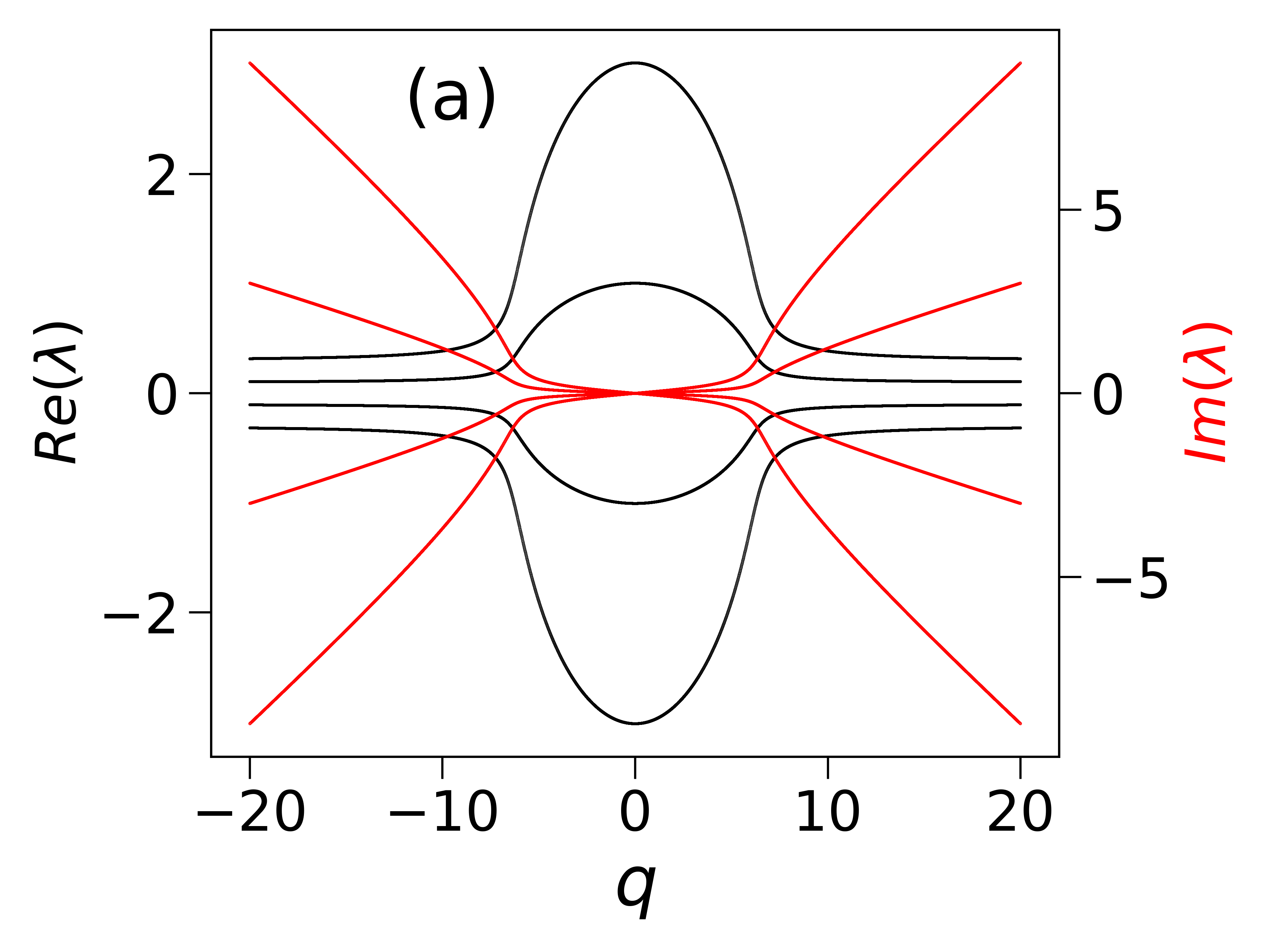}
    \includegraphics[width=0.23\linewidth]{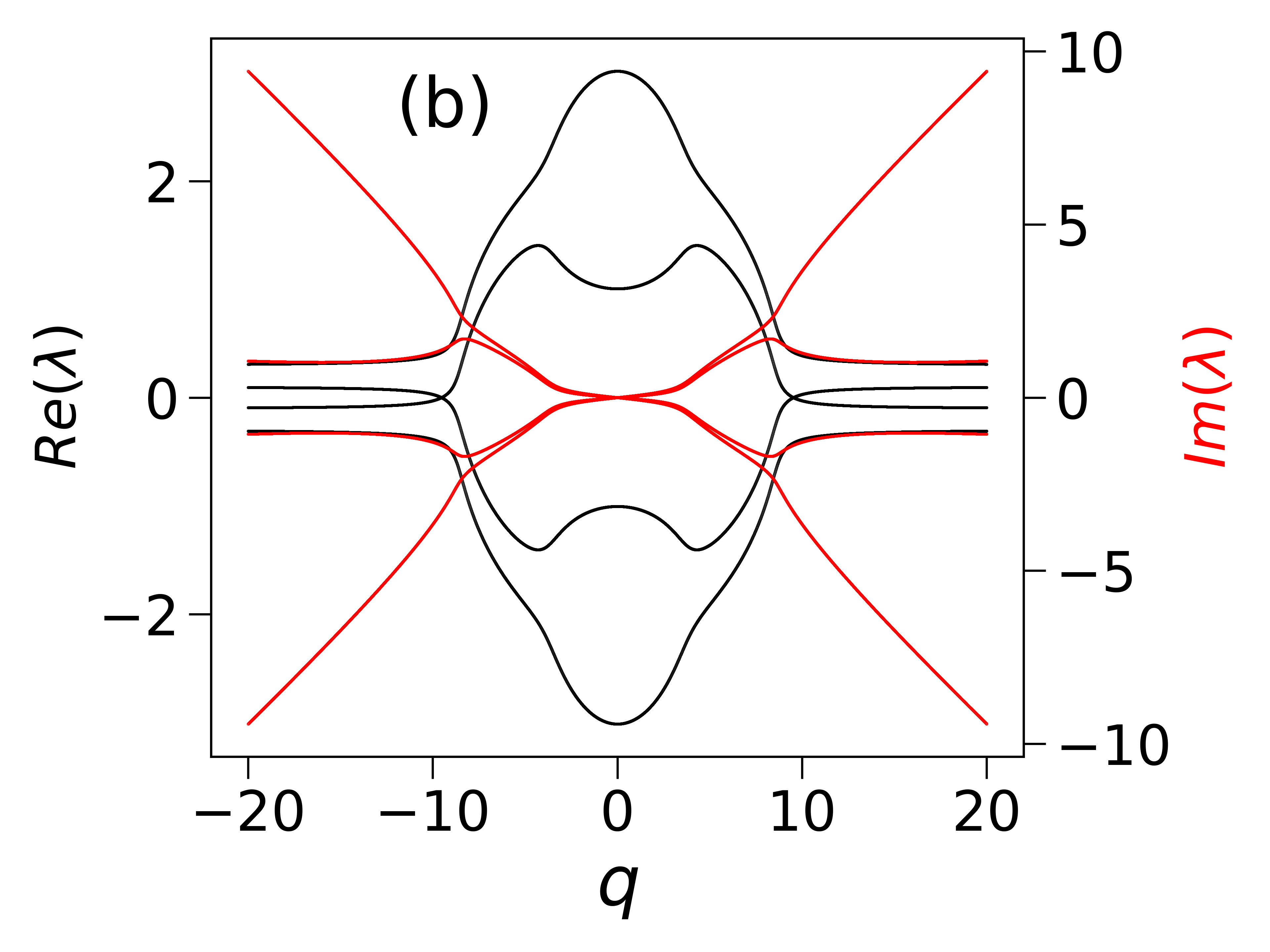}
    \includegraphics[width=0.23\linewidth]{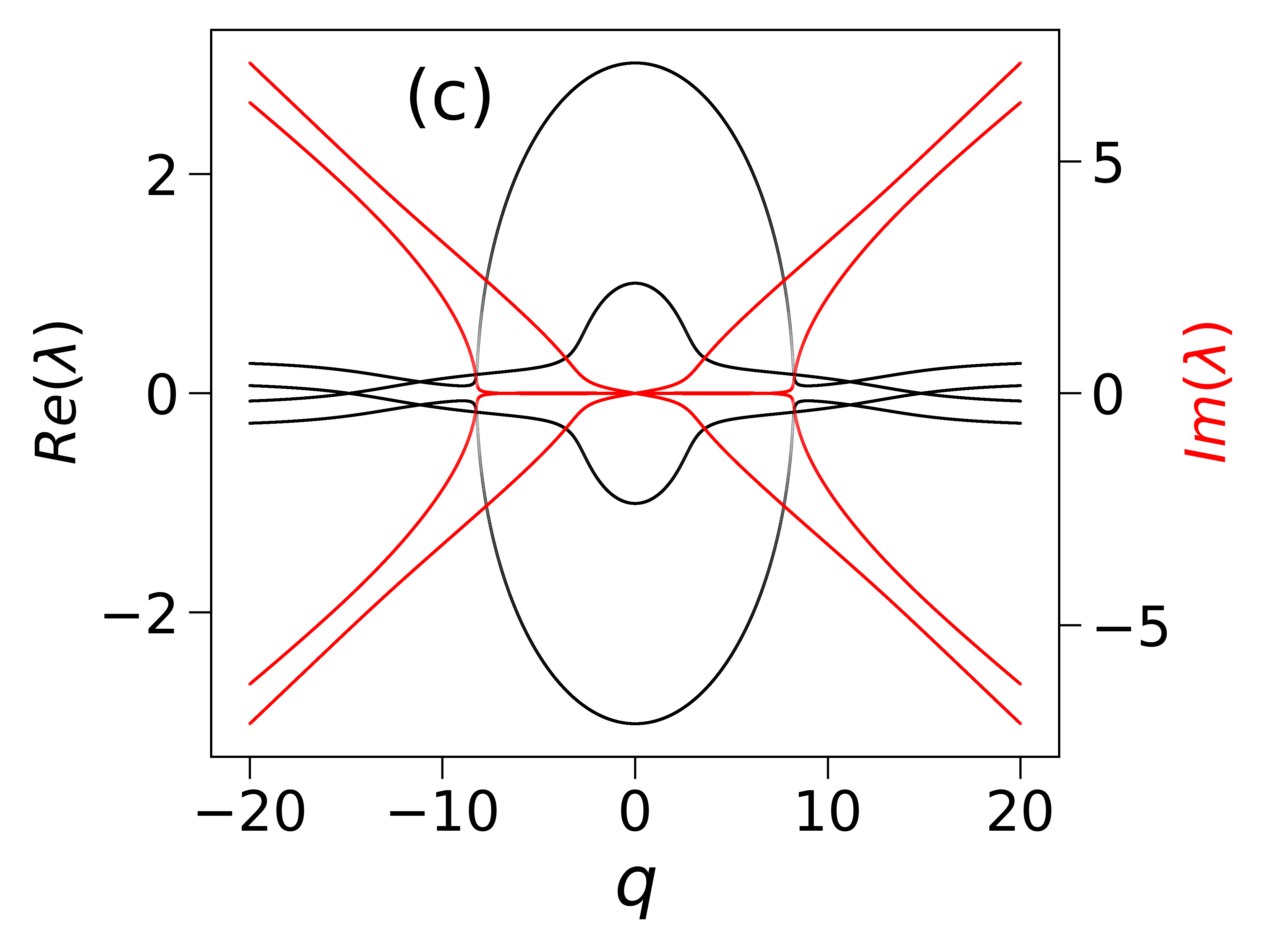}
    \includegraphics[width=0.23\linewidth]{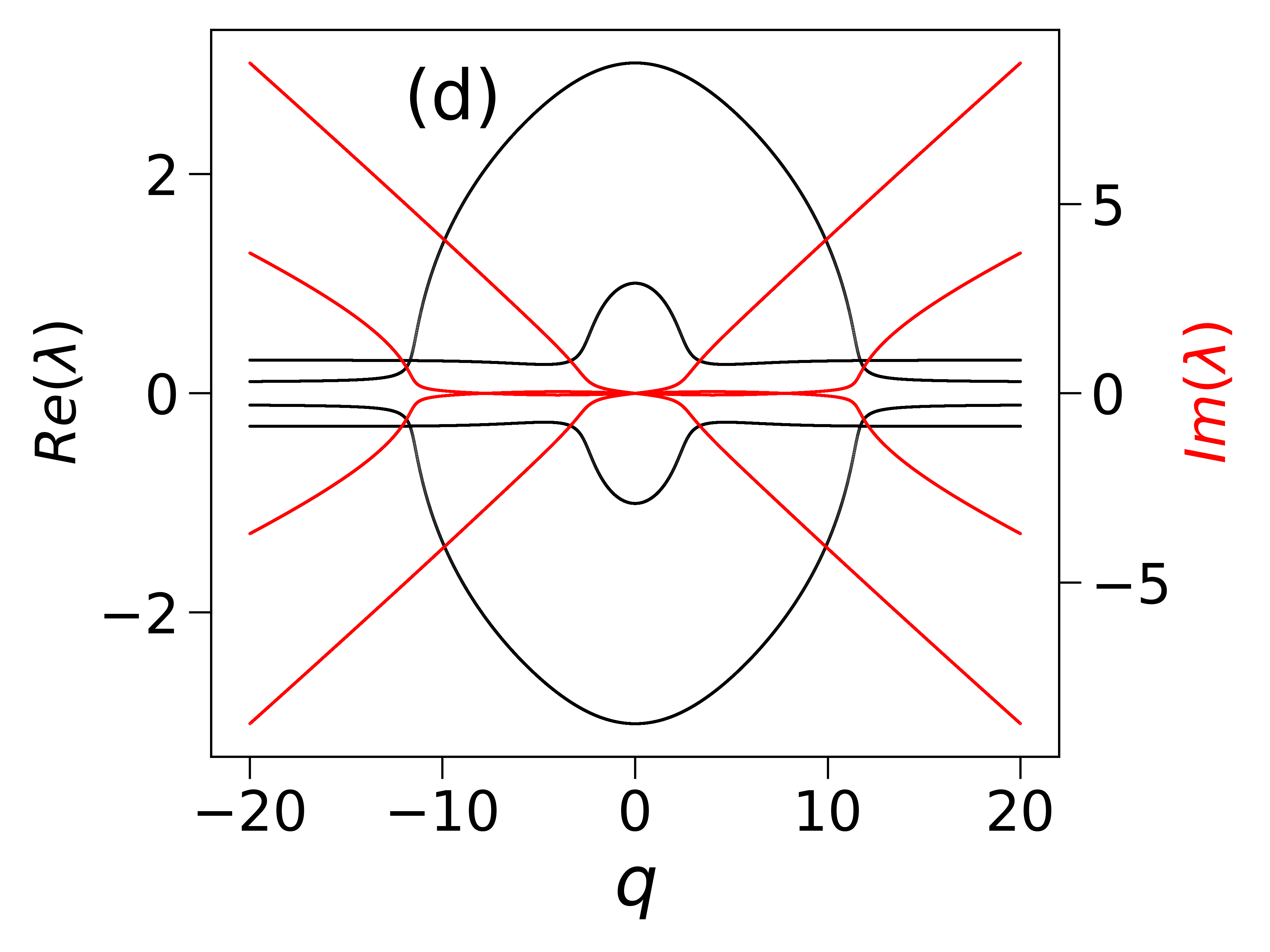}
    \caption{Real and imaginary parts of the eigenvalues of the effective Hamiltonian as a function of parameter $q$ for transformations \eqref{eq-lines-transform} (omitting the term proportional to the identity) with $\theta\in \{0,\,3\pi/8,\,5\pi/8,\,6\pi/8\}$ in figures (a-d), respectively. On-site interaction strengths are $U_1=-U_2=0.1$.}
    \label{fig-lines-through-origin-U1=-U2=0p1}
\end{figure}
\begin{figure}[t]
    \includegraphics[width=0.23\linewidth]{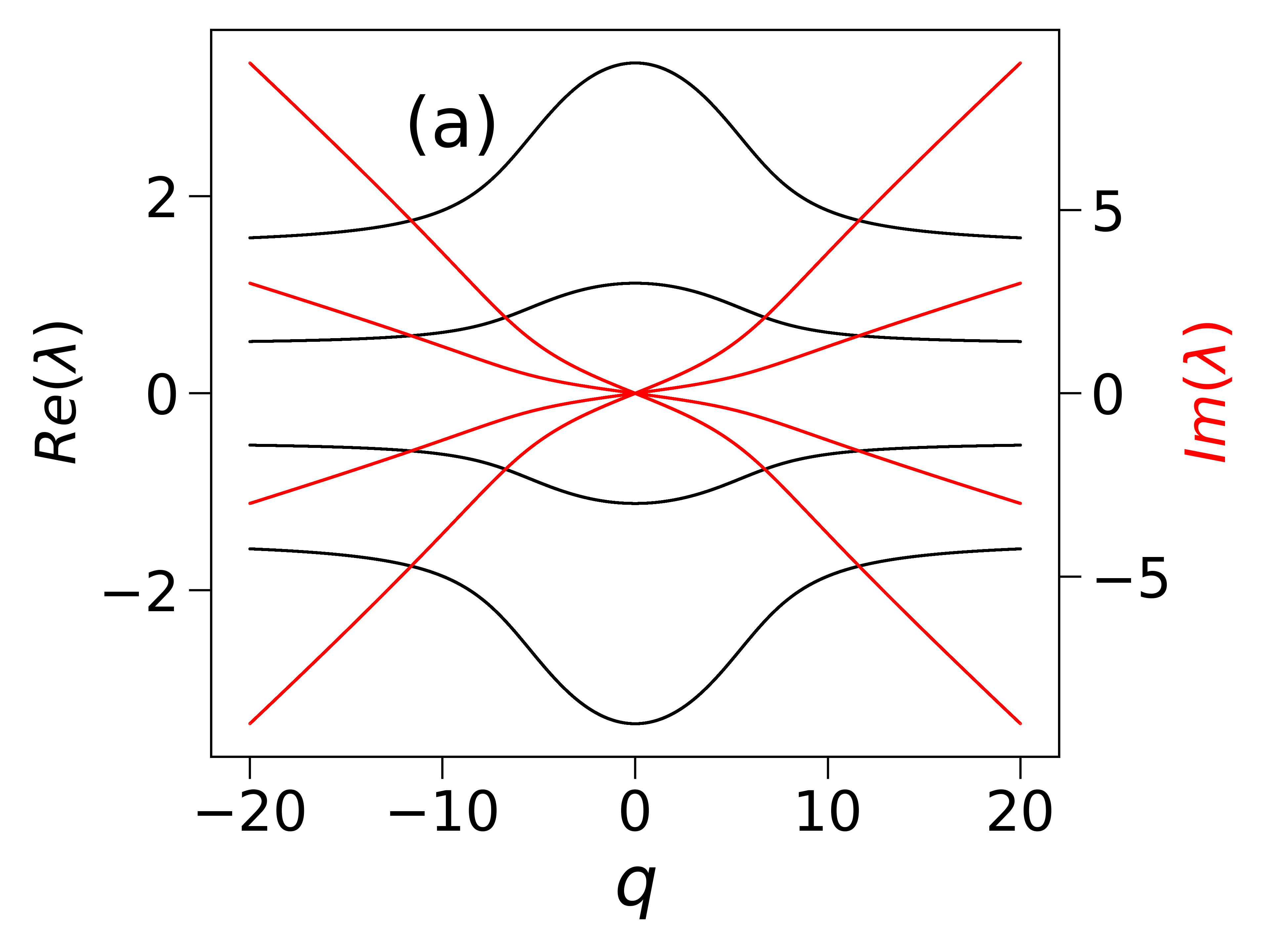}
    \includegraphics[width=0.23\linewidth]{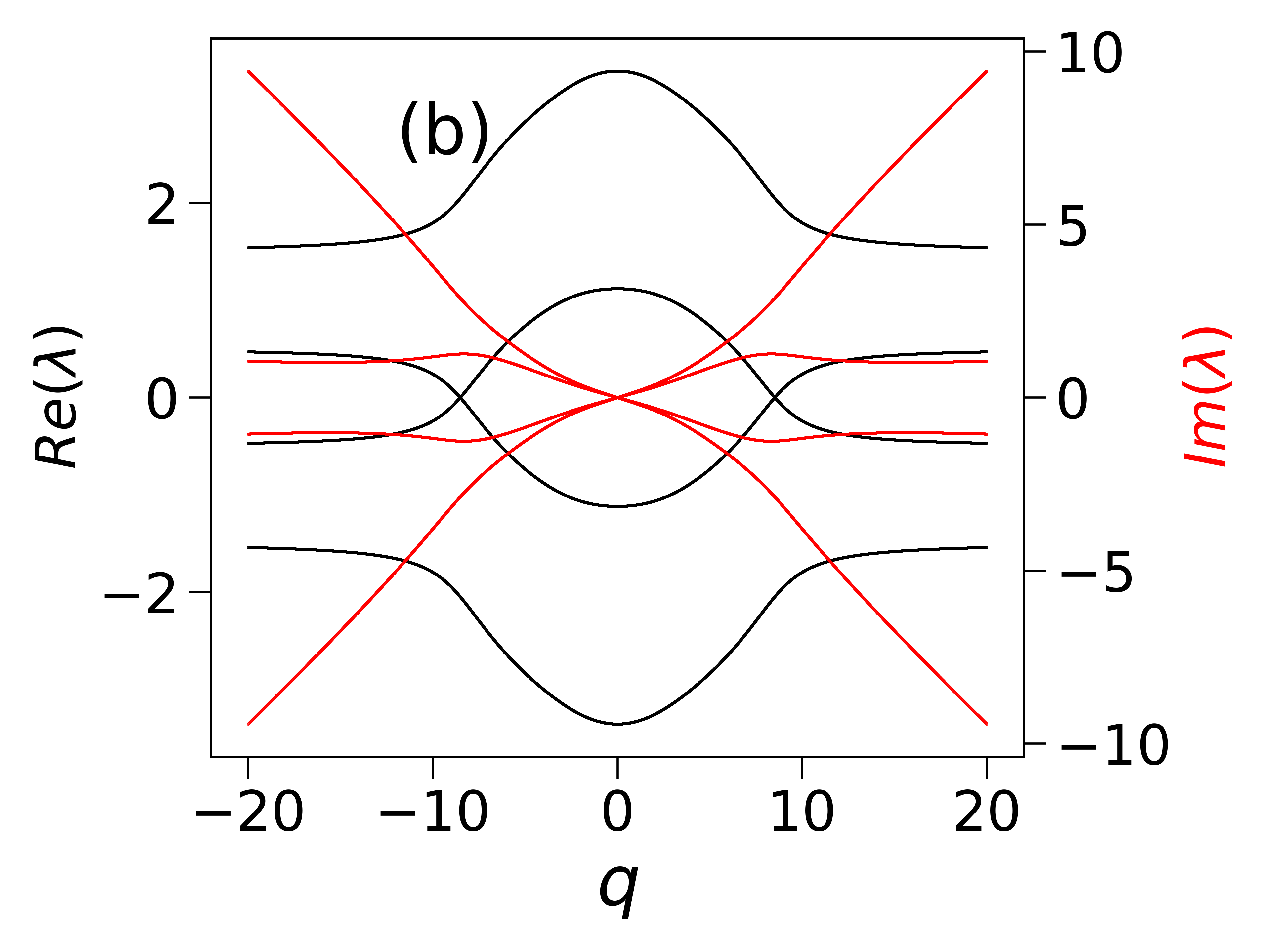}
    \includegraphics[width=0.23\linewidth]{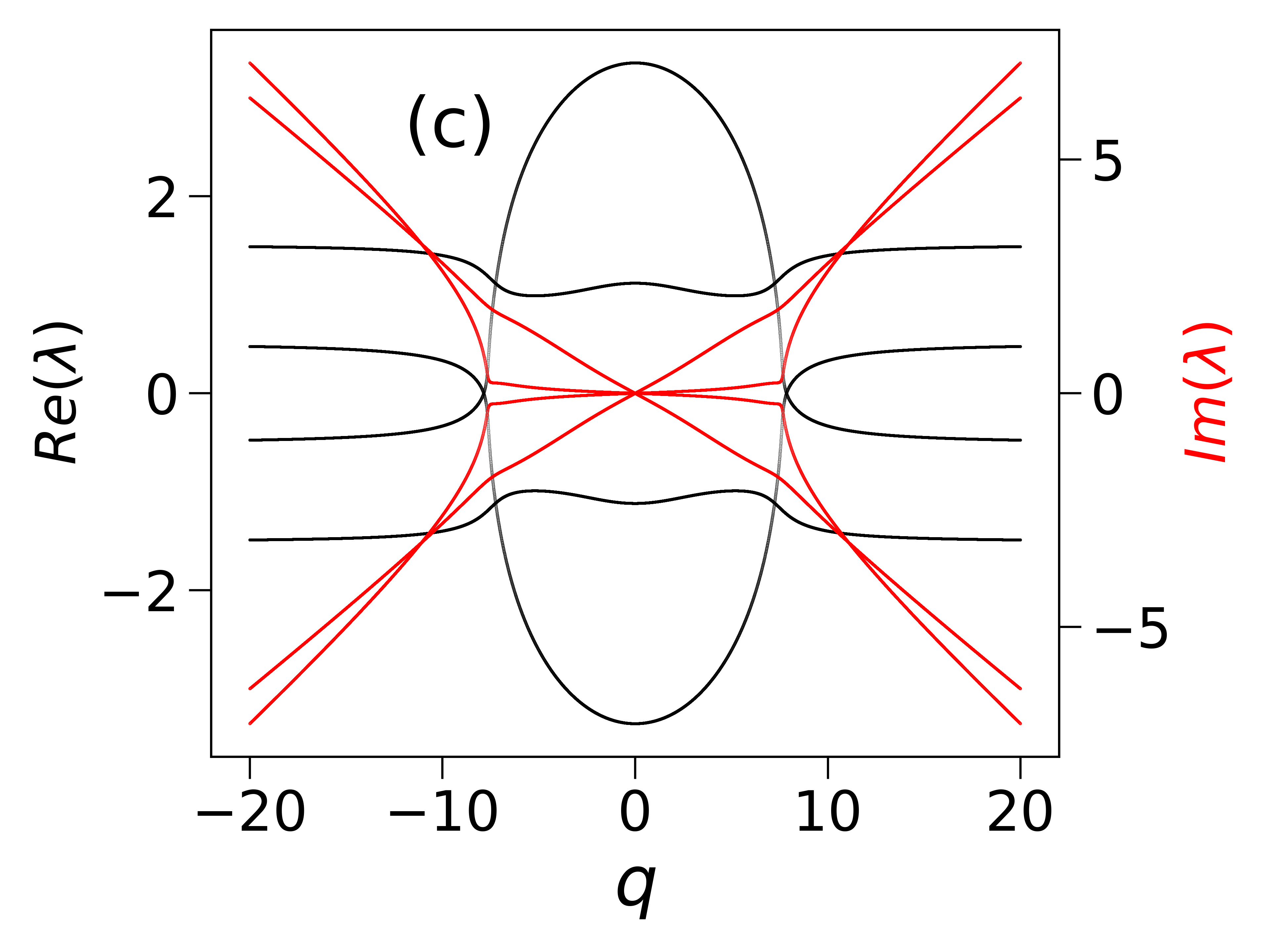}
    \includegraphics[width=0.23\linewidth]{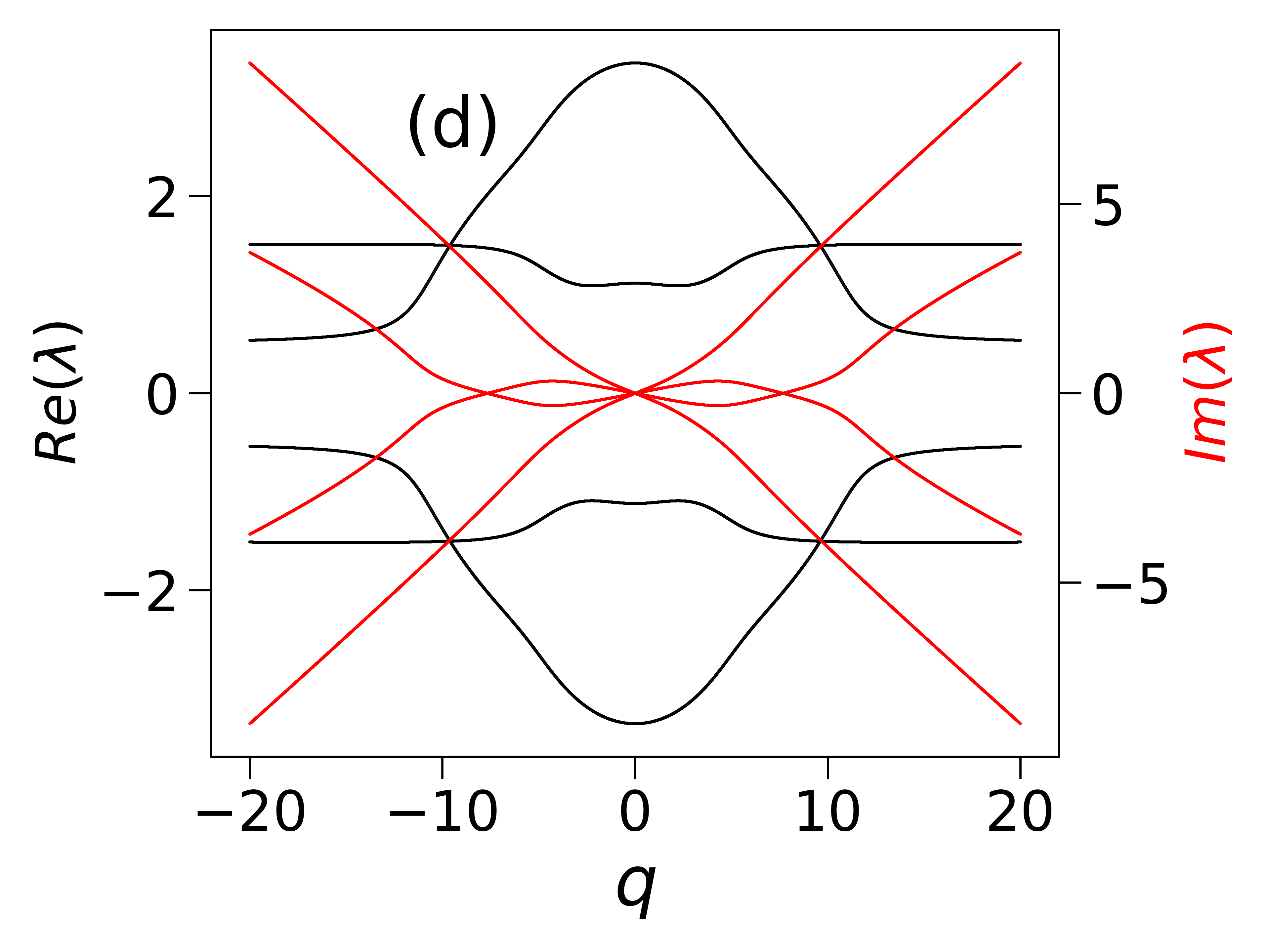}
    \caption{Real and imaginary parts of the eigenvalues of the effective Hamiltonian as a function of parameter $q$ for transformations \eqref{eq-lines-transform} (omitting the term proportional to the identity) with $\theta\in \{0,\,3\pi/8,\,5\pi/8,\,6\pi/8\}$ in figures (a-d), respectively. On-site interaction strengths are $U_1=-U_2=0.5$.}
    \label{fig-lines-through-origin-U1=-U2=0p5}
\end{figure}

Last we investigate the inhomogeneous on-site interaction defined by the Hermitian Hamiltonian
\begin{equation}
    H=-J(a_1^\dagger a_2 + a_2^\dagger a_1) + \frac{U_1}{2}a_1^\dagger a_1^\dagger a_1 a_1 + \frac{U_2}{2}a_2^\dagger a_2^\dagger a_2 a_2
    \label{eq:H_non_homogeneous}
\end{equation}
with $U_1 = -U_2\ne 0$. Negative on-site interaction can be achieved in ultracold atoms confined in optical lattices through Feshbach resonances \cite{Timmermans1999,Cornish2006,Chin2010}. Such Hamiltonian is no longer $\mathcal{PT}$-symmetric, although it still has the $\mathcal{PD}$-symmetry defined in Eq. \eqref{eq-chiral-symmetry}. We now study one-dimensional parameter trajectories instead of loops. In Figs. \ref{fig-lines-through-origin-U1=-U2=0p1} and \ref{fig-lines-through-origin-U1=-U2=0p5} we plot the spectrum of the non-Hermitian Hamiltonian obtained from Eq.~\eqref{eq:H_non_homogeneous} with the transformation  (\ref{eq-lines-transform}); the interaction strengths are chosen to be $U_1=-U_2=0.1$ and $U_1=-U_2=0.5$. The symmetry $\lambda\mapsto -\lambda$ in the spectrum originates from the $\mathcal{PD}$-symmetry. Comparing Fig.~\ref{fig-lines-through-origin-U1=-U2=0p1} to the vanishing interaction case in Fig.~\ref{fig-lines-through-origin}, we notice that a relatively small interaction breaks the $\mathcal{PT}$-symmetry drastically, and the way the degeneracies break depends significantly on the angle $\theta$ used in the transformation. In Fig.~\ref{fig-lines-through-origin-U1=-U2=0p5}, the differences are predictably larger, and we notice that for some values of the angle $\theta$ the separation between the imaginary values of the eigenvalue of the least decaying eigenstate and next to least decaying eigenstate grows quicker with growing $|q|$. Numerics shows that the least decaying eigenstate is different for $q<0$ and $q>0$. This motivates us to design parameter trajectories along such lines to obtain chiral state transfer quicker with relatively small changes to the parameters. 

\begin{figure}[t]
    \includegraphics[width=0.23\linewidth]{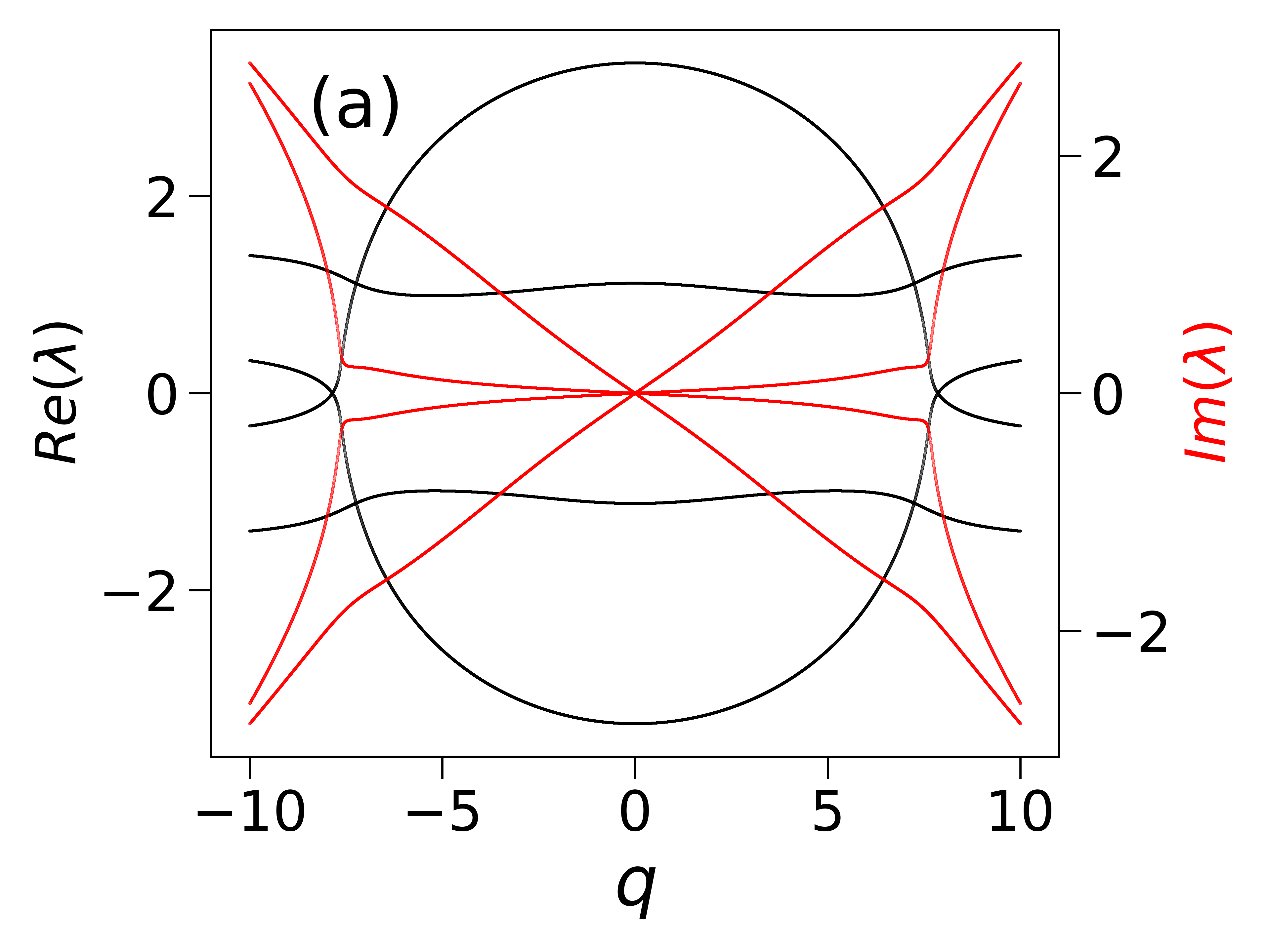}
    \includegraphics[width=0.23\linewidth]{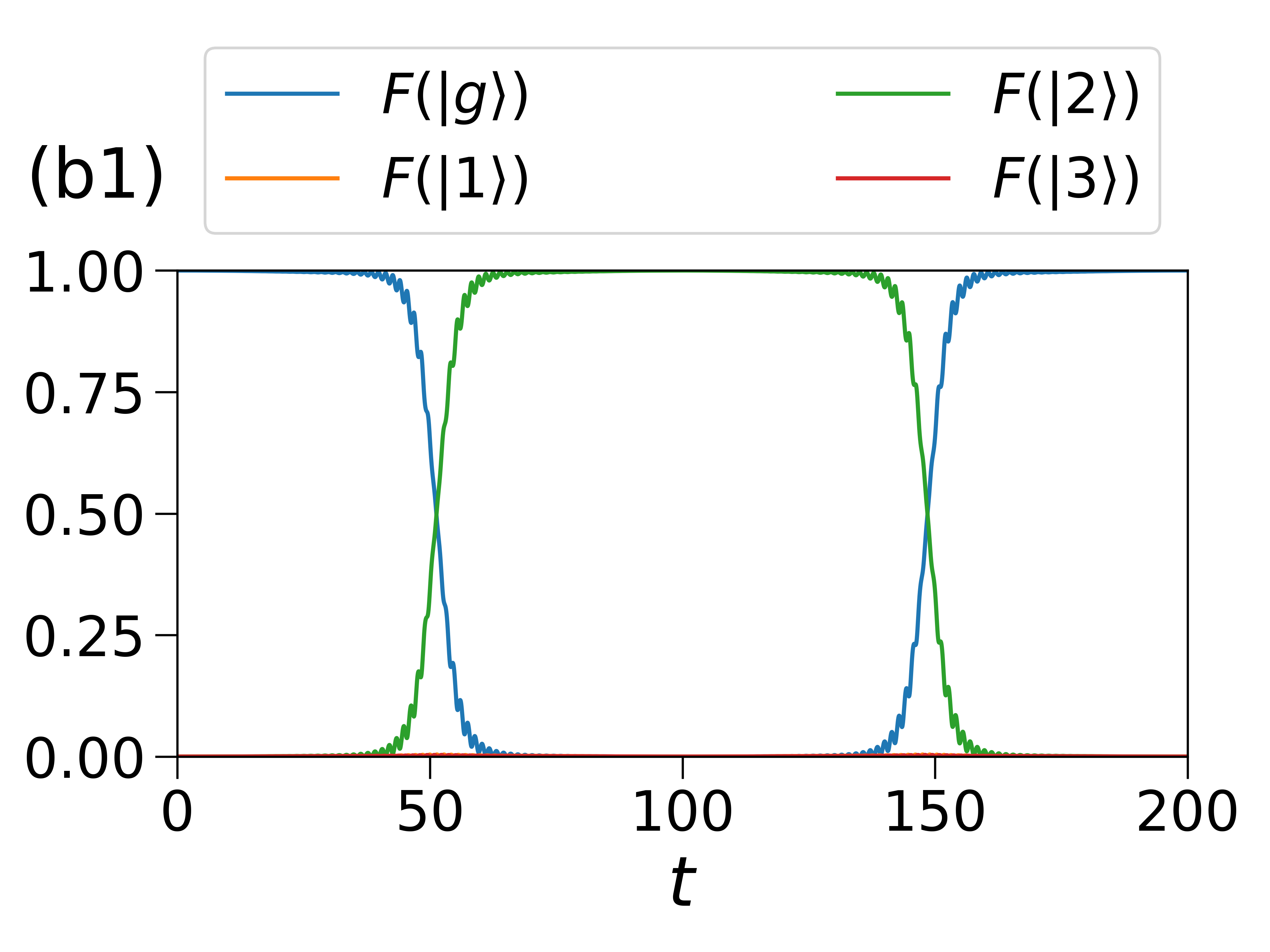}
    \includegraphics[width=0.23\linewidth]{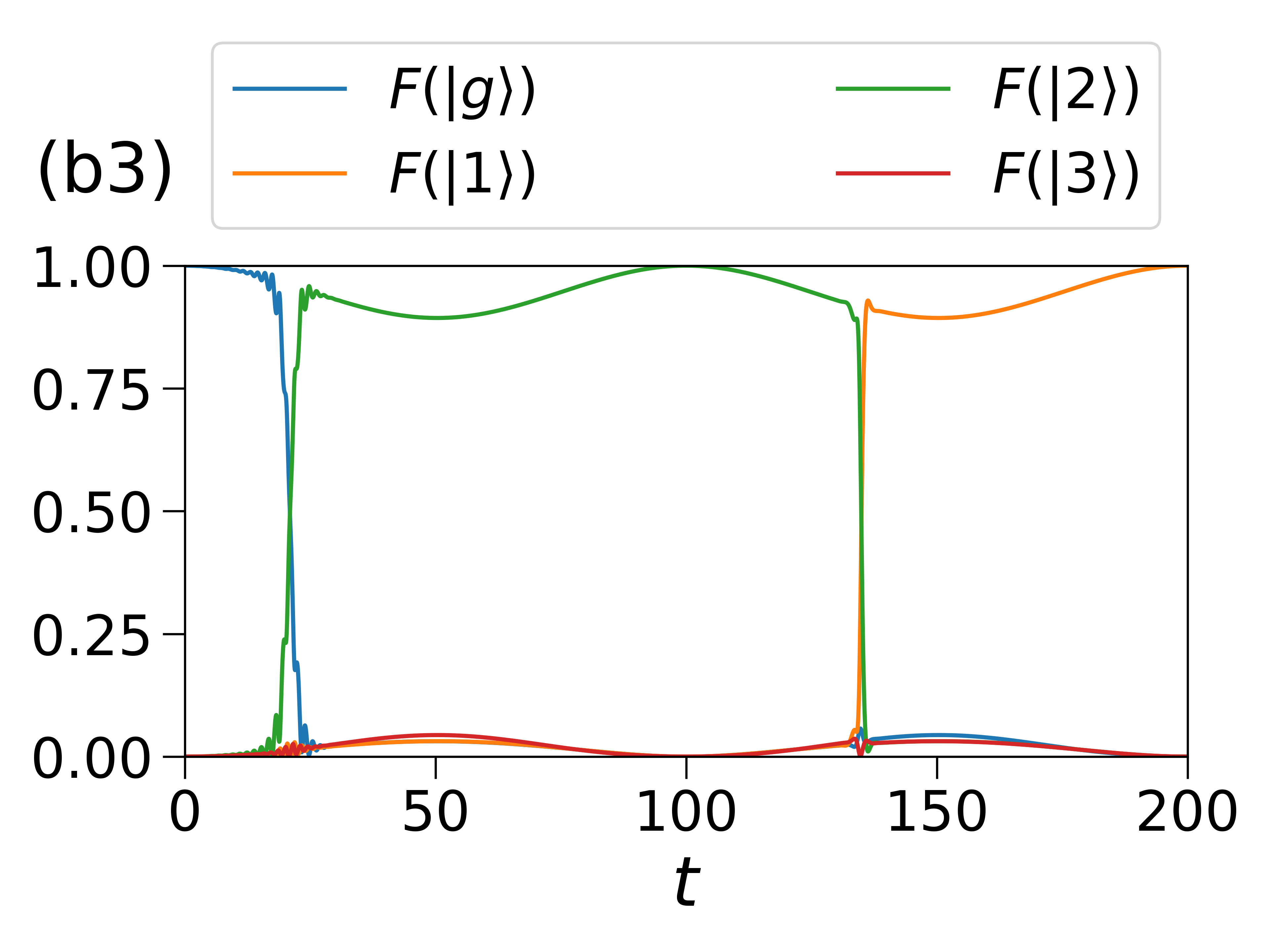} \\
    \hspace*{0.231\linewidth}
    \includegraphics[width=0.23\linewidth]{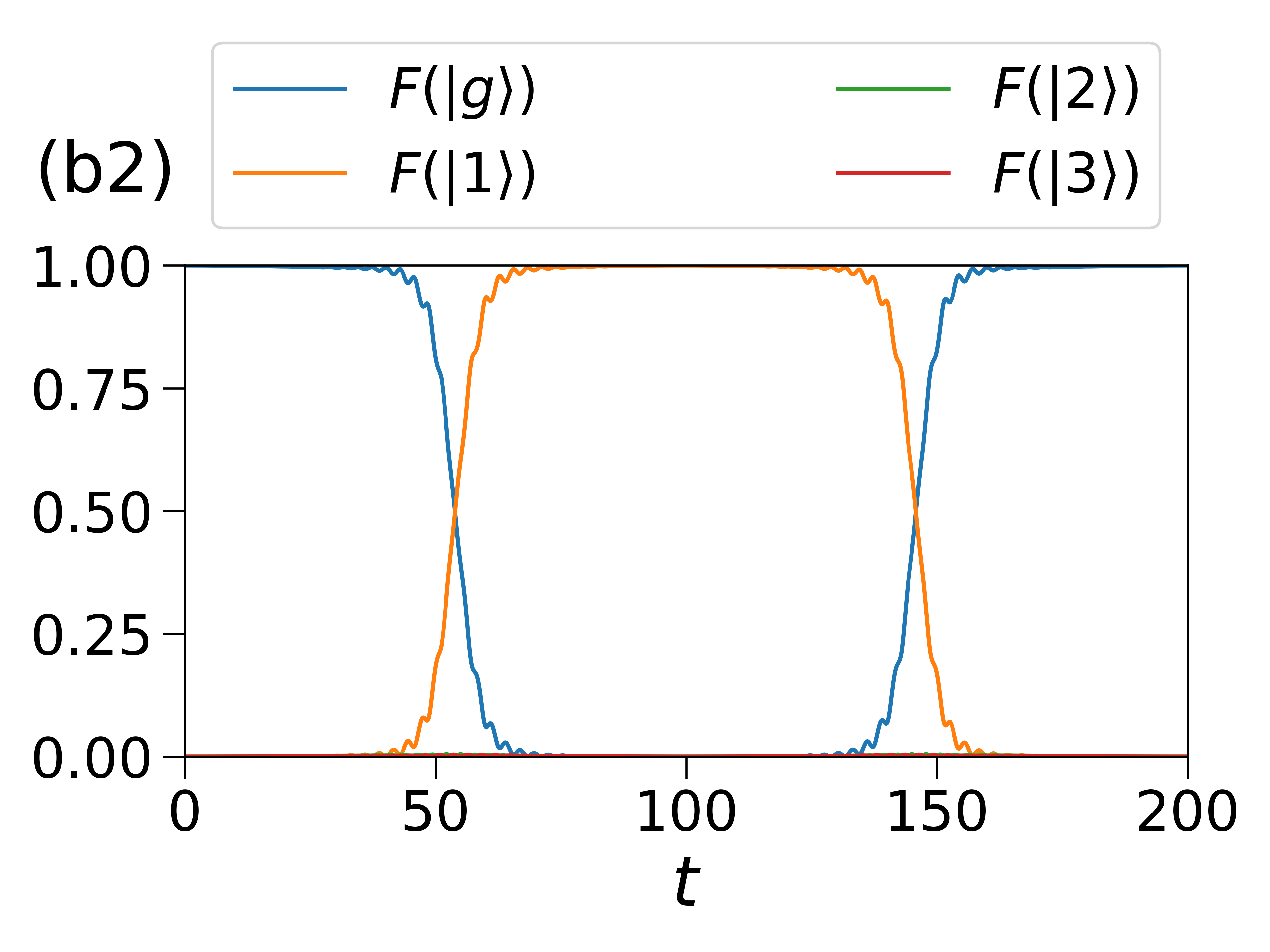}
    \includegraphics[width=0.23\linewidth]{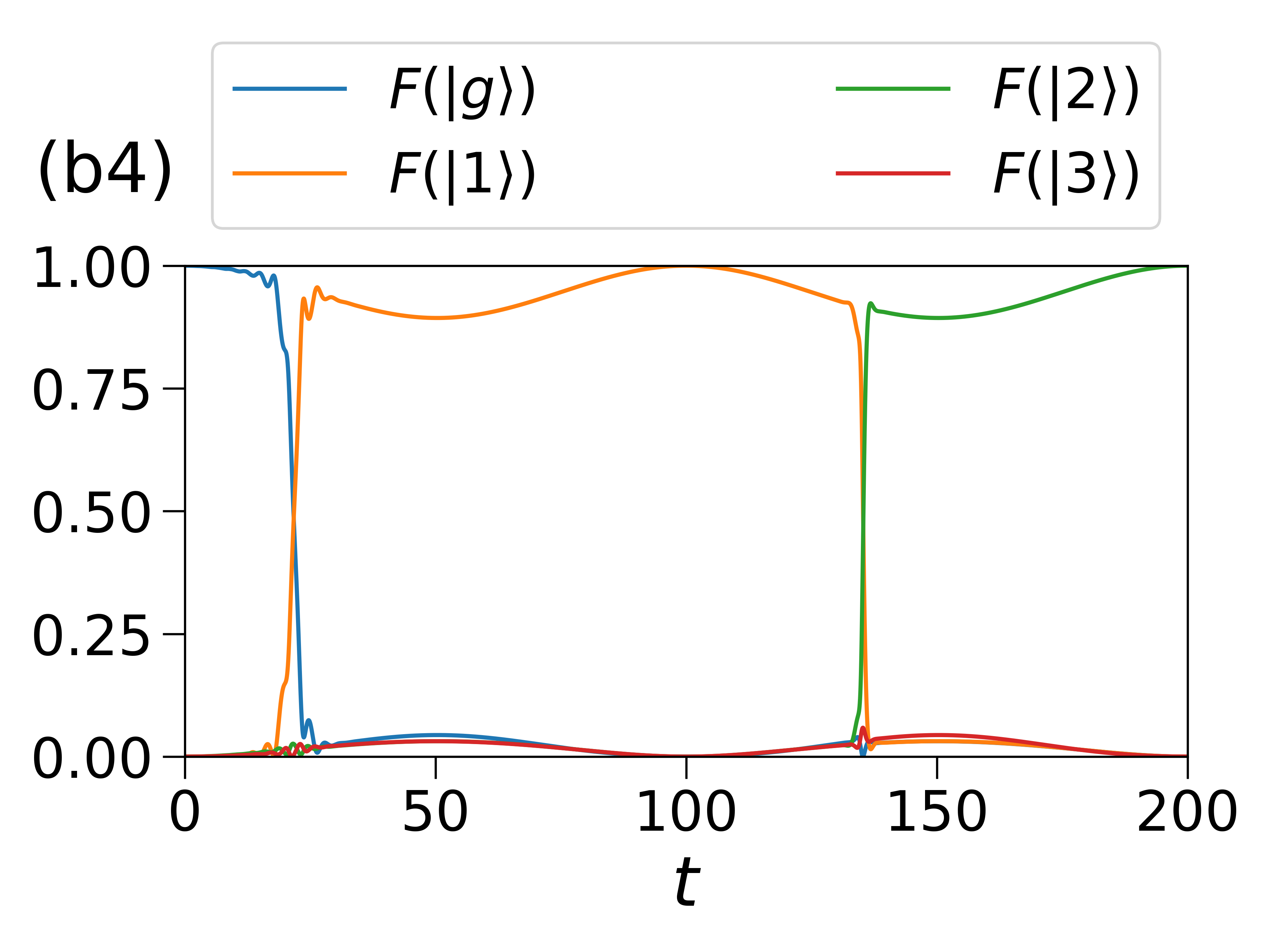} \\
    \hspace*{0.231\linewidth}
    \includegraphics[width=0.23\linewidth]{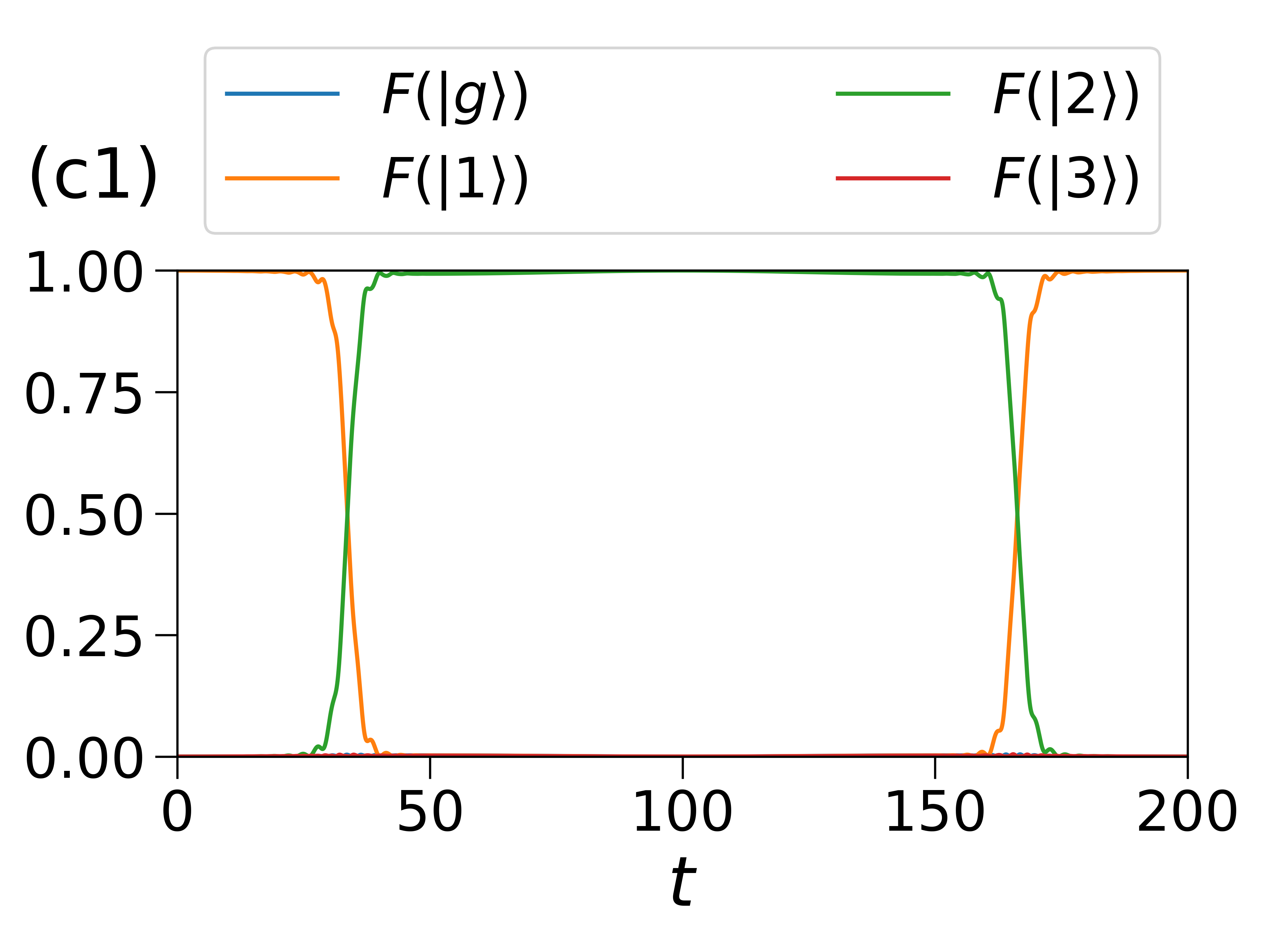}
    \includegraphics[width=0.23\linewidth]{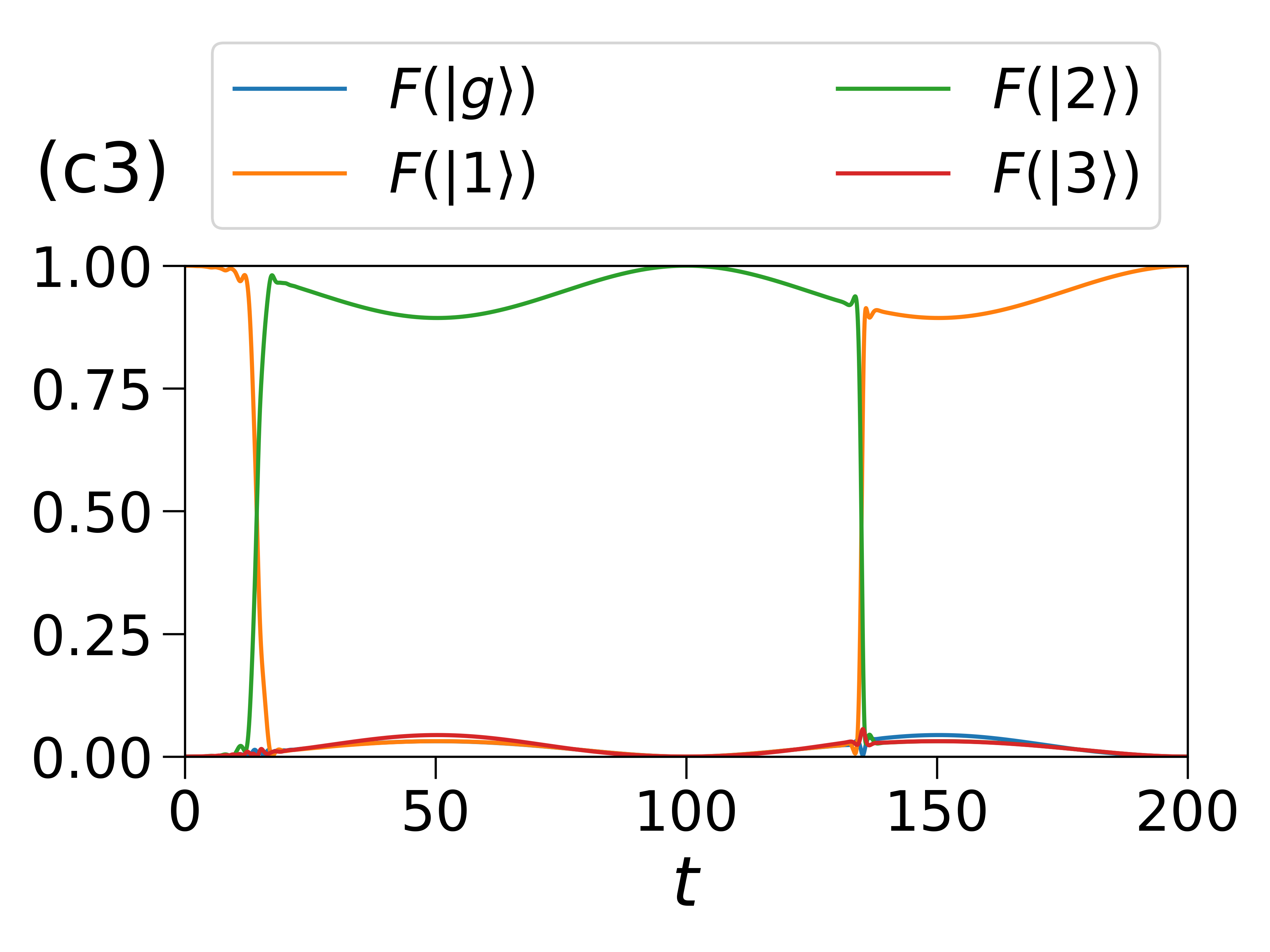} \\
    \hspace*{0.231\linewidth}
    \includegraphics[width=0.23\linewidth]{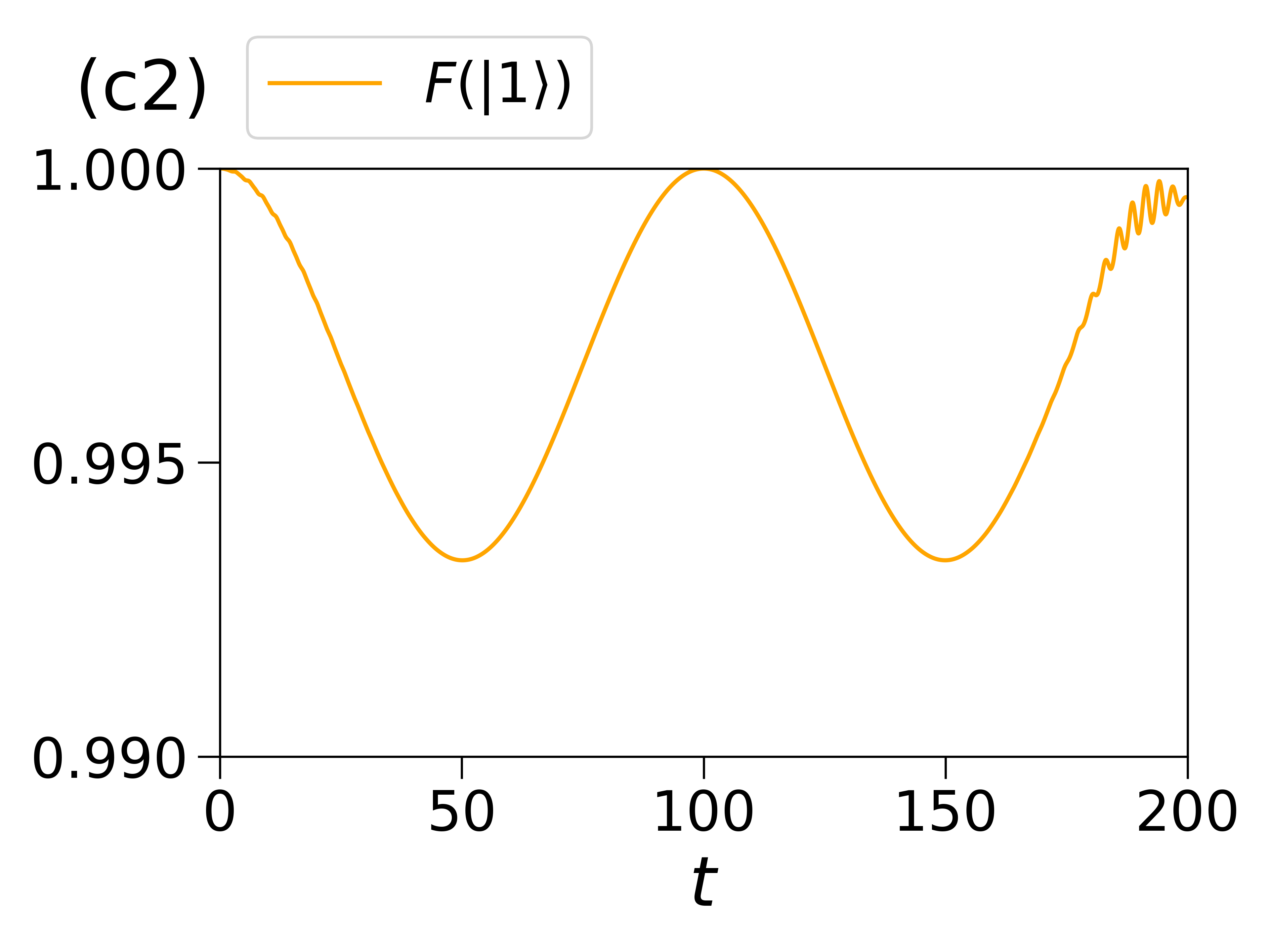}
    \includegraphics[width=0.23\linewidth]{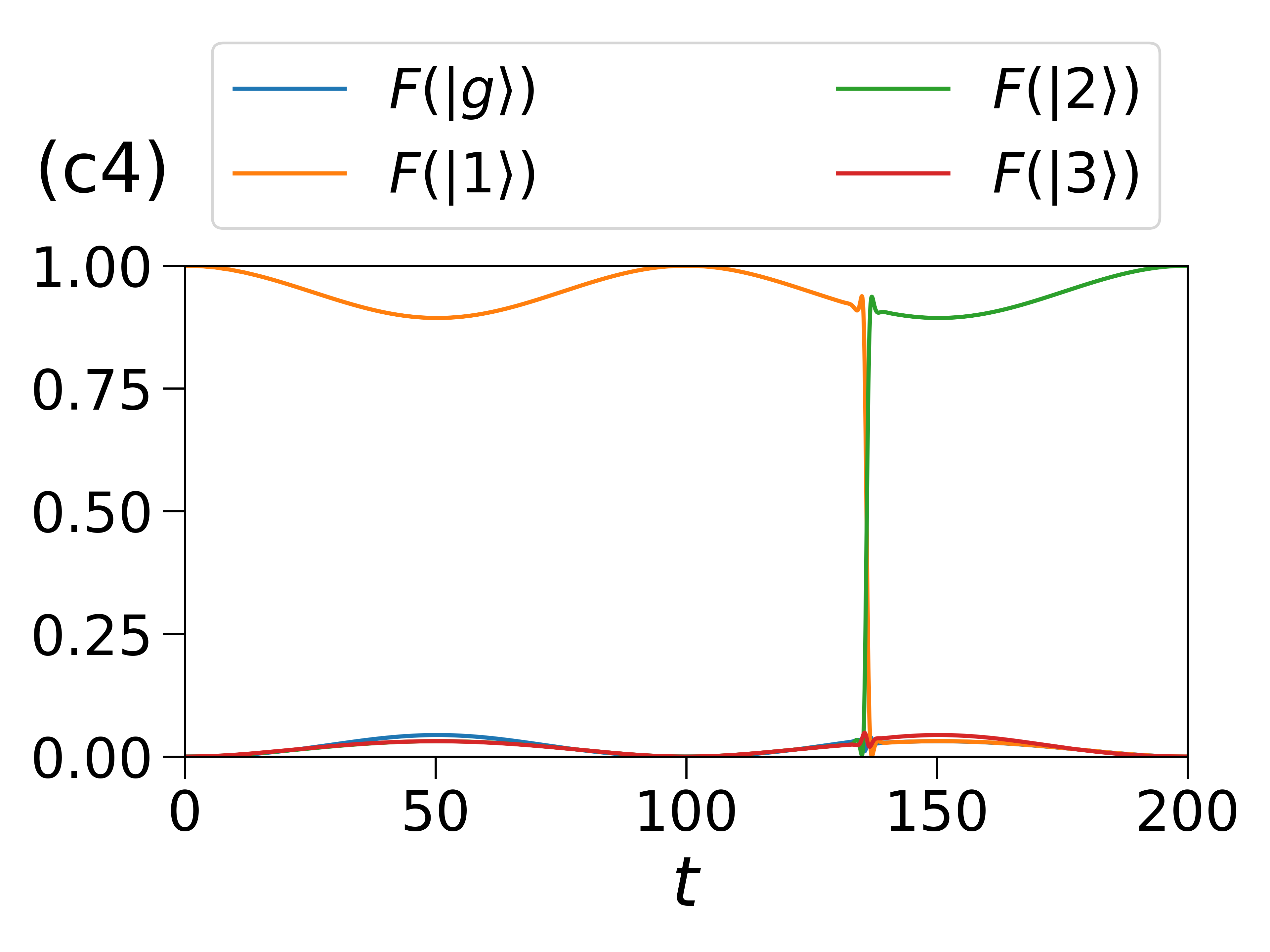}
    \caption{Deterministic evolution of fidelities of the eigenstates of the initial Hamiltonian using time-dependent transformation \eqref{eq-U1=-U2-transformation} with $\theta=5\pi /8$. $U_1=-U_2=0.5$, $T=200$. Initial state in figures (b1-b4) is the ground state of the initial Hamiltonian and in figures (c1-c4) it is the first excited state of the Hamiltonian. Value of $B$ is 1 on the left column and 4 on the right column. Figures numbered 1 and 3 use $+$-sign and figures numbered 2 and 4 use $-$-sign. Notice different ordinate on figure (c2).}
    \label{dephasing-detev-loops-T=all-U1=-U2-fid}
\end{figure}

We study parameter trajectories of the form
\begin{align}
    C_{B,\theta}(t)=B\sin(\pm2\pi t/T)\left[\cos(\theta)(3/\sqrt{10}L_{30,03}+1/\sqrt{10}L_{21,12})+\sin(\theta)(-1/\sqrt{10}L_{30,03}+3/\sqrt{10}L_{21,12})\right] + h(t)I,\label{eq-U1=-U2-transformation}
\end{align}
where constant $B$ controls the length of the path, the sign controls the traversal direction and the angle $\theta$ sets the orientation of the path and is fixed to $\theta=5\pi/8$. The function $h(t)$ is again chosen such that the rate operator is positive semi-definite, but it has no effect on the dynamics. These trajectories are no longer simple loops and the only relevant quality of the time-dependent Hamiltonian is the least decaying instantaneous eigenstate. To study traversal direction dependent state transfer, segments of lines in the parameter space $(l,\,k)$ will suffice. In Fig.~\ref{dephasing-detev-loops-T=all-U1=-U2-fid} we plot the fidelities of the time evolving state with respect to the eigenstates of the initial Hamiltonian $K(0)$ during the time evolution using the transformation (\ref{eq-U1=-U2-transformation}). The initial states are the ground state and the first excited state of the initial Hamiltonian. When the parameter trajectory returns to the starting point for the first time at $t=T/2$, the system reaches the same state for all initial states, the eigenstate related to the least decaying instantaneous eigenstate of the Hamiltonian $K(t)$ in the interval $0<t<T/2$, independently of the parameter $B$. But depending on the value of $B$, the final state after finishing the other half of the evolution, the system either returns to the state it started in or reaches the eigenstate related to the least decaying instantaneous eigenstate of the Hamiltonian $K(t)$ in the interval $T/2<t<T$. Instead of changing $B$, the similar results are obtained when changing $T$. We note that state transfer without encircling an EP has been observed also in \cite{Hassan2017,Svegborn2026}.

The main factor that affects whether state transfer happens or not is non-adiabatic transport between instantaneous eigenstates. For larger values of $B$ we expect to have larger non-adiabatic effects as the Hamiltonian is varying faster for smaller $T/A$ and the Hamiltonian has a larger non-Hermitian contribution for some parts of the trajectory when $B$ is larger. We can see in Fig.~\ref{dephasing-detev-loops-T=all-U1=-U2-fid} that when $B$ is large enough, state transfer does indeed occur. This is due to the non-adiabatic transport between the instantaneous eigenstate populations which eventually leads to the instantaneous least decaying eigenstate gaining more population over the initial state. After testing different values for $T$ and $B$, it was noticed that for large enough $B\approx4$ even small $T\approx10$ leads to state transfer, while small $B\approx1$ always requires large values of $T\gtrsim300$. As mentioned in \cite{Kumar2026}, errors originating from finite timestep $dt$ also lead to population transfer to the least decaying eigenstate. This error then grows exponentially and can sometimes have a significant contribution to state transfer. The simulations here were checked with different values of $dt$ and they were consistent with the results presented. Moreover, as noted in \cite{Kumar2026}, it is not necessarily important to find quantitatively accurate values for required value $T$ to see chiral state transfer in a theoretical model as any physical system would include perturbations that would contribute to population transfer to the least decaying state and thus speed up the state evolution to the least decaying state.

We have additionally studied rate operator transformations when the jump operators are $L_i=a_i$. $\mathcal{PT}$-symmetry in this case was not found to be relevant, because terms proportional to the annihilation operators $a_i$ have no effect on the spectrum the Hamiltonian. Instead changes in the dynamics are explained by mixing of eigenstates between different particle sectors. Results for this case are presented in Appendix \ref{Partdiss}.

\section{Conclusion and discussion \label{CandD}}

We have applied the rate operator quantum jumps formalism to engineer the non-Hermitian dynamics of the open Bose-Hubbard model with three particles subject to non-local dephasing. We have demonstrated that the rate operator transformations can be used to design time-dependent $\mathcal{PT}$-symmetric non-Hermitian Hamiltonians that can freely traverse part of the parameter space of the Hamiltonian, passing between unbroken and broken $\mathcal{PT}$-symmetry phases and circling exceptional points over time. This also shows that instead of varying parameters of system or the environment, one can instead vary the measurement scheme to form time-dependent non-Hermitian Hamiltonians.

We described the symmetry structure of the Bose-Hubbard model with two lattice sites and three particles with a general form of a diagonal $\mathcal{PT}$-symmetric non-Hermitian part for non-interacting and interacting cases. Specifically the non-interacting case has an interesting phase structure that was found to lead to chiral state transfer in closed loops, in which the final state was a different superposition of the eigenstates of the Hermitian Bose-Hubbard Hamiltonian depending on the traversal direction of the loop. Moreover, the final state was found to be dependent on the initial state on some parameter trajectories, which can be explained with the larger number of symmetries in the spectrum of the non-Hermitian Hamiltonian without on-site interaction. For non-zero interaction the region of unbroken $\mathcal{PT}$-symmetry was found to become narrower in one direction of the studied parameter space and stay larger in the perpendicular direction. Moreover, the final state at the end of the loop stayed close to the least decaying state of the broken $\mathcal{PT}$-symmetry phase even for large total evolution times.

In the case of inhomogeneous on-site interactions $U_1=-U_2$ we studied sinusoidal single-parameter trajectories over one period along a line in parameter space. There were no exceptional points on the parameter trajectory and only one crossing of the imaginary parts of the eigenvalues. Analogous behavior to chiral state transfer when traversing the trajectory in different directions was found in the model studied here. Whether this dynamical direction-dependence emerges was found to depend on the length of the trajectory traveled and on the total time of the evolution.

In general, we have shown that diverse modifications of non-Hermitian dynamics can be accomplished through switching of the measurement scheme defined by an unraveling of the master equation while keeping the physical parameters fixed. Here we used the description given by the rate operator quantum jumps unraveling. This opens options for future theoretical and experimental research in control of non-Hermitian dynamics using different measurement schemes instead of varying physical parameters. An interesting and more complete treatment of the rate operator quantum jumps would also include the quantum jumps in the evolution and an interpolation between the Hamiltonian dynamics and the complete GKSL dynamics.

\section*{Acknowledgements}
J.L. gratefully acknowledges financial support from Research Council of Finland through
project 353772. K.L. and J.L. gratefully acknowledge financial support from the Emil Aaltonen Foundation. J.L. thanks Eva and Niilo Pakkala Fund for travel grants.
F.S. acknowledges support from Magnus Ehrnrooth Foundation.

\bibliography{Bibliography}

\pagebreak
\appendix

\section{Particle dissipation\label{Partdiss}}

Here we study the effect of rate operator transformations when the system undergoes particle dissipation to the environment defined by Lindblad operators $L_i=a_i$. We use the transformation $C=k(a_1-a_2) + hI$ and fix $N=2$. The master equation is now of the form
\begin{align}
    \dot{\rho} = -i[H,\rho] + \gamma\sum_{i=1}^2 \left(a_i \rho a_i^\dagger - \frac{1}{2}(a_i^\dagger a_i\rho + \rho a_i^\dagger a_i)\right),
\end{align}
where the Hamiltonian $H$ is the same as in the main text. We set $U=0$, $J=1$, $\gamma=0.1$ and $h=|k|^2/(2\gamma)$, where $h$ is chosen again to be the smallest value that keeps the rate operator positive-semidefinite. The non-Hermitian Hamiltonian reads as
\begin{align}
    K=-J(a_1^\dagger a_2 + a_2^\dagger a_1) - \frac{i}{2}\gamma(n_1+n_2) - \frac{i}{2}\left(k(a_1-a_2) + hI\right), \label{eq-dissipation-hamiltonian}
\end{align}

\begin{figure}[t]
    \centering
    \includegraphics[width=0.4\linewidth]{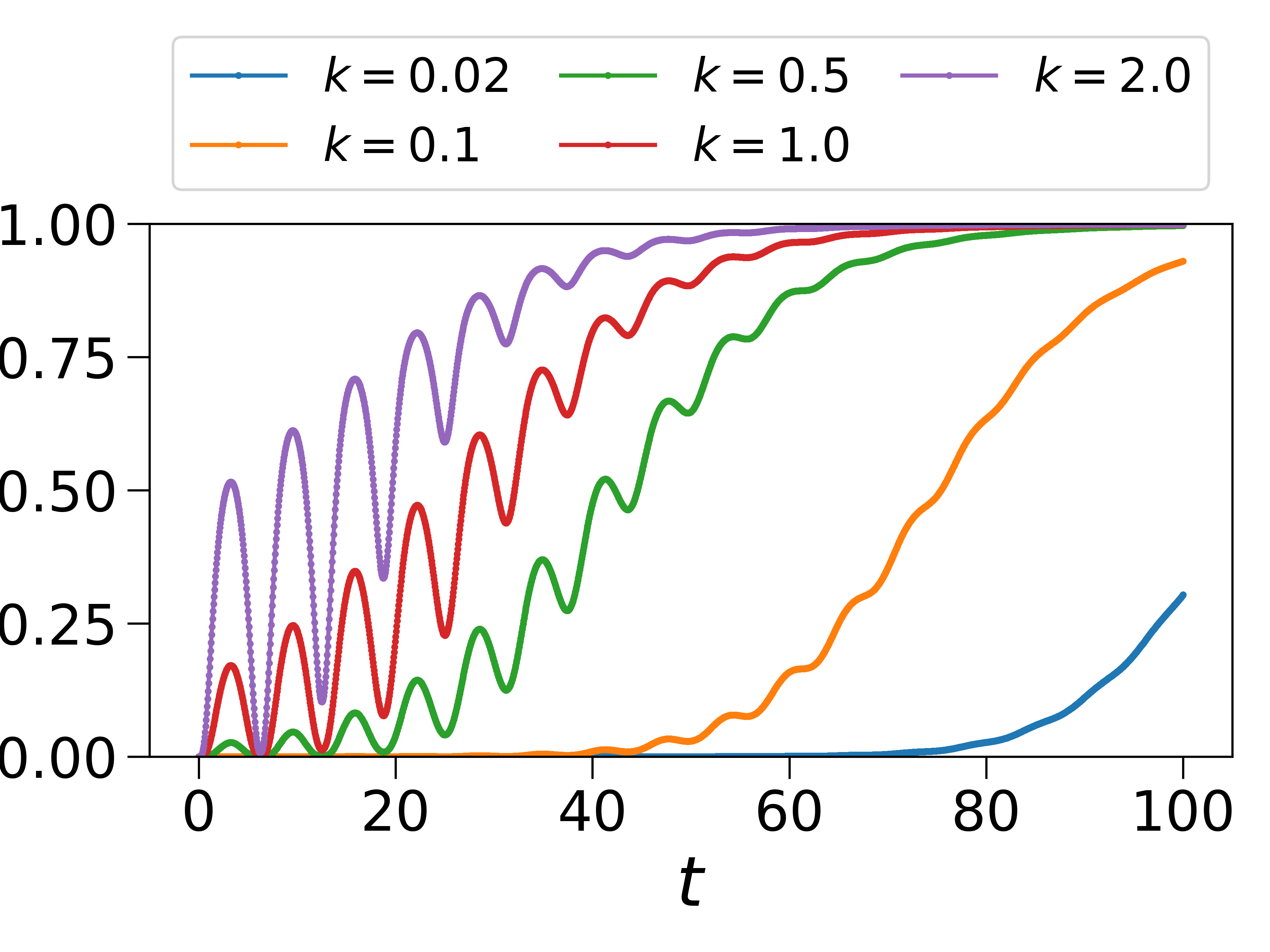}
    \caption{The absolute square of amplitude $\alpha_{00}$ for different values of the parameter $k$ in Eq. \eqref{eq-dissipation-hamiltonian}. $U=0$, $\gamma=0.1$ and the initial state is $\ket{\psi_0}=\ket{20}$.}
    \label{L=a-vacuum}
\end{figure}

In subspaces of fixed total number of excitations $N$, the term $\propto n_1+n_2$ acts as $NI$, so that it reduces the norm quicker from eigenstates with larger expectation values of total particle number. Without the rate operator transformations any superposition that includes a non-zero vacuum population would over time see its vacuum population monotonically increase to 1. Also, any terms linear in annihilation operators $a_i$ (them being strictly lower triangular) do not change the eigenvalues of a (finite dimensional) block diagonal matrix, but does in general change the eigenvectors by mixing different particle sectors. We can see the consequences of this in Fig.~\ref{L=a-vacuum}, where we plot the normalized evolution of the vacuum population generated by the Hamiltonian \eqref{eq-dissipation-hamiltonian}. We can see that although the Hamiltonian does not directly include creation operators the vacuum population oscillates. In terms of the eigenstates of the Hamiltonian \eqref{eq-dissipation-hamiltonian} this can be explained as interference between the different eigenstates that include non-zero amplitude for the vacuum.

\end{document}